\documentstyle[aps,prd]{revtex}
\begin{document}
\newcommand{\nc}{\newcommand}
\nc{\beq}{\begin{equation}}
\nc{\eeq}{\end{equation}}
\nc{\bea}{\begin{eqnarray}}
\nc{\eea}{\end{eqnarray}}
\nc{\ba}{\begin{array}}
\nc{\ea}{\end{array}}
\nc{\nn}{\nonumber}
\nc{\bpi}{\begin{picture}}
\nc{\epi}{\end{picture}}
\nc{\scs}{\scriptstyle}
\nc{\ts}{\textstyle}
\nc{\ds}{\displaystyle}
\nc{\unit}{{\mbox{\boldmath\large $1$}}}
\nc{\half}{{\ts\frac{1}{2}}}
\nc{\jb}{\bar{J}}
\nc{\jh}{\hat{J}}

\nc{\al}{\alpha}
\nc{\Ga}{\Gamma}
\nc{\de}{\delta}
\nc{\De}{\Delta}
\nc{\la}{\lambda}
\nc{\p}{\partial}

\begin{flushright}
\normalsize
HD-THEP-99-33\\
hep-th/9908172\\
August 1999\bigskip\\
\end{flushright}

\begin{center}
{\large\bf Recursive Graphical Construction of Feynman Diagrams\\
in $\phi^4$ Theory: Asymmetric Case and Effective Energy}\bigskip\bigskip\\
Boris Kastening\bigskip\\
\it Institut f\"ur Theoretische Physik\\
\it Universit\"at Heidelberg\\
\it Philosophenweg 16\\
\it D-69120 Heidelberg\\
\it Germany\bigskip\bigskip\\
\end{center}

\begin{center}
{\bf Abstract}\bigskip\\
\begin{minipage}{16cm}
The free energy of a multi-component scalar field theory is considered
as a functional $W[G,J]$ of the free correlation function $G$ and an
external current $J$.
It obeys non-linear functional differential equations which are turned
into recursion relations for the connected Greens functions in
a loop expansion.
These relations amount to a simple proof that $W[G,J]$ generates only
connected graphs and can be used to find all such graphs with their
combinatoric weights.

A Legendre transformation with respect to the external current converts
the functional differential equations for the free energy into those for
the effective energy $\Ga[G,\Phi]$, which is considered as a functional
of the free correlation function $G$ and the field expectation $\Phi$.
These equations are turned into recursion relations for the one-particle
irreducible Greens functions.
These relations amount to a simple proof that $\Ga[G,J]$ generates only
one-particle irreducible graphs and can be used to find all such graphs
with their combinatoric weights.

The techniques used also allow for a systematic investigation into
resummations of classes of graphs.
Examples are given for resumming one-loop and multi-loop tadpoles,
both through all orders of perturbation theory.

Since the functional differential equations derived are non-perturbative,
they constitute also a convenient starting point for other expansions
than those in numbers of loops or powers of coupling constants.

We work with general interactions through four powers in the field.
\end{minipage}
\bigskip\bigskip\\
\end{center}

\section{Introduction}
The free energy of a statistical or quantum field theory may be 
viewed as a functional of the free correlation functions.
It obeys functional differential equations which may be converted
into recursion relations for the connected vacuum graphs of the theory.
Subsequently, functional derivatives of $W$ with respect to the free
propagators or their inverses can be taken to generate the Feynman
diagrams of all connected Greens functions.
This program was developed a long time ago by Kleinert
\cite{Kleinert1,Kleinert2}, but used only recently for a systematic
generation of all Feynman diagrams of a multi-component
$\phi^4$- and $\phi^2A$-theory \cite{KPKB}, and of QED \cite{BKP}.
For $\phi^4$ theory, only the symmetric case was treated.

However, both in statistical physics and particle theory, this symmetry 
is often broken. 
For this reason we generalize the symmetric treatment of \cite{KPKB}, 
and allow for interactions of all powers of the field through four.
We introduce an external source $J$ to be able to generate also Greens 
functions with odd numbers of external legs as derivatives of $W$.
In contrast to \cite{KPKB}, this also enables us to generate connected
Feynman diagrams for the $n$-point functions through $L$ loops without
having to generate any diagrams with more than $L$ loops first.
As a byproduct, we get an alternative proof to the one found in 
\cite{Kleinert1,Kleinert2} that $W$ generates only connected Greens
functions.

We then Legendre transform the functional differential equations for
$W[G,J]$ into ones for the effective energy (or effective action in
quantum theory) $\Ga[G,\Phi]$ and derive from these recursion relations
for the one-particle irreducible (1PI) Feynman diagrams representing the
proper vertices of the theory.
No graphs beyond $L$ loops have to be considered to generate proper
$n$-point vertices through $L$ loops.
As a byproduct, we get an alternative proof that $\Ga[G,\Phi]$ generates
only 1PI Greens functions, similar to the one found in
\cite{Kleinert1,Kleinert2}.

By using $G$ as a functional argument, and, to the extent possible,
derivatives with respect to $G$ instead of $J$ or $\Phi$, we keep
the identities for $W$ and $\Ga$ and the recursion relations for the
connected and 1PI Greens functions simple.
In contrast to \cite{KPKB}, we do not use the technique of ``cutting''
free correlation functions, but always ``amputate'' them.
As in \cite{KPKB}, the graphical operations necessary to solve the
recursion relations can be implemented on a computer for an efficient
generation of higher order graphs.

Formally, we consider all our calculations for a statistical theory in
$d$ Euclidean dimensions, but with trivial changes of factors $i$, all
results are valid as well for a quantum field theory in Minkowski space
and for quantum mechanics.
In this work, where we often deal with more than one interaction term,
our ordering principle is always the number of loops and not powers
of coupling constants.

The structure of the paper is as follows:

In Section \ref{sym} we repeat the steps that led to a functional
identity for $W[G]$ and a recursion relation for its perturbative
coefficients in \cite{Kleinert1,Kleinert2,KPKB}.
This gives us the opportunity to specify our slightly different conventions.
Going beyond the considerations in \cite{KPKB}, we treat part of the
quadratic term as a perturbation.
This can be used to cancel one-loop tadpole corrections which drastically
reduces the number of vacuum graphs for the free energy, as utilized before
in \cite{4loops,5loops}.

In Section \ref{general} we treat the asymmetric case for the free energy
$W[G,J]$.
We derive identities for $W[G,J]$ and recursion relations for the Feynman
diagrams representing the connected Greens functions.

In Section \ref{effen} we translate the identities for $W[G,J]$ into
identities for the effective energy $\Ga[G,\Phi]$  and subsequently
derive recursion relations for the one-particle irreducible (1PI)
Feynman diagrams representing the proper vertices of the theory.
We finally present how part of the quadratic term can be treated as
a perturbation to cancel all tadpole corrections to propagators in
graphs needed for the proper vertices with one or more external legs,
thereby drastically reducing the number of diagrams.

Section \ref{summary} contains a summary of our results and an
outlook.

\section{Symmetric Case}
\label{sym}
\subsection{Definitions}
Consider a scalar field $\phi$ with $N$ components 
in $d$ Euclidean dimensions whose thermal fluctuations are controlled
by the energy functional
\beq
\label{esym2}
E[\phi,G,\De,L]
=
\frac{1}{2}\int_{12}\left(G^{-1}_{12}+\De_{12}\right)\phi_1\phi_2
+\frac{1}{24}\int_{1234}L_{1234}\phi_1\phi_2\phi_3\phi_4,
\eeq
where $L_{1234}$ is a self-coupling and where we keep the option open to
treat a part $\De_{12}$ of the quadratic term in $E$ as a perturbation.
The numerical indices of $\int$, $G^{-1}$, $\De$ and $L$ are meant
as a short-hand and represent spatial as well as tensorial arguments,
\beq
1\equiv\{x_1,\al_1\},\;\; 
\int_1\equiv\sum_{\al_1}\int d^dx_1,\;\;
\phi_1\equiv\phi_{\al_1}(x_1),
\eeq
\beq
G^{-1}_{12}\equiv G^{-1}_{\al_1\al_2}(x_1,x_2),\;\; 
\De_{12}\equiv\De_{\al_1\al_2}(x_1,x_2),\;\;
L_{1234}\equiv L_{\al_1\al_2\al_3\al_4}(x_1,x_2,x_3,x_4).
\eeq

For example, in standard $\phi^4$ theory we would have
\beq
G^{-1}_{\al_1\al_2}(x_1,x_2)=\de_{\al_1\al_2}\de(x_1-x_2)
\left(\p_1\cdot\p_2+m^2\right),
\eeq
\beq
\De_{\al_1\al_2}(x_1,x_2)=\de m^2\de_{\al_1\al_2}\de(x_1-x_2),
\eeq
\beq
L_{\al_1\al_2\al_3\al_4}(x_1,x_2,x_3,x_4)=\frac{1}{3}\la
(\de_{\al_1\al_2}\de_{\al_3\al_4}+\de_{\al_1\al_3}\de_{\al_2\al_4}
+\de_{\al_1\al_4}\de_{\al_2\al_3})
\de(x_1-x_2)\de(x_1-x_3)\de(x_1-x_4),
\eeq
where $\de m^2$ could represent a modification of $m^2$ that we want
to treat perturbatively.

Using natural units, where the Boltzmann constant $k_B$ times the
temperature $T$ equals unity, the partition function $Z$ and the
negative free energy $W$ are given by a functional integral over
the Boltzmann weight $\exp(-E[\phi])$,
\beq
Z[G,\De,L]=\exp(W[G,\De,L])=\int D\phi\exp(-E[\phi,G,\De,L]).
\eeq

\subsection{Identity for $W_I$}
Regarding $W$ as a functional of the kernel $G^{-1}$, we can derive a
functional differential equation for $W$.
Our starting point is the identity
\beq
\int D\phi\frac{\de}{\de\phi_1}\{\phi_2\exp(-E[\phi,G,\De,L])\}=0,
\eeq
which follows from functional partial integration and the vanishing of
the exponential at infinite fields.

Carrying out the $\phi_1$-derivative, replacing appearances of $\phi_i$
with appropriate derivatives with respect to $G^{-1}$ and finally using the
results of appendix \ref{symg} to translate all such derivatives into
derivatives with respect to $G$ yields an identity for $W$,
\bea
\label{w0symidentity}
\lefteqn{\de_{12}-2\int_3G_{23}\frac{\de W}{\de G_{13}}
-2\int_{345}\De_{13}G_{24}G_{35}\frac{\de W}{\de G_{45}}
-\frac{2}{3}\int_{345}L_{1345}
\left[\int_{67}\left(G_{23}G_{46}G_{57}+G_{26}G_{37}G_{45}\right)
\frac{\de W}{\de G_{67}}\right]}
\nn\\
&&
-\frac{2}{3}\int_{345}L_{1345}\left[
\int_{6789}G_{26}G_{37}G_{48}G_{59}\frac{\de^2W}{\de G_{67}\de G_{89}}
\right]
-\frac{2}{3}\int_{345}L_{1345}\left[
\int_{6789}G_{26}G_{37}\frac{\de W}{\de G_{67}}
G_{48}G_{59}\frac{\de W}{\de G_{89}}\right]=0.
\eea

Split $W$ into a free and an interacting part,
\beq
W=W_0+W_I\equiv W|_{\De,L=0}+W_I.
\eeq
For $W_0$, (\ref{w0symidentity}) reduces to
\beq
\label{w0functeq}
\de_{12}-2\int_3G_{23}\frac{\de W_0}{\de G_{13}}=0,
\eeq
so that, using also the results of appendix \ref{symg}, we get the
useful relations
\beq
\label{dw0dg}
\frac{\de W_0}{\de G_{12}}=\frac{1}{2}G_{12}^{-1}
\eeq
and
\beq
\label{d2w0dgdg}
\frac{\de^2W_0}{\de G_{12}\de G_{34}}
=-\frac{1}{4}\left(G_{13}^{-1}G_{24}^{-1}+G_{14}^{-1}G_{23}^{-1}\right).
\eeq
Up to an additive constant, which we assume to be adjusted to zero
by an appropriate normalization of the path integral measure $D\phi$,
$W_0$ itself is given as usual by
\beq
\label{w0h}
W_0[G]=-\frac{1}{2}\int_1(\ln G^{-1})_{11}
=\frac{1}{2}
\rule[-10pt]{0pt}{26pt}
\bpi(26,0)
\put(13,3){\circle{16}}
\epi,
\eeq
where he have introduced a graphical representation for $W_0$.

Subtracting (\ref{w0functeq}) from (\ref{w0symidentity}), using
(\ref{dw0dg}) and (\ref{d2w0dgdg}), setting $x_2=x_1$ and integrating
over $x_1$ gives a non-linear functional differential equation for $W_I$,
$$
\int_{12}G_{12}\frac{\de W_I}{\de G_{12}}
+\frac{1}{4}\int_{1234}L_{1234}G_{12}G_{34}
+\frac{1}{2}\int_{12}\De_{12}G_{12}
+\int_{1234}\De_{12}G_{13}G_{24}\frac{\de W_I}{\de G_{34}}
+\int_{123456}L_{1234}G_{12}G_{35}G_{46}\frac{\de W_I}{\de G_{56}}
$$
\beq
\label{wisymfuncteq}
+\frac{1}{3}\int_{12345678}L_{1234}G_{15}G_{26}G_{37}G_{48}
\frac{\de^2W_I}{\de G_{56}\de G_{78}}
+\frac{1}{3}\int_{12345678}L_{1234}G_{15}G_{26}G_{37}G_{48}
\frac{\de W_I}{\de G_{56}}\frac{\de W_I}{\de G_{78}}=0.
\eeq
For $\De=0$, this reduces to equation (2.58) in \cite{KPKB}.

To represent (\ref{wisymfuncteq}) graphically, write for the derivatives
of $W_I$ with respect to $G$
\beq
\frac{\de W_I}{\de G_{12}}=
\rule[-18pt]{0pt}{42pt}
\bpi(50,0)(-10,0)
\put(19,3){\circle{32}}
\put(19,3){\makebox(1,0){$W_I$}}
\put(-0.5,13){\makebox(0,0){$\scs 1$}}
\put(-0.5,-7){\makebox(0,0){$\scs 2$}}
\put(5.77,12){\circle*{4}}
\put(5.77,-6){\circle*{4}}
\epi,
\;\;\;\;
\frac{\de^2 W_I}{\de G_{12}\de G_{34}}=
\rule[-18pt]{0pt}{42pt}
\bpi(47,0)(-7,0)
\put(19,3){\circle{32}}
\put(19,3){\makebox(1,0){$W_I$}}
\put(6,21){\makebox(0,0){$\scs 1$}}
\put(0,15){\makebox(0,0){$\scs 2$}}
\put(0,-9){\makebox(0,0){$\scs 3$}}
\put(6,-15){\makebox(0,0){$\scs 4$}}
\put(5.22,11.4){\circle*{4}}
\put(10.87,17.1){\circle*{4}}
\put(5.22,-5.13){\circle*{4}}
\put(10.87,-10.78){\circle*{4}}
\epi
\eeq
and use the vertices
\beq
-\De_{12}=
\rule[-7pt]{0pt}{20pt}
\bpi(58,20)(-4,0)
\put(5,3){\line(1,0){15}}
\put(30,3){\line(1,0){15}}
\put(20,-2){\line(1,0){10}}
\put(20,-2){\line(0,1){10}}
\put(20,8){\line(1,0){10}}
\put(30,-2){\line(0,1){10}}
\put(25,3){\makebox(1,0){$\scs\De$}}
\put(1,4){\makebox(0,0){$\scs 1$}}
\put(49,4){\makebox(0,0){$\scs 2$}}
\epi,
\;\;\;\;
-L_{1234}=
\rule[-14pt]{0pt}{34pt}
\bpi(34,0)(-4,0)
\put(5,-5){\line(1,1){16}}
\put(5,11){\line(1,-1){16}}
\put(13,3){\circle*{4}}
\put(25,15){\makebox(0,0){$\scs 1$}}
\put(1,15){\makebox(0,0){$\scs 2$}}
\put(1,-9){\makebox(0,0){$\scs 3$}}
\put(25,-9){\makebox(0,0){$\scs 4$}}
\epi.
\eeq
Lines that are connected at both ends are propagators $G$.
All space arguments that are not indicated by numbers are integrated over.
Now (\ref{wisymfuncteq}) reads
\beq
\label{wisym}
\rule[-18pt]{0pt}{42pt}
\bpi(51.23,0)(-13.23,0)
\put(19,3){\circle{32}}
\put(19,3){\makebox(1,0){$W_I$}}
\put(5.77,12){\circle*{4}}
\put(5.77,-6){\circle*{4}}
\put(5.77,3){\oval(28,18)[l]}
\epi
=
\frac{1}{4}
\rule[-10pt]{0pt}{26pt}
\bpi(42,0)
\put(21,3){\circle*{4}}
\put(13,3){\circle{16}}
\put(29,3){\circle{16}}
\epi
+
\frac{1}{2}
\rule[-11pt]{0pt}{28pt}
\bpi(33,0)
\put(19,8){\oval(18,8)[t]}
\put(19,-2){\oval(18,8)[b]}
\put(28,-2){\line(0,1){10}}
\put(5,-2){\line(1,0){10}}
\put(5,-2){\line(0,1){10}}
\put(5,8){\line(1,0){10}}
\put(15,-2){\line(0,1){10}}
\put(10,3){\makebox(1,0){$\scs\De$}}
\epi
+
\rule[-18pt]{0pt}{42pt}
\bpi(58.23,0)(-18.23,0)
\put(19,3){\circle{32}}
\put(19,3){\makebox(1,0){$W_I$}}
\put(5.77,12){\circle*{4}}
\put(5.77,-6){\circle*{4}}
\put(5.77,8){\oval(28,8)[tl]}
\put(5.77,-2){\oval(28,8)[bl]}
\put(-13.23,-2){\line(1,0){10}}
\put(-13.23,-2){\line(0,1){10}}
\put(-13.23,8){\line(1,0){10}}
\put(-3.23,-2){\line(0,1){10}}
\put(-8.23,3){\makebox(1,0){$\scs\De$}}
\epi
+
\rule[-18pt]{0pt}{42pt}
\bpi(69.23,0)(-29.23,0)
\put(19,3){\circle{32}}
\put(19,3){\makebox(1,0){$W_I$}}
\put(5.77,12){\circle*{4}}
\put(5.77,-6){\circle*{4}}
\put(5.77,3){\oval(28,18)[l]}
\put(-8.23,3){\circle*{4}}
\put(-16.23,3){\circle{16}}
\epi
+\frac{1}{3}
\rule[-18pt]{0pt}{42pt}
\bpi(54.78,0)(-14.78,0)
\put(19,3){\circle{32}}
\put(19,3){\makebox(1,0){$W_I$}}
\put(5.22,11.4){\circle*{4}}
\put(10.87,17.1){\circle*{4}}
\put(5.22,-5.13){\circle*{4}}
\put(10.87,-10.78){\circle*{4}}
\put(10.87,3){\oval(41.3,27.88)[l]}
\put(5.22,3){\oval(30,16.53)[l]}
\put(-9.78,3){\circle*{4}}
\epi
+\frac{1}{3}
\rule[-18pt]{0pt}{42pt}
\bpi(96.46,0)(-56.46,0)
\put(19,3){\circle{32}}
\put(5.77,12){\circle*{4}}
\put(5.77,-6){\circle*{4}}
\put(5.77,3){\oval(28,18)[l]}
\put(-8.23,3){\circle*{4}}
\put(-22.23,3){\oval(28,18)[r]}
\put(-22.23,12){\circle*{4}}
\put(-22.23,-6){\circle*{4}}
\put(-35.46,3){\circle{32}}
\put(-35.46,3){\makebox(1,0){$W_I$}}
\put(19,3){\makebox(1,0){$W_I$}}
\epi\!.
\eeq
Note that a derivative with respect to $G$ graphically means removing
(``amputating'') a line from a Feynman graph (for details see \cite{KPKB}
and Appendix \ref{symg}).
This will be important when we represent $W_I$ as a sum of Feynman graphs
in the next section.
For example, the operation on $W_I$ on the left hand side of (\ref{wisym}) 
multiplies each graph in $W_I$ by the number of its lines.

\subsection{Recursion Relation}
Now split $W_I$ and $\De$ into different loop orders,
\beq
W_I\equiv
\rule[-18pt]{0pt}{42pt}
\bpi(40,0)(0,0)
\put(19,3){\circle{32}}
\put(19,3){\makebox(1,0){$W_I$}}
\epi
=
\sum_{L=2}^\infty W^{(L)}
\equiv
\sum_{L=2}^\infty
\rule[-18pt]{0pt}{42pt}
\bpi(40,0)(0,0)
\put(19,3){\circle{32}}
\put(19,3){\makebox(1,0){$L$}}
\epi,
\eeq
and
\beq
\label{de}
-\De\equiv
\rule[-7pt]{0pt}{20pt}
\bpi(50,20)
\put(5,3){\line(1,0){15}}
\put(30,3){\line(1,0){15}}
\put(20,-2){\line(1,0){10}}
\put(20,-2){\line(0,1){10}}
\put(20,8){\line(1,0){10}}
\put(30,-2){\line(0,1){10}}
\put(25,3){\makebox(1,0){$\scs\De$}}
\epi
=
-\sum_{L=1}^\infty\De^{(L)}
\equiv
\sum_{L=1}^\infty
\rule[-7pt]{0pt}{20pt}
\bpi(50,20)
\put(5,3){\line(1,0){15}}
\put(30,3){\line(1,0){15}}
\put(20,-2){\line(1,0){10}}
\put(20,-2){\line(0,1){10}}
\put(20,8){\line(1,0){10}}
\put(30,-2){\line(0,1){10}}
\put(25,3){\makebox(1,0){$\scs L$}}
\epi,
\eeq
where $\De^{(L)}$ counts formally for $L$ intrinsic loops.

Eq.\ (\ref{wisym}) then splits into
\beq
\rule[-18pt]{0pt}{42pt}
\bpi(53.23,0)(-13.23,0)
\put(19,3){\circle{32}}
\put(19,3){\makebox(1,0){$2$}}
\put(5.77,12){\circle*{4}}
\put(5.77,-6){\circle*{4}}
\put(5.77,3){\oval(28,18)[l]}
\epi
=
\frac{1}{4}
\rule[-10pt]{0pt}{26pt}
\bpi(42,0)
\put(21,3){\circle*{4}}
\put(13,3){\circle{16}}
\put(29,3){\circle{16}}
\epi
+
\frac{1}{2}
\rule[-11pt]{0pt}{28pt}
\bpi(33,0)
\put(19,8){\oval(18,8)[t]}
\put(19,-2){\oval(18,8)[b]}
\put(28,-2){\line(0,1){10}}
\put(5,-2){\line(1,0){10}}
\put(5,-2){\line(0,1){10}}
\put(5,8){\line(1,0){10}}
\put(15,-2){\line(0,1){10}}
\put(10,3){\makebox(0,0){$\scs 1$}}
\epi,
\eeq
from which follows
\beq
\label{w1sym}
\rule[-18pt]{0pt}{42pt}
\bpi(40,0)(0,0)
\put(19,3){\circle{32}}
\put(19,3){\makebox(1,0){$2$}}
\epi
=
\frac{1}{8}
\rule[-10pt]{0pt}{26pt}
\bpi(42,0)
\put(21,3){\circle*{4}}
\put(13,3){\circle{16}}
\put(29,3){\circle{16}}
\epi
+\frac{1}{2}
\rule[-11pt]{0pt}{28pt}
\bpi(33,0)
\put(19,8){\oval(18,8)[t]}
\put(19,-2){\oval(18,8)[b]}
\put(28,-2){\line(0,1){10}}
\put(5,-2){\line(1,0){10}}
\put(5,-2){\line(0,1){10}}
\put(5,8){\line(1,0){10}}
\put(15,-2){\line(0,1){10}}
\put(10,3){\makebox(0,0){$\scs 1$}}
\epi,
\eeq
and
\beq
\label{wsymrecrel}
\rule[-18pt]{0pt}{42pt}
\bpi(53.23,0)(-13.23,0)
\put(19,3){\circle{32}}
\put(19,3){\makebox(1,0){$L$}}
\put(5.77,12){\circle*{4}}
\put(5.77,-6){\circle*{4}}
\put(5.77,3){\oval(28,18)[l]}
\epi
=
\frac{1}{2}
\rule[-11pt]{0pt}{28pt}
\bpi(37,0)(-4,0)
\put(19,8){\oval(18,8)[t]}
\put(19,-2){\oval(18,8)[b]}
\put(28,-2){\line(0,1){10}}
\put(0,-2){\line(1,0){18}}
\put(0,-2){\line(0,1){10}}
\put(0,8){\line(1,0){18}}
\put(18,-2){\line(0,1){10}}
\put(9,3){\makebox(0,0){$\scs L-1$}}
\epi
+
\sum_{l=1}^{L-2}
\rule[-18pt]{0pt}{42pt}
\bpi(58.23,0)(-18.23,0)
\put(19,3){\circle{32}}
\put(19,3){\makebox(1,0){$L-l$}}
\put(5.77,12){\circle*{4}}
\put(5.77,-6){\circle*{4}}
\put(5.77,8){\oval(28,8)[tl]}
\put(5.77,-2){\oval(28,8)[bl]}
\put(-13.23,-2){\line(1,0){10}}
\put(-13.23,-2){\line(0,1){10}}
\put(-13.23,8){\line(1,0){10}}
\put(-3.23,-2){\line(0,1){10}}
\put(-8.23,3){\makebox(0,0){$\scs l$}}
\epi
+
\rule[-18pt]{0pt}{42pt}
\bpi(69.23,0)(-29.23,0)
\put(19,3){\circle{32}}
\put(19,3){\makebox(1,0){$L-1$}}
\put(5.77,12){\circle*{4}}
\put(5.77,-6){\circle*{4}}
\put(5.77,3){\oval(28,18)[l]}
\put(-8.23,3){\circle*{4}}
\put(-16.23,3){\circle{16}}
\epi
+
\frac{1}{3}
\rule[-18pt]{0pt}{42pt}
\bpi(54.78,0)(-14.78,0)
\put(19,3){\circle{32}}
\put(19,3){\makebox(1,0){$L-1$}}
\put(5.22,11.4){\circle*{4}}
\put(10.87,17.1){\circle*{4}}
\put(5.22,-5.13){\circle*{4}}
\put(10.87,-10.78){\circle*{4}}
\put(10.87,3){\oval(41.3,27.88)[l]}
\put(5.22,3){\oval(30,16.53)[l]}
\put(-9.78,3){\circle*{4}}
\epi
+
\frac{1}{3}\sum_{l=2}^{L-2}
\rule[-18pt]{0pt}{42pt}
\bpi(96.46,0)(-56.46,0)
\put(19,3){\circle{32}}
\put(19,3){\makebox(1,0){$L-l$}}
\put(5.77,12){\circle*{4}}
\put(5.77,-6){\circle*{4}}
\put(5.77,3){\oval(28,18)[l]}
\put(-8.23,3){\circle*{4}}
\put(-22.23,3){\oval(28,18)[r]}
\put(-22.23,12){\circle*{4}}
\put(-22.23,-6){\circle*{4}}
\put(-35.46,3){\circle{32}}
\put(-35.46,3){\makebox(1,-2){$l$}}
\epi
\eeq
for $L>2$.

Let us now derive a recursion relation for the $W^{(L)}$ themselves instead
of their derivatives with respect to $G$.
First note that since $W$ depends only on $G^{-1}$ and $\De$ only through
the combination $G^{-1}+\De$ we have
\beq
\frac{\de W}{\de\De_{12}}
=
\frac{\de W}{\de G_{12}^{-1}}
\nn\\
=
-\int_{34}G_{13}G_{24}\frac{\de W}{\de G_{34}},
\eeq
where we have used (\ref{ddginv}).

Because of (\ref{de}) we can write then for any $L$
\beq
\label{dedwkde}
\int_{12}\De_{12}^{(L)}\frac{\de W}{\de\De_{12}^{(L)}}
=-\int_{1234}G_{31}\De_{12}^{(L)}G_{24}\frac{\de W}{\de G_{34}}
\equiv
\rule[-18pt]{0pt}{42pt}
\bpi(58.23,0)(-18.23,0)
\put(19,3){\circle{32}}
\put(19,3){\makebox(1,0){$W$}}
\put(5.77,12){\circle*{4}}
\put(5.77,-6){\circle*{4}}
\put(5.77,8){\oval(28,8)[tl]}
\put(5.77,-2){\oval(28,8)[bl]}
\put(-13.23,-2){\line(1,0){10}}
\put(-13.23,-2){\line(0,1){10}}
\put(-13.23,8){\line(1,0){10}}
\put(-3.23,-2){\line(0,1){10}}
\put(-8.23,3){\makebox(0,0){$\scs L$}}
\epi.
\eeq
Splitting up into loop orders gives
\beq
\label{ddwid1}
\int_{12}\De_{12}^{(l)}\frac{\de W^{(L)}}{\de\De_{12}^{(l)}}
=-\int_{1234}\De_{12}^{(l)}G_{13}G_{24}\frac{\de W^{(L-l)}}{\de G_{34}}
\equiv
\rule[-18pt]{0pt}{42pt}
\bpi(58.23,0)(-18.23,0)
\put(19,3){\circle{32}}
\put(19,3){\makebox(1,0){$L-l$}}
\put(5.77,12){\circle*{4}}
\put(5.77,-6){\circle*{4}}
\put(5.77,8){\oval(28,8)[tl]}
\put(5.77,-2){\oval(28,8)[bl]}
\put(-13.23,-2){\line(1,0){10}}
\put(-13.23,-2){\line(0,1){10}}
\put(-13.23,8){\line(1,0){10}}
\put(-3.23,-2){\line(0,1){10}}
\put(-8.23,3){\makebox(0,0){$\scs l$}}
\epi
\eeq
for $1\leq l\leq L-2$, and, using (\ref{dw0dg}),
\beq
\label{ddwid2}
\int_{12}\De_{12}^{(L-1)}\frac{\de W^{(L)}}{\de\De_{12}^{(L-1)}}
=-\int_{1234}\De_{12}^{(L-1)}G_{13}G_{24}\frac{\de W_0}{\de G_{34}}
=-\frac{1}{2}\int_{12}\De_{12}^{(L-1)}G_{12}
\equiv
\frac{1}{2}
\rule[-11pt]{0pt}{28pt}
\bpi(37,0)(-4,0)
\put(19,8){\oval(18,8)[t]}
\put(19,-2){\oval(18,8)[b]}
\put(28,-2){\line(0,1){10}}
\put(0,-2){\line(1,0){18}}
\put(0,-2){\line(0,1){10}}
\put(0,8){\line(1,0){18}}
\put(18,-2){\line(0,1){10}}
\put(9,3){\makebox(0,0){$\scs L-1$}}
\epi
\eeq
for $L\geq1$.
Since an $L$-loop diagram without two-point insertions contains $2(L-1)$
propagators and since an $l$-loop two-point insertion causes a reduction
in the number of propagators by $2l-1$, the following relation for the
$L$-loop contribution to $W$ holds for $L\geq 2$:
\beq
\int_{12}\left[G_{12}\frac{\de}{\de G_{12}}
+\sum_{l=1}^{L-1}(2l-1)\De_{12}^{(l)}\frac{\de}{\de\De_{12}^{(l)}}\right]W^{(L)}
=2(L-1)W^{(L)}.
\eeq
Making use of (\ref{ddwid1}) and (\ref{ddwid2}) this can
be rewritten as
\beq
\left[\int_{12}G_{12}\frac{\de W^{(L)}}{\de G_{12}}
-\frac{2L-3}{2}\int_{12}\De_{12}^{(L-1)}G_{12}
-\sum_{l=1}^{L-2}(2l-1)
\int_{123}\De_{12}^{(l)}G_{13}G_{24}\frac{\de W^{(L-l)}}{\de G_{34}}
\right]=2(L-1)W^{(L)}
\eeq
or
\beq
\rule[-18pt]{0pt}{42pt}
\bpi(53.23,0)(-13.23,0)
\put(19,3){\circle{32}}
\put(19,3){\makebox(1,0){$L$}}
\put(5.77,12){\circle*{4}}
\put(5.77,-6){\circle*{4}}
\put(5.77,3){\oval(28,18)[l]}
\epi
+
\frac{2L-3}{2}
\rule[-11pt]{0pt}{28pt}
\bpi(37,0)(-4,0)
\put(19,8){\oval(18,8)[t]}
\put(19,-2){\oval(18,8)[b]}
\put(28,-2){\line(0,1){10}}
\put(0,-2){\line(1,0){18}}
\put(0,-2){\line(0,1){10}}
\put(0,8){\line(1,0){18}}
\put(18,-2){\line(0,1){10}}
\put(9,3){\makebox(0,0){$\scs L-1$}}
\epi
+
\sum_{l=1}^{L-2}(2l-1)
\rule[-18pt]{0pt}{42pt}
\bpi(58.23,0)(-18.23,0)
\put(19,3){\circle{32}}
\put(19,3){\makebox(1,0){$L-l$}}
\put(5.77,12){\circle*{4}}
\put(5.77,-6){\circle*{4}}
\put(5.77,8){\oval(28,8)[tl]}
\put(5.77,-2){\oval(28,8)[bl]}
\put(-13.23,-2){\line(1,0){10}}
\put(-13.23,-2){\line(0,1){10}}
\put(-13.23,8){\line(1,0){10}}
\put(-3.23,-2){\line(0,1){10}}
\put(-8.23,3){\makebox(0,0){$\scs l$}}
\epi
=
2(L-1)
\rule[-18pt]{0pt}{42pt}
\bpi(42,0)(-2,0)
\put(19,3){\circle{32}}
\put(19,3){\makebox(1,0){$L$}}
\epi.
\eeq
Rewriting the first term using the recursion relation (\ref{wsymrecrel})
gives
\bea
\label{newrecrel}
\lefteqn{
\rule[-18pt]{0pt}{42pt}
\bpi(40,0)(0,0)
\put(19,3){\circle{32}}
\put(19,3){\makebox(1,0){$L$}}
\epi
=
\frac{1}{2}
\rule[-11pt]{0pt}{28pt}
\bpi(37,0)(-4,0)
\put(19,8){\oval(18,8)[t]}
\put(19,-2){\oval(18,8)[b]}
\put(28,-2){\line(0,1){10}}
\put(0,-2){\line(1,0){18}}
\put(0,-2){\line(0,1){10}}
\put(0,8){\line(1,0){18}}
\put(18,-2){\line(0,1){10}}
\put(9,3){\makebox(0,0){$\scs L-1$}}
\epi
}
\nn\\
&&
+\frac{1}{L-1}\left[\frac{1}{6}
\rule[-18pt]{0pt}{42pt}
\bpi(54.78,0)(-14.78,0)
\put(19,3){\circle{32}}
\put(19,3){\makebox(1,0){$L-1$}}
\put(5.22,11.4){\circle*{4}}
\put(10.87,17.1){\circle*{4}}
\put(5.22,-5.13){\circle*{4}}
\put(10.87,-10.78){\circle*{4}}
\put(10.87,3){\oval(41.3,27.88)[l]}
\put(5.22,3){\oval(30,16.53)[l]}
\put(-9.78,3){\circle*{4}}
\epi
+
\sum_{l=1}^{L-2}l
\rule[-18pt]{0pt}{42pt}
\bpi(58.23,0)(-18.23,0)
\put(19,3){\circle{32}}
\put(19,3){\makebox(1,0){$L-l$}}
\put(5.77,12){\circle*{4}}
\put(5.77,-6){\circle*{4}}
\put(5.77,8){\oval(28,8)[tl]}
\put(5.77,-2){\oval(28,8)[bl]}
\put(-13.23,-2){\line(1,0){10}}
\put(-13.23,-2){\line(0,1){10}}
\put(-13.23,8){\line(1,0){10}}
\put(-3.23,-2){\line(0,1){10}}
\put(-8.23,3){\makebox(0,0){$\scs l$}}
\epi
+\frac{1}{2}
\rule[-18pt]{0pt}{42pt}
\bpi(69.23,0)(-29.23,0)
\put(19,3){\circle{32}}
\put(19,3){\makebox(1,0){$L-1$}}
\put(5.77,12){\circle*{4}}
\put(5.77,-6){\circle*{4}}
\put(5.77,3){\oval(28,18)[l]}
\put(-8.23,3){\circle*{4}}
\put(-16.23,3){\circle{16}}
\epi
+\frac{1}{6}\sum_{l=2}^{L-2}
\rule[-18pt]{0pt}{42pt}
\bpi(96.46,0)(-56.46,0)
\put(19,3){\circle{32}}
\put(19,3){\makebox(1,0){$L-l$}}
\put(5.77,12){\circle*{4}}
\put(5.77,-6){\circle*{4}}
\put(5.77,3){\oval(28,18)[l]}
\put(-8.23,3){\circle*{4}}
\put(-22.23,3){\oval(28,18)[r]}
\put(-22.23,12){\circle*{4}}
\put(-22.23,-6){\circle*{4}}
\put(-35.46,3){\circle{32}}
\put(-35.46,3){\makebox(1,0){$l$}}
\epi
\right]
\eea
for $L>2$.
For $\De=0$ and appropriately adjusted conventions this reduces to
eq.\ (2.64) in \cite{KPKB}.

We have used (\ref{w1sym}) and (\ref{newrecrel}) to determine all
vacuum graphs and their weights (i.e.\ combinatorial prefactors)
through five loops for the case $\De=0$ and listed them in
Table \ref{allgraphs}.
\begin{table}[hbt]
\begin{center}
\begin{tabular}{c}
\begin{tabular}{|c|c|}
\hline
\begin{tabular}{c}number\\of loops\end{tabular}
&
diagrams with their weights
\\
\hline
1&
\hspace{10pt}
$\frac{1}{2}$
\hspace{-7pt}
\rule[-10pt]{0pt}{26pt}
\bpi(26,12)
\put(13,3){\circle{16}}
\epi
\\
\hline
2&
\hspace{10pt}
$\frac{1}{8}$
\hspace{-7pt}
\rule[-10pt]{0pt}{26pt}
\bpi(42,12)
\put(13,3){\circle{16}}
\put(29,3){\circle{16}}
\put(21,3){\circle*{4}}
\epi
\\
\hline
3&
\hspace{10pt}
$\frac{1}{48}$
\hspace{-5pt}
\rule[-14pt]{0pt}{34pt}
\bpi(34,12)
\put(17,3){\circle{24}}
\put(17,3){\oval(24,8)}
\put(5,3){\circle*{4}}
\put(29,3){\circle*{4}}
\epi
\hspace{10pt}
$\frac{1}{16}$
\hspace{-7pt}
\rule[-10pt]{0pt}{26pt}
\bpi(58,12)
\put(13,3){\circle{16}}
\put(29,3){\circle{16}}
\put(45,3){\circle{16}}
\put(21,3){\circle*{4}}
\put(37,3){\circle*{4}}
\epi
\\
\hline
4&
\hspace{10pt}
$\frac{1}{48}$
\hspace{-7pt}
\rule[-14pt]{0pt}{34pt}
\bpi(34,12)
\put(17,3){\circle{24}}
\put(6.6,9){\line(1,0){20.8}}
\put(6.6,9){\line(3,-5){10.4}}
\put(27.4,9){\line(-3,-5){10.4}}
\put(6.6,9){\circle*{4}}
\put(27.4,9){\circle*{4}}
\put(17,-9){\circle*{4}}
\epi
\hspace{10pt}
$\frac{1}{24}$
\hspace{-5pt}
\rule[-14pt]{0pt}{50pt}
\bpi(34,12)
\put(17,3){\circle{24}}
\put(17,3){\oval(24,8)}
\put(17,23){\circle{16}}
\put(5,3){\circle*{4}}
\put(29,3){\circle*{4}}
\put(17,15){\circle*{4}}
\epi
\hspace{10pt}
$\frac{1}{32}$
\hspace{-7pt}
\rule[-10pt]{0pt}{26pt}
\bpi(74,12)
\put(13,3){\circle{16}}
\put(29,3){\circle{16}}
\put(45,3){\circle{16}}
\put(61,3){\circle{16}}
\put(21,3){\circle*{4}}
\put(37,3){\circle*{4}}
\put(53,3){\circle*{4}}
\epi
\hspace{10pt}
$\frac{1}{48}$
\hspace{-7pt}
\rule[-18pt]{0pt}{50pt}
\bpi(53.7,12)
\put(26.85,3){\circle{16}}
\put(26.85,19){\circle{16}}
\put(13,-5){\circle{16}}
\put(40.7,-5){\circle{16}}
\put(26.85,11){\circle*{4}}
\put(19.95,-1){\circle*{4}}
\put(33.75,-1){\circle*{4}}
\epi
\\
\hline
5&
\hspace{10pt}
$\frac{1}{128}$
\hspace{-7pt}
\rule[-14pt]{0pt}{34pt}
\bpi(34,12)
\put(17,3){\circle{24}}
\put(8.5,-5.5){\line(1,0){17}}
\put(8.5,-5.5){\line(0,1){17}}
\put(8.5,11.5){\line(1,0){17}}
\put(25.5,-5.5){\line(0,1){17}}
\put(8.5,-5.5){\circle*{4}}
\put(8.5,11.5){\circle*{4}}
\put(25.5,-5.5){\circle*{4}}
\put(25.5,11.5){\circle*{4}}
\epi
\hspace{10pt}
$\frac{1}{144}$
\hspace{-7pt}
\rule[-22pt]{0pt}{50pt}
\bpi(42,12)
\put(21,-9){\circle{16}}
\put(21,15){\circle{16}}
\put(13,-9){\line(1,0){16}}
\put(13,15){\line(1,0){16}}
\put(13,3){\oval(16,24)[l]}
\put(29,3){\oval(16,24)[r]}
\put(13,-9){\circle*{4}}
\put(13,15){\circle*{4}}
\put(29,-9){\circle*{4}}
\put(29,15){\circle*{4}}
\epi
\hspace{10pt}
$\frac{1}{32}$
\hspace{-5pt}
\rule[-18pt]{0pt}{42pt}
\bpi(58,12)
\put(13,3){\circle{16}}
\put(45,3){\circle{16}}
\put(5,-5){\line(0,1){16}}
\put(25,-5){\oval(40,16)[b]}
\put(25,11){\oval(40,16)[t]}
\put(45,3){\oval(48,16)[l]}
\put(5,3){\circle*{4}}
\put(21,3){\circle*{4}}
\put(45,-5){\circle*{4}}
\put(45,11){\circle*{4}}
\epi
\hspace{10pt}
$\frac{1}{16}$
\hspace{-7pt}
\rule[-14pt]{0pt}{50pt}
\bpi(34,12)
\put(17,3){\circle{24}}
\put(17,23){\circle{16}}
\put(6.6,9){\line(1,0){20.8}}
\put(6.6,9){\line(3,-5){10.4}}
\put(27.4,9){\line(-3,-5){10.4}}
\put(6.6,9){\circle*{4}}
\put(27.4,9){\circle*{4}}
\put(17,-9){\circle*{4}}
\put(17,15){\circle*{4}}
\epi
\hspace{10pt}
$\frac{1}{48}$
\hspace{-5pt}
\rule[-14pt]{0pt}{47.3pt}
\bpi(46,12)
\put(23,3){\circle{24}}
\put(13,20.3){\circle{16}}
\put(33,20.3){\circle{16}}
\put(23,3){\oval(24,8)}
\put(11,3){\circle*{4}}
\put(35,3){\circle*{4}}
\put(17,13.4){\circle*{4}}
\put(29,13.4){\circle*{4}}
\epi
\hspace{10pt}
$\frac{1}{32}$
\hspace{-5pt}
\rule[-30pt]{0pt}{66pt}
\bpi(34,12)
\put(17,3){\circle{24}}
\put(17,-17){\circle{16}}
\put(17,23){\circle{16}}
\put(17,3){\oval(24,8)}
\put(5,3){\circle*{4}}
\put(17,-9){\circle*{4}}
\put(17,15){\circle*{4}}
\put(29,3){\circle*{4}}
\epi
\\
&
\hspace{10pt}
$\frac{1}{48}$
\hspace{-5pt}
\rule[-14pt]{0pt}{66pt}
\bpi(34,12)
\put(17,3){\circle{24}}
\put(17,23){\circle{16}}
\put(17,39){\circle{16}}
\put(17,3){\oval(24,8)}
\put(5,3){\circle*{4}}
\put(17,15){\circle*{4}}
\put(17,31){\circle*{4}}
\put(29,3){\circle*{4}}
\epi
\hspace{10pt}
$\frac{1}{64}$
\hspace{-10pt}
\rule[-7pt]{0pt}{26pt}
\bpi(90,12)
\put(13,3){\circle{16}}
\put(29,3){\circle{16}}
\put(45,3){\circle{16}}
\put(61,3){\circle{16}}
\put(77,3){\circle{16}}
\put(21,3){\circle*{4}}
\put(37,3){\circle*{4}}
\put(53,3){\circle*{4}}
\put(69,3){\circle*{4}}
\epi
\hspace{10pt}
$\frac{1}{32}$
\hspace{-7pt}
\rule[-23.85pt]{0pt}{53.7pt}
\bpi(66,12)
\put(13,3){\circle{16}}
\put(29,3){\circle{16}}
\put(45,3){\circle{16}}
\put(53,-10.85){\circle{16}}
\put(53,16.85){\circle{16}}
\put(21,3){\circle*{4}}
\put(37,3){\circle*{4}}
\put(49,-3.9){\circle*{4}}
\put(49,9.9){\circle*{4}}
\epi
\hspace{10pt}
$\frac{1}{128}$
\hspace{-7pt}
\rule[-26pt]{0pt}{58pt}
\bpi(58,12)
\put(13,3){\circle{16}}
\put(29,-13){\circle{16}}
\put(29,3){\circle{16}}
\put(29,19){\circle{16}}
\put(45,3){\circle{16}}
\put(21,3){\circle*{4}}
\put(29,-5){\circle*{4}}
\put(29,11){\circle*{4}}
\put(37,3){\circle*{4}}
\epi
\\
\hline
\end{tabular}
\end{tabular}
\end{center}
\caption{\label{allgraphs}
Vacuum diagrams with their weights through five loops.}
\end{table}
With a one-loop correction
\beq
\label{oneloopcorr}
\De=\De^{(1)}
=
\rule[-7pt]{0pt}{20pt}
\bpi(50,20)
\put(5,3){\line(1,0){15}}
\put(30,3){\line(1,0){15}}
\put(20,-2){\line(1,0){10}}
\put(20,-2){\line(0,1){10}}
\put(20,8){\line(1,0){10}}
\put(30,-2){\line(0,1){10}}
\put(25,3){\makebox(1,0){$\scs 1$}}
\epi
\equiv
\rule[-2pt]{0pt}{10pt}
\bpi(40,0)
\put(5,3){\line(1,0){30}}
\put(20,3){\circle*{4}}
\epi
\eeq
we get the additional graphs listed in Table \ref{additionalgraphs}.
\begin{table}
\begin{center}
\begin{tabular}{c}
\begin{tabular}{|c|c|}
\hline
\begin{tabular}{c}number\\of loops\end{tabular}
&
additional diagrams with their weights
\\
\hline
$2$
&
\hspace{10pt}
$\frac{1}{2}$
\hspace{-7pt}
\rule[-10pt]{0pt}{26pt}
\bpi(26,12)
\put(13,3){\circle{16}}
\put(21,3){\circle*{4}}
\epi
\\
\hline
$3$
&
\hspace{10pt}
$\frac{1}{4}$
\rule[-10pt]{0pt}{26pt}
\hspace{-5pt}
\bpi(26,12)
\put(13,3){\circle{16}}
\put(5,3){\circle*{4}}
\put(21,3){\circle*{4}}
\epi
\hspace{10pt}
$\frac{1}{4}$
\rule[-10pt]{0pt}{26pt}
\hspace{-7pt}
\bpi(42,12)
\put(13,3){\circle{16}}
\put(29,3){\circle{16}}
\put(21,3){\circle*{4}}
\put(37,3){\circle*{4}}
\epi
\\
\hline
$4$
&
\hspace{10pt}
$\frac{1}{6}$
\rule[-10pt]{0pt}{26pt}
\hspace{-7pt}
\bpi(26,12)
\put(13,3){\circle{16}}
\put(13,11){\circle*{4}}
\put(6.1,-1){\circle*{4}}
\put(19.9,-1){\circle*{4}}
\epi
\hspace{10pt}
$\frac{1}{8}$
\rule[-10pt]{0pt}{26pt}
\hspace{-5pt}
\bpi(42,12)
\put(13,3){\circle{16}}
\put(29,3){\circle{16}}
\put(5,3){\circle*{4}}
\put(21,3){\circle*{4}}
\put(37,3){\circle*{4}}
\epi
\hspace{10pt}
$\frac{1}{4}$
\hspace{-7pt}
\rule[-10pt]{0pt}{26pt}
\bpi(42,12)
\put(13,3){\circle{16}}
\put(29,3){\circle{16}}
\put(21,3){\circle*{4}}
\put(33,-3.9){\circle*{4}}
\put(33,9.9){\circle*{4}}
\epi
\\
&\hspace{10pt}
$\frac{1}{12}$
\rule[-14pt]{0pt}{34pt}
\hspace{-5pt}
\bpi(34,12)
\put(17,3){\circle{24}}
\put(17,3){\oval(24,8)}
\put(5,3){\circle*{4}}
\put(17,15){\circle*{4}}
\put(29,3){\circle*{4}}
\epi
\hspace{10pt}
$\frac{1}{8}$
\rule[-10pt]{0pt}{26pt}
\hspace{-7pt}
\bpi(58,12)
\put(13,3){\circle{16}}
\put(29,3){\circle{16}}
\put(45,3){\circle{16}}
\put(21,3){\circle*{4}}
\put(29,11){\circle*{4}}
\put(37,3){\circle*{4}}
\epi
\hspace{10pt}
$\frac{1}{8}$
\rule[-10pt]{0pt}{26pt}
\hspace{-7pt}
\bpi(58,12)
\put(13,3){\circle{16}}
\put(29,3){\circle{16}}
\put(45,3){\circle{16}}
\put(21,3){\circle*{4}}
\put(37,3){\circle*{4}}
\put(53,3){\circle*{4}}
\epi
\\
\hline
$5$
&
\hspace{10pt}
$\frac{1}{8}$
\rule[-10pt]{0pt}{26pt}
\hspace{-5pt}
\bpi(26,12)
\put(13,3){\circle{16}}
\put(5,3){\circle*{4}}
\put(13,-5){\circle*{4}}
\put(13,11){\circle*{4}}
\put(21,3){\circle*{4}}
\epi
\hspace{10pt}
$\frac{1}{4}$
\rule[-10pt]{0pt}{26pt}
\hspace{-7pt}
\bpi(42,12)
\put(13,3){\circle{16}}
\put(29,3){\circle{16}}
\put(21,3){\circle*{4}}
\put(29,-5){\circle*{4}}
\put(29,11){\circle*{4}}
\put(37,3){\circle*{4}}
\epi
\hspace{10pt}
$\frac{1}{4}$
\rule[-10pt]{0pt}{26pt}
\hspace{-5pt}
\bpi(42,12)
\put(13,3){\circle{16}}
\put(29,3){\circle{16}}
\put(5,3){\circle*{4}}
\put(21,3){\circle*{4}}
\put(33,-3.9){\circle*{4}}
\put(33,9.9){\circle*{4}}
\epi
\hspace{10pt}
$\frac{1}{12}$
\rule[-14pt]{0pt}{34pt}
\hspace{-5pt}
\bpi(34,12)
\put(17,3){\circle{24}}
\put(17,3){\oval(24,8)}
\put(5,3){\circle*{4}}
\put(29,3){\circle*{4}}
\put(11,13.4){\circle*{4}}
\put(23,13.4){\circle*{4}}
\epi
\hspace{10pt}
$\frac{1}{8}$
\rule[-14pt]{0pt}{34pt}
\hspace{-5pt}
\bpi(34,12)
\put(17,3){\circle{24}}
\put(17,3){\oval(24,8)}
\put(5,3){\circle*{4}}
\put(17,-9){\circle*{4}}
\put(17,15){\circle*{4}}
\put(29,3){\circle*{4}}
\epi
\\
&
\hspace{10pt}
$\frac{1}{16}$
\rule[-10pt]{0pt}{26pt}
\hspace{-7pt}
\bpi(58,12)
\put(13,3){\circle{16}}
\put(29,3){\circle{16}}
\put(45,3){\circle{16}}
\put(21,3){\circle*{4}}
\put(29,11){\circle*{4}}
\put(29,-5){\circle*{4}}
\put(37,3){\circle*{4}}
\epi
\hspace{10pt}
$\frac{1}{8}$
\rule[-10pt]{0pt}{42pt}
\hspace{-7pt}
\bpi(42,12)
\put(13,3){\circle{16}}
\put(29,3){\circle{16}}
\put(29,19){\circle{16}}
\put(21,3){\circle*{4}}
\put(29,-5){\circle*{4}}
\put(29,11){\circle*{4}}
\put(37,3){\circle*{4}}
\epi
\hspace{10pt}
$\frac{1}{8}$
\rule[-10pt]{0pt}{26pt}
\hspace{-7pt}
\bpi(58,12)
\put(13,3){\circle{16}}
\put(29,3){\circle{16}}
\put(45,3){\circle{16}}
\put(21,3){\circle*{4}}
\put(37,3){\circle*{4}}
\put(49,-3.9){\circle*{4}}
\put(49,9.9){\circle*{4}}
\epi
\hspace{10pt}
$\frac{1}{16}$
\rule[-10pt]{0pt}{26pt}
\hspace{-5pt}
\bpi(58,12)
\put(13,3){\circle{16}}
\put(29,3){\circle{16}}
\put(45,3){\circle{16}}
\put(5,3){\circle*{4}}
\put(21,3){\circle*{4}}
\put(37,3){\circle*{4}}
\put(53,3){\circle*{4}}
\epi
\hspace{10pt}
$\frac{1}{4}$
\rule[-10pt]{0pt}{39.85pt}
\hspace{-5pt}
\bpi(50,12)
\put(13,3){\circle{16}}
\put(29,3){\circle{16}}
\put(37,16.85){\circle{16}}
\put(5,3){\circle*{4}}
\put(21,3){\circle*{4}}
\put(33,-3.9){\circle*{4}}
\put(33,9.9){\circle*{4}}
\epi
\\
&
\hspace{10pt}
$\frac{1}{8}$
\rule[-14pt]{0pt}{34pt}
\hspace{-7pt}
\bpi(34,12)
\put(17,3){\circle{24}}
\put(6.6,9){\line(1,0){20.8}}
\put(6.6,9){\line(3,-5){10.4}}
\put(27.4,9){\line(-3,-5){10.4}}
\put(6.6,9){\circle*{4}}
\put(27.4,9){\circle*{4}}
\put(17,-9){\circle*{4}}
\put(17,15){\circle*{4}}
\epi
\hspace{10pt}
$\frac{1}{8}$
\rule[-14pt]{0pt}{50pt}
\hspace{-5pt}
\bpi(34,12)
\put(17,3){\circle{24}}
\put(17,3){\oval(24,8)}
\put(17,23){\circle{16}}
\put(5,3){\circle*{4}}
\put(17,-9){\circle*{4}}
\put(17,15){\circle*{4}}
\put(29,3){\circle*{4}}
\epi
\hspace{10pt}
$\frac{1}{12}$
\rule[-14pt]{0pt}{47.3pt}
\hspace{-5pt}
\bpi(46,12)
\put(23,3){\circle{24}}
\put(13,20.3){\circle{16}}
\put(23,3){\oval(24,8)}
\put(11,3){\circle*{4}}
\put(35,3){\circle*{4}}
\put(17,13.4){\circle*{4}}
\put(29,13.4){\circle*{4}}
\epi
\hspace{10pt}
$\frac{1}{24}$
\rule[-14pt]{0pt}{50pt}
\hspace{-5pt}
\bpi(34,12)
\put(17,3){\circle{24}}
\put(17,3){\oval(24,8)}
\put(17,23){\circle{16}}
\put(5,3){\circle*{4}}
\put(17,15){\circle*{4}}
\put(17,31){\circle*{4}}
\put(29,3){\circle*{4}}
\epi
\\
&
\hspace{10pt}
$\frac{1}{8}$
\rule[-10pt]{0pt}{39.85pt}
\hspace{-7pt}
\bpi(66,12)
\put(13,3){\circle{16}}
\put(29,3){\circle{16}}
\put(45,3){\circle{16}}
\put(53,16.85){\circle{16}}
\put(21,3){\circle*{4}}
\put(37,3){\circle*{4}}
\put(49,-3.9){\circle*{4}}
\put(49,9.9){\circle*{4}}
\epi
\hspace{10pt}
$\frac{1}{16}$
\rule[-10pt]{0pt}{26pt}
\hspace{-7pt}
\bpi(74,12)
\put(13,3){\circle{16}}
\put(29,3){\circle{16}}
\put(45,3){\circle{16}}
\put(61,3){\circle{16}}
\put(21,3){\circle*{4}}
\put(37,3){\circle*{4}}
\put(53,3){\circle*{4}}
\put(69,3){\circle*{4}}
\epi
\hspace{10pt}
$\frac{1}{16}$
\rule[-10pt]{0pt}{42pt}
\hspace{-7pt}
\bpi(58,12)
\put(13,3){\circle{16}}
\put(29,3){\circle{16}}
\put(29,19){\circle{16}}
\put(45,3){\circle{16}}
\put(21,3){\circle*{4}}
\put(29,-5){\circle*{4}}
\put(29,11){\circle*{4}}
\put(37,3){\circle*{4}}
\epi
\hspace{10pt}
$\frac{1}{16}$
\rule[-23.85pt]{0pt}{53.7pt}
\hspace{-5pt}
\bpi(50,12)
\put(13,3){\circle{16}}
\put(29,3){\circle{16}}
\put(37,-10.85){\circle{16}}
\put(37,16.85){\circle{16}}
\put(5,3){\circle*{4}}
\put(21,3){\circle*{4}}
\put(33,-3.9){\circle*{4}}
\put(33,9.9){\circle*{4}}
\epi
\\
\hline
\end{tabular}
\end{tabular}
\end{center}
\caption{\protect\label{additionalgraphs}
Additional vacuum diagrams through five loops
caused by the one-loop insertion
$
\protect\rule[-7pt]{0pt}{20pt}
\protect\begin{picture}(50,20)
\protect\put(5,3){\protect\line(1,0){15}}
\protect\put(30,3){\protect\line(1,0){15}}
\protect\put(20,-2){\protect\line(1,0){10}}
\protect\put(20,-2){\protect\line(0,1){10}}
\protect\put(20,8){\protect\line(1,0){10}}
\protect\put(30,-2){\protect\line(0,1){10}}
\protect\put(25,3){\protect\makebox(0,0){$\scs 1$}}
\protect\end{picture}
\equiv
\protect\rule[-2pt]{0pt}{10pt}
\protect\begin{picture}(40,0)
\protect\put(5,3){\protect\line(1,0){30}}
\protect\put(20,3){\protect\circle*{4}}
\protect\end{picture}
$
and their weights.}
\end{table}

\subsection{One-Loop Resummation}
\label{oneloop}
Let us now try to adjust the one-loop two-point insertion
(\ref{oneloopcorr}) so that it cancels the trivial but ubiquitous
one-loop fluctuation
$\bpi(40,0)(0,5)
\put(5,3){\line(1,0){30}}
\put(20,8){\circle{10}}
\put(20,3){\circle*{4}}
\epi$,
present in most diagrams in Table \ref{allgraphs}.
For this purpose, define
\beq
\label{circle1}
\rule[-12pt]{0pt}{30pt}
\bpi(50,0)
\put(5,3){\line(1,0){15}}
\put(25,3){\circle{10}}
\put(30,3){\line(1,0){15}}
\put(25,3){\makebox(0,0){$\scs 1$}}
\epi
=
\rule[-7pt]{0pt}{20pt}
\bpi(50,20)
\put(5,3){\line(1,0){15}}
\put(30,3){\line(1,0){15}}
\put(20,-2){\line(1,0){10}}
\put(20,-2){\line(0,1){10}}
\put(20,8){\line(1,0){10}}
\put(30,-2){\line(0,1){10}}
\put(25,3){\makebox(0,0){$\scs 1$}}
\epi
+
\frac{1}{2}
\rule[-2pt]{0pt}{27pt}
\bpi(40,0)
\put(5,3){\line(1,0){30}}
\put(20,11){\circle{16}}
\put(20,3){\circle*{4}}
\epi
\eeq
and set
\beq
\rule[-7pt]{0pt}{20pt}
\bpi(50,20)
\put(5,3){\line(1,0){15}}
\put(30,3){\line(1,0){15}}
\put(20,-2){\line(1,0){10}}
\put(20,-2){\line(0,1){10}}
\put(20,8){\line(1,0){10}}
\put(30,-2){\line(0,1){10}}
\put(25,3){\makebox(0,0){$\scs L$}}
\epi
=0
\eeq
for $L>1$.
Then (\ref{newrecrel}) becomes
\bea
\label{rec1rel}
\rule[-18pt]{0pt}{42pt}
\bpi(40,0)(0,0)
\put(19,3){\circle{32}}
\put(19,3){\makebox(1,0){$L$}}
\epi
=
\frac{1}{L-1}\left[\frac{1}{6}
\rule[-18pt]{0pt}{42pt}
\bpi(54.78,0)(-14.78,0)
\put(19,3){\circle{32}}
\put(19,3){\makebox(1,0){$L-1$}}
\put(5.22,11.4){\circle*{4}}
\put(10.87,17.1){\circle*{4}}
\put(5.22,-5.13){\circle*{4}}
\put(10.87,-10.78){\circle*{4}}
\put(10.87,3){\oval(41.3,27.88)[l]}
\put(5.22,3){\oval(30,16.53)[l]}
\put(-9.78,3){\circle*{4}}
\epi
+
\rule[-18pt]{0pt}{42pt}
\bpi(58.23,0)(-18.23,0)
\put(19,3){\circle{32}}
\put(19,3){\makebox(1,0){$L-1$}}
\put(5.77,12){\circle*{4}}
\put(5.77,-6){\circle*{4}}
\put(5.77,8){\oval(28,8)[tl]}
\put(5.77,-2){\oval(28,8)[bl]}
\put(-8.23,3){\circle{10}}
\put(-8.23,3){\makebox(0,0){$\scs 1$}}
\epi
+\frac{1}{6}\sum_{l=2}^{L-2}
\rule[-18pt]{0pt}{42pt}
\bpi(96.46,0)(-56.46,0)
\put(19,3){\circle{32}}
\put(19,3){\makebox(1,0){$L-l$}}
\put(5.77,12){\circle*{4}}
\put(5.77,-6){\circle*{4}}
\put(5.77,3){\oval(28,18)[l]}
\put(-8.23,3){\circle*{4}}
\put(-22.23,3){\oval(28,18)[r]}
\put(-22.23,12){\circle*{4}}
\put(-22.23,-6){\circle*{4}}
\put(-35.46,3){\circle{32}}
\put(-35.46,3){\makebox(1,0){$l$}}
\epi
\right]
\eea
for $L>2$.

Now we show (i) that $W^{(3)}$ contains the two terms
on the right hand side of (\ref{circle1}) only in this combination
and (ii) that if $W^{(L)}$ with $L>2$ contains the two terms on the right
hand side of (\ref{circle1}) only in this combination, then this is
also true for $W^{(L+1)}$.

Using (\ref{w1sym}) and (\ref{rec1rel}) gives
\beq
\label{w3sym}
\rule[-18pt]{0pt}{42pt}
\bpi(42,0)(-2,0)
\put(19,3){\circle{32}}
\put(19,3){\makebox(1,0){$3$}}
\epi
=
\frac{1}{48}
\rule[-18pt]{0pt}{42pt}
\bpi(42,0)(-2,0)
\put(19,3){\circle{32}}
\put(19,3){\oval(32,10)}
\put(3,3){\circle*{4}}
\put(35,3){\circle*{4}}
\epi
+\frac{1}{4}
\rule[-11pt]{0pt}{28pt}
\bpi(38,0)
\put(10,3){\circle{10}}
\put(28,3){\circle{10}}
\put(19,8){\oval(18,8)[t]}
\put(19,-2){\oval(18,8)[b]}
\put(10,3){\makebox(0,0){$\scs 1$}}
\put(28,3){\makebox(0,0){$\scs 1$}}
\epi,
\eeq
which proves (i).

The only terms on the right hand side of (\ref{rec1rel}) that could
potentially violate (ii) are the terms with $l=2$ and/or $L-l=2$ in
the sum.
If both $l=2$ and $L-l=2$ (i.e.\ for $L=4$), the only term in the
sum is
\bea
\rule[-18pt]{0pt}{42pt}
\bpi(96.46,0)(-56.46,0)
\put(19,3){\circle{32}}
\put(19,3){\makebox(1,0){$2$}}
\put(5.77,12){\circle*{4}}
\put(5.77,-6){\circle*{4}}
\put(5.77,3){\oval(28,18)[l]}
\put(-8.23,3){\circle*{4}}
\put(-22.23,3){\oval(28,18)[r]}
\put(-22.23,12){\circle*{4}}
\put(-22.23,-6){\circle*{4}}
\put(-35.46,3){\circle{32}}
\put(-35.46,3){\makebox(1,0){$2$}}
\epi
&=&
\frac{1}{16}
\rule[-10pt]{0pt}{16pt}
\bpi(74,0)
\put(13,3){\circle{16}}
\put(29,3){\circle{16}}
\put(45,3){\circle{16}}
\put(61,3){\circle{16}}
\put(21,3){\circle*{4}}
\put(37,3){\circle*{4}}
\put(53,3){\circle*{4}}
\epi
+\frac{1}{8}
\rule[-11pt]{0pt}{17pt}
\bpi(63,0)
\put(13,3){\circle{16}}
\put(29,3){\circle{16}}
\put(45,8){\oval(16,8)[t]}
\put(45,-2){\oval(16,8)[b]}
\put(37,-2){\line(0,1){10}}
\put(21,3){\circle*{4}}
\put(37,3){\circle*{4}}
\put(48,-2){\line(1,0){10}}
\put(48,-2){\line(0,1){10}}
\put(48,8){\line(1,0){10}}
\put(58,-2){\line(0,1){10}}
\put(53,3){\makebox(0,0){$\scs 1$}}
\epi
+\frac{1}{8}
\rule[-11pt]{0pt}{17pt}
\bpi(63,0)
\put(34,3){\circle{16}}
\put(50,3){\circle{16}}
\put(18,8){\oval(16,8)[t]}
\put(18,-2){\oval(16,8)[b]}
\put(26,-2){\line(0,1){10}}
\put(26,3){\circle*{4}}
\put(42,3){\circle*{4}}
\put(5,-2){\line(1,0){10}}
\put(5,-2){\line(0,1){10}}
\put(5,8){\line(1,0){10}}
\put(15,-2){\line(0,1){10}}
\put(10,3){\makebox(0,0){$\scs 1$}}
\epi
+\frac{1}{4}
\rule[-11pt]{0pt}{17pt}
\bpi(63,0)
\put(18,8){\oval(16,8)[t]}
\put(18,-2){\oval(16,8)[b]}
\put(34,8){\oval(16,8)[t]}
\put(34,-2){\oval(16,8)[b]}
\put(26,-2){\line(0,1){10}}
\put(26,3){\circle*{4}}
\put(5,-2){\line(1,0){10}}
\put(5,-2){\line(0,1){10}}
\put(5,8){\line(1,0){10}}
\put(15,-2){\line(0,1){10}}
\put(10,3){\makebox(0,0){$\scs 1$}}
\put(37,-2){\line(1,0){10}}
\put(37,-2){\line(0,1){10}}
\put(37,8){\line(1,0){10}}
\put(47,-2){\line(0,1){10}}
\put(42,3){\makebox(0,0){$\scs 1$}}
\epi
\nn\\
&=&
\frac{1}{4}
\rule[-11pt]{0pt}{17pt}
\bpi(63,0)
\put(18,8){\oval(16,8)[t]}
\put(18,-2){\oval(16,8)[b]}
\put(34,8){\oval(16,8)[t]}
\put(34,-2){\oval(16,8)[b]}
\put(26,-2){\line(0,1){10}}
\put(26,3){\circle*{4}}
\put(10,3){\circle{10}}
\put(10,3){\makebox(0,0){$\scs 1$}}
\put(42,3){\circle{10}}
\put(42,3){\makebox(0,0){$\scs 1$}}
\epi.
\eea
If only one of $l$ and $L-l$ equals $2$ (i.e.\ for $L>4$), the potentially
dangerous terms in the sum are of the form
\bea
\rule[-18pt]{0pt}{42pt}
\bpi(96.46,0)(-56.46,0)
\put(19,3){\circle{32}}
\put(19,3){\makebox(1,0){$L-2$}}
\put(5.77,12){\circle*{4}}
\put(5.77,-6){\circle*{4}}
\put(5.77,3){\oval(28,18)[l]}
\put(-8.23,3){\circle*{4}}
\put(-22.23,3){\oval(28,18)[r]}
\put(-22.23,12){\circle*{4}}
\put(-22.23,-6){\circle*{4}}
\put(-35.46,3){\circle{32}}
\put(-35.46,3){\makebox(1,0){$2$}}
\epi
&=&
\frac{1}{4}
\rule[-18pt]{0pt}{42pt}
\bpi(85.23,0)
\put(13,3){\circle{16}}
\put(29,3){\circle{16}}
\put(21,3){\circle*{4}}
\put(37,3){\circle*{4}}
\put(64.23,3){\circle{32}}
\put(64.23,3){\makebox(1,0){$L-2$}}
\put(51,12){\circle*{4}}
\put(51,-6){\circle*{4}}
\put(51,3){\oval(28,18)[l]}
\epi
+\frac{1}{2}
\rule[-18pt]{0pt}{42pt}
\bpi(77.23,0)(8,0)
\put(28,-2){\oval(18,8)[b]}
\put(28,8){\oval(18,8)[t]}
\put(37,-2){\line(0,1){10}}
\put(37,3){\circle*{4}}
\put(64.23,3){\circle{32}}
\put(64.23,3){\makebox(1,0){$L-2$}}
\put(51,12){\circle*{4}}
\put(51,-6){\circle*{4}}
\put(51,3){\oval(28,18)[l]}
\put(13,-2){\line(1,0){10}}
\put(13,-2){\line(0,1){10}}
\put(13,8){\line(1,0){10}}
\put(23,-2){\line(0,1){10}}
\put(18,3){\makebox(0,0){$\scs 1$}}
\epi
=
\frac{1}{2}
\rule[-18pt]{0pt}{42pt}
\bpi(77.23,0)(8,0)
\put(28,-2){\oval(18,8)[b]}
\put(28,8){\oval(18,8)[t]}
\put(37,-2){\line(0,1){10}}
\put(37,3){\circle*{4}}
\put(64.23,3){\circle{32}}
\put(64.23,3){\makebox(1,0){$L-2$}}
\put(51,12){\circle*{4}}
\put(51,-6){\circle*{4}}
\put(51,3){\oval(28,18)[l]}
\put(18,3){\circle{10}}
\put(18,3){\makebox(0,0){$\scs 1$}}
\epi.
\eea
That is, in both cases only the combination on the right hand side of
(\ref{circle1}) appears.
This finishes the proof of (ii).

Once we have used the recursion relation (\ref{rec1rel}) to compute
any $W^{(L)}$, we can set
\beq
\label{adjust1}
\rule[-7pt]{0pt}{20pt}
\bpi(50,20)
\put(5,3){\line(1,0){15}}
\put(30,3){\line(1,0){15}}
\put(20,-2){\line(1,0){10}}
\put(20,-2){\line(0,1){10}}
\put(20,8){\line(1,0){10}}
\put(30,-2){\line(0,1){10}}
\put(25,3){\makebox(0,0){$\scs 1$}}
\epi
=
-\frac{1}{2}
\rule[-2pt]{0pt}{27pt}
\bpi(40,0)
\put(5,3){\line(1,0){30}}
\put(20,11){\circle{16}}
\put(20,3){\circle*{4}}
\epi
\eeq
such that
\beq
\label{adjust2}
\rule[-12pt]{0pt}{30pt}
\bpi(50,0)
\put(5,3){\line(1,0){15}}
\put(25,3){\circle{10}}
\put(30,3){\line(1,0){15}}
\put(25,3){\makebox(0,0){$\scs 1$}}
\epi
=0.
\eeq
This drastically reduces the number of diagrams in any given order,
since for $L>2$ no one-loop mass corrections is present anymore.
However, we are not allowed to use the result as an input for our
recursion relation (\ref{rec1rel}), since the two terms on the right
hand side of (\ref{circle1}) behave differently in the recursion
relation.

Note that now
\beq
\rule[-18pt]{0pt}{42pt}
\bpi(40,0)(0,0)
\put(19,3){\circle{32}}
\put(19,3){\makebox(1,0){$2$}}
\epi
=
-\frac{1}{8}
\rule[-10pt]{0pt}{26pt}
\bpi(42,0)
\put(21,3){\circle*{4}}
\put(13,3){\circle{16}}
\put(29,3){\circle{16}}
\epi
+\frac{1}{2}
\rule[-11pt]{0pt}{28pt}
\bpi(33,0)
\put(19,8){\oval(18,8)[t]}
\put(19,-2){\oval(18,8)[b]}
\put(28,-2){\line(0,1){10}}
\put(10,3){\circle{10}}
\put(10,3){\makebox(0,0){$\scs 1$}}
\epi.
\eeq
With the condition (\ref{adjust2}) this becomes
\beq
\label{twoloopvac}
\rule[-18pt]{0pt}{42pt}
\bpi(40,0)(0,0)
\put(19,3){\circle{32}}
\put(19,3){\makebox(1,0){$2$}}
\epi
=
-\frac{1}{8}
\rule[-10pt]{0pt}{26pt}
\bpi(42,0)
\put(21,3){\circle*{4}}
\put(13,3){\circle{16}}
\put(29,3){\circle{16}}
\epi,
\eeq
which is the only diagram left with a one-loop mass correction.

\begin{table}
\begin{center}
\begin{tabular}{c}
\begin{tabular}{|c|c|}
\hline
\begin{tabular}{c}number\\of loops\end{tabular}
&
remaining diagrams with their weights
\\
\hline
1,2,3,4&
\hspace{10pt}
$\frac{1}{2}$
\rule[-10pt]{0pt}{26pt}
\hspace{-7pt}
\bpi(26,12)
\put(13,3){\circle{16}}
\epi
\hspace{10pt}
$-\frac{1}{8}$
\rule[-10pt]{0pt}{26pt}
\hspace{-7pt}
\bpi(42,12)
\put(13,3){\circle{16}}
\put(29,3){\circle{16}}
\put(21,3){\circle*{4}}
\epi
\hspace{10pt}
$\frac{1}{48}$
\rule[-14pt]{0pt}{34pt}
\hspace{-5pt}
\bpi(34,12)
\put(17,3){\circle{24}}
\put(17,3){\oval(24,8)}
\put(5,3){\circle*{4}}
\put(29,3){\circle*{4}}
\epi
\hspace{10pt}
$\frac{1}{48}$
\rule[-14pt]{0pt}{34pt}
\hspace{-7pt}
\bpi(34,12)
\put(17,3){\circle{24}}
\put(6.6,9){\line(1,0){20.8}}
\put(6.6,9){\line(3,-5){10.4}}
\put(27.4,9){\line(-3,-5){10.4}}
\put(6.6,9){\circle*{4}}
\put(27.4,9){\circle*{4}}
\put(17,-9){\circle*{4}}
\epi
\\
\hline
5&
\hspace{10pt}
$\frac{1}{128}$
\rule[-14pt]{0pt}{34pt}
\hspace{-7pt}
\bpi(34,12)
\put(17,3){\circle{24}}
\put(8.5,-5.5){\line(1,0){17}}
\put(8.5,-5.5){\line(0,1){17}}
\put(8.5,11.5){\line(1,0){17}}
\put(25.5,-5.5){\line(0,1){17}}
\put(8.5,-5.5){\circle*{4}}
\put(8.5,11.5){\circle*{4}}
\put(25.5,-5.5){\circle*{4}}
\put(25.5,11.5){\circle*{4}}
\epi
\hspace{10pt}
$\frac{1}{144}$
\rule[-22pt]{0pt}{50pt}
\hspace{-7pt}
\bpi(42,12)
\put(21,-9){\circle{16}}
\put(21,15){\circle{16}}
\put(13,-9){\line(1,0){16}}
\put(13,15){\line(1,0){16}}
\put(13,3){\oval(16,24)[l]}
\put(29,3){\oval(16,24)[r]}
\put(13,-9){\circle*{4}}
\put(13,15){\circle*{4}}
\put(29,-9){\circle*{4}}
\put(29,15){\circle*{4}}
\epi
\hspace{10pt}
$\frac{1}{32}$
\rule[-18pt]{0pt}{42pt}
\hspace{-5pt}
\bpi(58,12)
\put(13,3){\circle{16}}
\put(45,3){\circle{16}}
\put(5,-5){\line(0,1){16}}
\put(25,-5){\oval(40,16)[b]}
\put(25,11){\oval(40,16)[t]}
\put(45,3){\oval(48,16)[l]}
\put(5,3){\circle*{4}}
\put(21,3){\circle*{4}}
\put(45,-5){\circle*{4}}
\put(45,11){\circle*{4}}
\epi
\\
\hline
6&
\hspace{10pt}
$\frac{1}{320}$
\rule[-14pt]{0pt}{34pt}
\hspace{-7pt}
\bpi(34,12)
\put(17,3){\circle{24}}
\put(17,15){\circle*{4}}
\put(5.59,6.71){\circle*{4}}
\put(28.41,6.71){\circle*{4}}
\put(9.95,-6.71){\circle*{4}}
\put(24.05,-6.71){\circle*{4}}
\put(9.95,-6.71){\line(1,0){14.1}}
\put(9.95,-6.71){\line(-1,3){4.36}}
\put(24.04,-6.71){\line(1,3){4.36}}
\put(5.59,6.71){\line(4,3){11.41}}
\put(28.41,6.71){\line(-4,3){11.41}}
\epi
\hspace{10pt}
$\frac{1}{288}$
\rule[-14pt]{0pt}{34pt}
\hspace{-7pt}
\bpi(58,12)
\put(17,3){\circle{24}}
\put(41,3){\circle{24}}
\put(17,3){\oval(8,24)}
\put(41,3){\oval(8,24)}
\put(29,3){\circle*{4}}
\put(17,15){\circle*{4}}
\put(17,-9){\circle*{4}}
\put(41,15){\circle*{4}}
\put(41,-9){\circle*{4}}
\epi
\hspace{10pt}
$\frac{1}{48}$
\rule[-18pt]{0pt}{42pt}
\hspace{-7pt}
\bpi(66,12)
\put(13,3){\circle{16}}
\put(13,-5){\line(0,1){16}}
\put(37,11){\oval(48,16)[t]}
\put(37,-5){\oval(48,16)[b]}
\put(45,-5){\circle{16}}
\put(45,11){\circle{16}}
\put(61,-5){\line(0,1){16}}
\put(13,11){\circle*{4}}
\put(13,-5){\circle*{4}}
\put(45,-13){\circle*{4}}
\put(45,3){\circle*{4}}
\put(45,19){\circle*{4}}
\epi
\\
&
\hspace{10pt}
$\frac{1}{32}$
\rule[-18pt]{0pt}{42pt}
\hspace{-5pt}
\bpi(74,12)
\put(13,3){\circle{16}}
\put(29,3){\circle{16}}
\put(61,3){\circle{16}}
\put(5,-5){\line(0,1){16}}
\put(33,-5){\oval(56,16)[b]}
\put(33,11){\oval(56,16)[t]}
\put(61,3){\oval(48,16)[l]}
\put(5,3){\circle*{4}}
\put(21,3){\circle*{4}}
\put(37,3){\circle*{4}}
\put(61,-5){\circle*{4}}
\put(61,11){\circle*{4}}
\epi
\hspace{10pt}
$\frac{1}{16}$
\rule[-18pt]{0pt}{42pt}
\hspace{-7pt}
\bpi(74,12)
\put(13,3){\circle{16}}
\put(61,3){\circle{16}}
\put(13,3){\oval(48,16)[r]}
\put(61,3){\oval(48,16)[l]}
\put(37,11){\oval(48,16)[t]}
\put(37,-5){\oval(48,16)[b]}
\put(13,11){\circle*{4}}
\put(13,-5){\circle*{4}}
\put(37,3){\circle*{4}}
\put(61,11){\circle*{4}}
\put(61,-5){\circle*{4}}
\epi
\hspace{10pt}
$\frac{1}{120}$
\rule[-14pt]{0pt}{34pt}
\hspace{-7pt}
\bpi(34,12)
\put(17,3){\circle{24}}
\put(17,15){\circle*{4}}
\put(5.59,6.71){\circle*{4}}
\put(28.41,6.71){\circle*{4}}
\put(9.95,-6.71){\circle*{4}}
\put(24.05,-6.71){\circle*{4}}
\put(5.59,6.71){\line(1,0){22.8}}
\put(9.95,-6.71){\line(1,3){7.05}}
\put(24.05,-6.71){\line(-1,3){7.05}}
\put(9.95,-6.71){\line(4,3){18.46}}
\put(24.05,-6.71){\line(-4,3){18.46}}
\epi
\\
\hline
\end{tabular}
\end{tabular}
\end{center}
\caption{\label{remaininggraphs}
Remaining diagrams with their weights through six loops with a one-loop
adjusted two-point insertion.}
\end{table}

In Table \ref{remaininggraphs} we list the diagrams with their weights
through six loops that are left after adjusting the two-point insertion
according to (\ref{adjust1}).
Through four and five loops this adjustment has been used in \cite{4loops}
and \cite{5loops}, respectively, to simplify the renormalization of the
vacuum energy in $\phi^4$ theory, which is used for the computation of
some universal critical amplitude ratios \cite{LMSD}.

As an alternative to explicitly constructing the graphs in Table
\ref{additionalgraphs} by recursion relations, we could replace
the propagator in the graphs of Table \ref{allgraphs} according to
\beq
G^{-1}\rightarrow G^{-1}+\De,
\eeq
i.e.
\beq
G\rightarrow(G^{-1}+\De)^{-1}
=G(\unit+\De G)^{-1}
=G+G\De G+G\De G\De G+\ldots
\eeq
with
\beq
\De_{12}=-\frac{1}{2}\int_{34}L_{1234}G_{34}.
\eeq
Diagrammatically, this amounts to replacing
\beq
\bpi(30,0)
\put(5,3){\line(1,0){20}}
\epi
\rightarrow
\bpi(30,0)
\put(5,3){\line(1,0){20}}
\epi
-\frac{1}{2}
\rule[-2pt]{0pt}{27pt}
\bpi(40,0)
\put(5,3){\line(1,0){30}}
\put(20,11){\circle{16}}
\put(20,3){\circle*{4}}
\epi
+\frac{1}{4}
\rule[-2pt]{0pt}{27pt}
\bpi(70,0)
\put(5,3){\line(1,0){60}}
\put(20,11){\circle{16}}
\put(20,3){\circle*{4}}
\put(50,11){\circle{16}}
\put(50,3){\circle*{4}}
\epi
-\frac{1}{8}
\rule[-2pt]{0pt}{27pt}
\bpi(100,0)
\put(5,3){\line(1,0){90}}
\put(20,11){\circle{16}}
\put(20,3){\circle*{4}}
\put(50,11){\circle{16}}
\put(50,3){\circle*{4}}
\put(80,11){\circle{16}}
\put(80,3){\circle*{4}}
\epi
+
\ldots
\eeq
and adding up the resulting graphs through the appropriate loop order.
The result is again the graphs in Table \ref{remaininggraphs} with the same
weights.

\section{General Case}
\label{general}
\subsection{Definitions}
Now let us generalize our treatment to the case with general interactions
through four powers in the field,
\beq
\label{egen}
E[\phi,C,J,G,K,L]=C+\int_1J_1\phi_1
+\frac{1}{2}\int_{12}G_{12}^{-1}\phi_1\phi_2
+\frac{1}{6}\int_{123}K_{123}\phi_1\phi_2\phi_3
+\frac{1}{24}\int_{1234}L_{1234}\phi_1\phi_2\phi_3\phi_4,
\eeq
where $G_{12}^{-1}$, $K_{123}$, $L_{1234}$ are symmetric in their
indices.
E.g., for a $Z_2$-symmetric single-component $\phi^4$ theory with
background field $\varphi$, i.e.\
\beq
\label{ephi}
E[\phi]=\int_1\left[\half\left(\p_\mu\varphi+\p_\mu\phi\right)^2
+\half m^2(\varphi+\phi)^2+{\ts\frac{1}{24}}\la(\varphi+\phi)^4+c\right]_1,
\eeq
we would have
\beq
\label{cexample}
C=\int_1\left[\half\left(\p_\mu\varphi\right)^2
+\half m^2\varphi^2+{\ts\frac{1}{24}}\la\varphi^4+c\right]_1,
\eeq
\beq
J_1=-\p^2\varphi_1+m^2\varphi_1+{\ts\frac{1}{6}}\la\varphi_1^3,
\eeq
\beq
G_{12}^{-1}=\de_{12}\left(\p_1\cdot\p_2+m^2+\half\la\varphi_1\varphi_2\right),
\eeq
\beq
K_{123}=\de_{12}\de_{13}\la\varphi_1,
\eeq
\beq
\label{lexample}
L_{1234}=\de_{12}\de_{13}\de_{14}\la.
\eeq

The partition function $Z$ and the negative free energy $W$ are given by
\beq
\label{zwgeneral}
Z[C,J,G,K,L]=\exp(W[C,J,G,K,L])=\int D\phi\exp(-E[\phi,C,J,G,K,L]).
\eeq

The energy is now regarded as a function of $C$ and a functional of
$\phi$, $J$, $G$, $K$, $L$.
We will mainly be interested in its dependence on $\phi$, $J$, $G$.

\subsection{Identities for $W_I$}
We continue with deriving identities similar to (\ref{w0symidentity}).
We now have the possibility to represent each occurrence of the field
$\phi$ by a derivative with respect to $J$.
We keep the number of these derivatives at a minimum and use
as much as possible derivatives with respect to $G$ to keep the
identities and the recursion relations derived from them as simple as
possible.

The identities we need are
\bea
\label{wssbidentity1}
0
&=&
\exp(-W[C,J,G,K,L])\int D\phi\frac{\de}{\de\phi_1}\exp(-E[\phi,C,J,G,K,L])
\nn\\
&=&
-J_1+\int_2G_{12}^{-1}\frac{\de W}{\de J_2}
+\int_{23}K_{123}\frac{\de W}{\de G_{23}^{-1}}
-\frac{1}{3}\int_{234}L_{1234}
\left(\frac{\de^2W}{\de J_2\de G_{34}^{-1}}
+\frac{\de W}{\de J_2}\frac{\de W}{\de G_{34}^{-1}}\right)
\eea
and
\bea
\label{wssbidentity2}
0
&=&
\exp(-W[C,J,G,K,L])\int D\phi
\frac{\de}{\de\phi_1}\{\phi_2\exp(-E[\phi,C,J,G,K,L])\}
\nn\\
&=&
\de_{12}+J_1\frac{\de W}{\de J_2}
+2\int_3G_{13}^{-1}\frac{\de W}{\de G_{23}^{-1}}
\nn\\
&&
-\int_{34}K_{134}\left(\frac{\de^2W}{\de J_2\de G_{34}^{-1}}
+\frac{\de W}{\de J_2}\frac{\de W}{\de G_{34}^{-1}}\right)
-\frac{2}{3}\int_{345}L_{1345}
\left(\frac{\de^2W}{\de G_{23}^{-1}\de G_{45}^{-1}}
+\frac{\de W}{\de G_{23}^{-1}}\frac{\de W}{\de G_{45}^{-1}}\right),
\eea
where we have followed similar steps as for the derivation of
(\ref{w0symidentity}) except that we have not yet replaced
derivatives with respect to $G^{-1}$ by those with respect to $G$.

Split $W$ again into a free and an interacting part,
\beq
\label{ww0wi}
W=W_0+W_I\equiv W|_{K,L=0}+W_I.
\eeq
For $W_0$, (\ref{wssbidentity1}) and (\ref{wssbidentity2}) reduce to
\beq
\label{w0ssbidentity1}
-J_1+\int_2G_{12}^{-1}\frac{\de W_0}{\de J_2}=0
\eeq
and
\beq
\label{w0ssbidentity2}
\de_{12}+J_1\frac{\de W_0}{\de J_2}
+2\int_3G_{13}^{-1}\frac{\de W_0}{\de G_{23}^{-1}}=0,
\eeq
respectively.
Combining them and using the results from appendix \ref{symg}, we get the
useful relations
\beq
\label{dw0ssbdj1}
\frac{\de W_0}{\de J_1}=\int_2G_{12}J_2,
\eeq

\beq
\label{dw0ssbdginv11}
\frac{\de W_0}{\de G_{12}^{-1}}
=-\frac{1}{2}\left(G_{12}+\int_{34}G_{13}J_3G_{24}J_4\right),
\eeq

\beq
\label{dw0ssbdginvdj}
\frac{\de^2W_0}{\de J_1\de G_{23}^{-1}}
=-\frac{1}{2}\int_4\left(G_{12}G_{34}+G_{13}G_{24}\right)J_4
\eeq
and
\beq
\label{d2w0ssbdginv112}
\frac{\de^2 W_0}{\de G_{12}^{-1}\de G_{34}^{-1}}
=\frac{1}{4}\bigg[G_{13}G_{24}+G_{14}G_{23}
+\int_{56}(G_{13}G_{25}G_{46}+G_{14}G_{25}G_{36}
+G_{23}G_{15}G_{46}+G_{24}G_{15}G_{36})J_5J_6\bigg].
\eeq

With the same normalization of the path integral measure $D\phi$ as
before we have
\bea
\label{w0ssb1}
W_0[C,J,G]
&=&
\ln\int D\phi\exp\left(-C-\int_1J_1\phi_1
-\frac{1}{2}\int_{12}G_{12}^{-1}\phi_1\phi_2\right)
\nn\\
&=&
\ln\int D\phi\exp
\left[-C-\int_1J_1\left(\phi_1-\int_2G_{12}J_2\right)
-\frac{1}{2}\int_{12}G_{12}^{-1}\left(\phi_1-\int_3G_{13}J_3\right)
\left(\phi_2-\int_4G_{24}J_4\right)\right]
\nn\\
&=&
-C+\frac{1}{2}\int_{12}G_{12}J_1J_2
+\ln\int D\phi\exp\left(-\frac{1}{2}\int_{12}G_{12}^{-1}\phi_1\phi_2\right)
\nn\\
&=&
-C+\frac{1}{2}\int_{12}G_{12}J_1J_2-\frac{1}{2}\int_1(\ln G^{-1})_{11}.
\eea

Subtracting (\ref{w0ssbidentity1}) from (\ref{wssbidentity1}),
multiplying with $\int_2G_{12}J_2$, integrating over $x_1$
and using (\ref{dw0ssbdj1})-(\ref{dw0ssbdginvdj}) and (\ref{ddginv}),
we get
\bea
\label{wissbid1c}
&&
\int_1J_1\frac{\de W_I}{\de J_1}
-\frac{1}{2}\int_{1234}K_{123}G_{12}G_{34}J_4
-\frac{1}{2}\int_{123456}K_{123}G_{14}J_4G_{25}J_5G_{36}J_6
\nn\\
&&
+\frac{1}{2}\int_{123456}L_{1234}G_{12}G_{35}J_5G_{46}J_6
+\frac{1}{6}\int_{12345678}L_{1234}G_{15}J_5G_{26}J_6G_{37}J_7G_{48}J_8
\nn\\
&&
-\int_{123456}K_{123}G_{14}J_4G_{25}G_{36}\frac{\de W_I}{\de G_{56}}
+\frac{1}{6}\int_{12345}L_{1234}G_{12}G_{35}J_5\frac{\de W_I}{\de J_4}
\nn\\
&&
+\frac{1}{6}
\int_{1234567}L_{1234}G_{15}J_5G_{26}J_6G_{37}J_7\frac{\de W_I}{\de J_4}
+\frac{1}{3}\int_{12345678}L_{1234}G_{15}J_5
G_{26}J_6G_{37}G_{48}\frac{\de W_I}{\de G_{78}}
\nn\\
&&
+\frac{1}{3}\int_{1234567}L_{1234}G_{15}J_5G_{36}G_{47}
\frac{\de^2W_I}{\de J_2\de G_{67}}
+\frac{1}{3}\int_{1234567}L_{1234}G_{15}J_5G_{36}G_{47}
\frac{\de W_I}{\de J_2}\frac{\de W_I}{\de G_{67}}=0.
\eea

Subtracting (\ref{w0ssbidentity2}) from (\ref{wssbidentity2}),
setting $x_2=x_1$, integrating over $x_1$ and using
(\ref{dw0ssbdj1})-(\ref{d2w0ssbdginv112}) and
(\ref{ddginv}) we get
\bea
\label{wissbid2c}
\lefteqn{\int_1J_1\frac{\de W_I}{\de J_1}
-2\int_{12}G_{12}\frac{\de W_I}{\de G_{12}}}
\nn\\
&&
+\frac{3}{2}\int_{1234}K_{123}G_{12}G_{34}J_4
+\frac{1}{2}\int_{123456}K_{123}G_{14}J_4G_{25}J_5G_{36}J_6
\nn\\
&&
-\frac{1}{2}\int_{1234}L_{1234}G_{12}G_{34}
-\int_{123456}L_{1234}G_{12}G_{35}J_5G_{46}J_6
-\frac{1}{6}\int_{12345678}L_{1234}G_{15}J_5G_{26}J_6G_{37}J_7G_{48}J_8
\nn\\
&&
+\frac{1}{2}\int_{123}K_{123}G_{12}\frac{\de W_I}{\de J_3}
+\frac{1}{2}\int_{12345}K_{123}G_{14}J_4G_{25}J_5\frac{\de W_I}{\de J_3}
+\int_{123456}K_{123}G_{14}J_4G_{25}G_{36}\frac{\de W_I}{\de G_{56}}
\nn\\
&&
+\int_{12345}K_{123}G_{24}G_{35}\frac{\de^2W_I}{\de J_1\de G_{45}}
+\int_{12345}K_{123}G_{24}G_{35}
\frac{\de W_I}{\de J_1}\frac{\de W_I}{\de G_{45}}
\nn\\
&&
-2\int_{123456}L_{1234}G_{12}G_{35}G_{46}
\frac{\de W_I}{\de G_{56}}
-\frac{2}{3}\int_{12345678}L_{1234}G_{15}J_5G_{26}J_6G_{37}G_{48}
\frac{\de W_I}{\de G_{78}}
\nn\\
&&
-\frac{2}{3}\int_{12345678}L_{1234}G_{15}G_{26}
G_{37}G_{48}\frac{\de^2W_I}{\de G_{56}\de G_{78}}
-\frac{2}{3}\int_{12345678}L_{1234}G_{15}G_{26}G_{37}G_{48}
\frac{\de W_I}{\de G_{56}}\frac{\de W_I}{\de G_{78}}=0.
\eea

\subsection{Change of Variables}
\label{chofvar}
Instead of representing (\ref{w0ssb1})-(\ref{wissbid2c}) graphically,
let us first perform a change of variables that reduces the amount
of work needed for solving the recursion relations to be derived.
Since $J$ is always connected to a free propagator $G$, we can as well
define a modified current
\beq
\jb_1= \int_2G_{12}J_2
\eeq
which already incorporates this propagator.
Then (\ref{w0ssb1}) can be rewritten as
\beq
\label{w0ssb2}
W_0[C,\jb,G]=-C+\frac{1}{2}\int_{12}G_{12}^{-1}\jb_1\jb_2
-\frac{1}{2}\int_1(\ln G^{-1})_{11}.
\eeq
Performing this change of variables introduces into the $W$ identities
double derivatives with respect to $\jb$ which we would like to
avoid in favor of derivatives with respect to free correlation function
$G$.
This can be achieved with the result
\beq
\label{wijbjbwijbwijb}
\left(\frac{\de^2W_I}{\de\jb_1\de\jb_2}\right)_G
+\left(\frac{\de W_I}{\de\jb_1}\right)_G
\left(\frac{\de W_I}{\de\jb_2}\right)_G
=
2\left(\frac{\de W_I}{\de G_{12}}\right)_{\jb}
\eeq
of appendix \ref{d2wdjbdjb}.

Eqs.\ (\ref{wissbid1c}) and (\ref{wissbid2c}) then become
\bea
\label{wissbid1e}
&&
\int_1\jb_1\frac{\de W_I}{\de\jb_1}
\nn\\
&&
-\frac{1}{2}\int_{123}K_{123}G_{12}\jb_3
-\frac{1}{2}\int_{123}K_{123}\jb_1\jb_2\jb_3
+\frac{1}{2}\int_{1234}L_{1234}G_{12}\jb_3\jb_4
+\frac{1}{6}\int_{1234}L_{1234}\jb_1\jb_2\jb_3\jb_4
\nn\\
&&
-\int_{12345}K_{123}\jb_1G_{24}G_{35}\frac{\de W_I}{\de G_{45}}
-\int_{1234}K_{123}\jb_1\jb_2G_{34}\frac{\de W_I}{\de\jb_4}
\nn\\
&&
+\frac{1}{2}\int_{12345}L_{1234}G_{12}\jb_3
G_{45}\frac{\de W_I}{\de\jb_5}
+\frac{1}{2}\int_{12345}L_{1234}\jb_1\jb_2\jb_3
G_{45}\frac{\de W_I}{\de\jb_5}
+\int_{123456}L_{1234}\jb_1\jb_2
G_{35}G_{46}\frac{\de W_I}{\de G_{56}}
\nn\\
&&
+\frac{1}{3}\int_{1234567}L_{1234}\jb_1
G_{25}G_{36}G_{47}\frac{\de^2W_I}{\de\jb_5\de G_{67}}
+\frac{1}{3}\int_{1234567}L_{1234}\jb_1G_{25}G_{36}G_{47}
\frac{\de W_I}{\de\jb_5}\frac{\de W_I}{\de G_{67}}
=0
\eea
and
\bea
\label{wissbid2e}
\lefteqn{2\int_{12}G_{12}\frac{\de W_I}{\de G_{12}}
+\int_1\jb_1\frac{\de W_I}{\de\jb_1}}
\nn\\
&&
-\frac{3}{2}\int_{123}K_{123}G_{12}\jb_3
-\frac{1}{2}\int_{123}K_{123}\jb_1\jb_2\jb_3
\nn\\
&&
+\frac{1}{2}\int_{1234}L_{1234}G_{12}G_{34}
+\int_{1234}L_{1234}G_{12}\jb_3\jb_4
+\frac{1}{6}\int_{1234}L_{1234}\jb_1\jb_2\jb_3\jb_4
\nn\\
&&
-\frac{3}{2}\int_{1234}K_{123}G_{12}G_{34}\frac{\de W_I}{\de\jb_4}
-\frac{3}{2}\int_{1234}K_{123}\jb_1\jb_2
G_{34}\frac{\de W_I}{\de\jb_4}
-3\int_{12345}K_{123}\jb_1G_{24}G_{35}\frac{\de W_I}{\de G_{45}}
\nn\\
&&
-\int_{123456}K_{123}G_{14}G_{25}G_{36}
\frac{\de^2W_I}{\de\jb_4\de G_{56}}
-\int_{123456}K_{123}G_{24}G_{35}
G_{16}\frac{\de W_I}{\de\jb_6}\frac{\de W_I}{\de G_{45}}
\nn\\
&&
+2\int_{123456}L_{1234}G_{12}G_{35}G_{46}
\frac{\de W_I}{\de G_{56}}
+2\int_{12345}L_{1234}G_{12}\jb_3G_{45}
\frac{\de W_I}{\de\jb_5}
\nn\\
&&
+2\int_{123456}L_{1234}\jb_1\jb_2
G_{35}G_{46}\frac{\de W_I}{\de G_{56}}
+\frac{2}{3}\int_{12345}L_{1234}\jb_1\jb_2\jb_3
G_{45}\frac{\de W_I}{\de\jb_5}
\nn\\
&&
+\frac{2}{3}\int_{12345678}L_{1234}G_{15}G_{26}
G_{37}G_{48}\frac{\de^2W_I}{\de G_{56}\de G_{78}}
+\frac{4}{3}\int_{1234567}L_{1234}\jb_1G_{25}G_{36}G_{47}
\frac{\de^2W_I}{\de G_{56}\de\jb_7}
\nn\\
&&
+\frac{2}{3}\int_{12345678}L_{1234}G_{15}G_{26}G_{37}G_{48}
\frac{\de W_I}{\de G_{56}}\frac{\de W_I}{\de G_{78}}
+\frac{4}{3}\int_{1234567}L_{1234}\jb_1G_{25}G_{36}G_{47}
\frac{\de W_I}{\de\jb_5}\frac{\de W_I}{\de G_{67}}
=0.\;\;\;\;
\eea

To represent (\ref{wissbid1e}) and (\ref{wissbid2e}) graphically, write
for the derivatives of $W_I$ with respect to $\jb$ and $G$
\beq
-\frac{\de W_I}{\de\jb_1}=
\rule[-18pt]{0pt}{42pt}
\bpi(55,0)(-23,0)
\put(8,3){\circle{32}}
\put(8,3){\makebox(1,0){$W_I$}}
\put(-8,3){\circle{4}}
\put(-18,1){$\scs 1$}
\epi,
\;\;\;\;
\frac{\de W_I}{\de G_{12}}=
\rule[-18pt]{0pt}{42pt}
\bpi(50,0)(-10,0)
\put(19,3){\circle{32}}
\put(19,3){\makebox(1,0){$W_I$}}
\put(-0.5,13){\makebox(0,0){$\scs 1$}}
\put(-0.5,-7){\makebox(0,0){$\scs 2$}}
\put(5.77,12){\circle*{4}}
\put(5.77,-6){\circle*{4}}
\epi,
\;\;\;\;
-\frac{\de^2W_I}{\de\jb_3\de G_{12}}=
\rule[-18pt]{0pt}{42pt}
\bpi(47,0)(-7,0)
\put(19,3){\circle{32}}
\put(19,3){\makebox(1,0){$W_I$}}
\put(6,21){\makebox(0,0){$\scs 1$}}
\put(0,15){\makebox(0,0){$\scs 2$}}
\put(2,-12){\makebox(0,0){$\scs 3$}}
\put(5.22,11.4){\circle*{4}}
\put(10.87,17.1){\circle*{4}}
\put(7.69,-8.31){\circle{4}}
\epi,
\;\;\;\;
\frac{\de^2W_I}{\de G_{12}\de G_{34}}=
\rule[-18pt]{0pt}{42pt}
\bpi(47,0)(-7,0)
\put(19,3){\circle{32}}
\put(19,3){\makebox(1,0){$W_I$}}
\put(6,21){\makebox(0,0){$\scs 1$}}
\put(0,15){\makebox(0,0){$\scs 2$}}
\put(0,-9){\makebox(0,0){$\scs 3$}}
\put(6,-15){\makebox(0,0){$\scs 4$}}
\put(5.22,11.4){\circle*{4}}
\put(10.87,17.1){\circle*{4}}
\put(5.22,-5.13){\circle*{4}}
\put(10.87,-10.78){\circle*{4}}
\epi
\eeq
and use the vertices
\beq
-L_{1234}=
\rule[-14pt]{0pt}{34pt}
\bpi(34,0)(-4,0)
\put(5,-5){\line(1,1){16}}
\put(5,11){\line(1,-1){16}}
\put(13,3){\circle*{4}}
\put(25,15){\makebox(0,0){$\scs 1$}}
\put(1,15){\makebox(0,0){$\scs 2$}}
\put(1,-9){\makebox(0,0){$\scs 3$}}
\put(25,-9){\makebox(0,0){$\scs 4$}}
\epi,
\;\;\;\;
-K_{123}=
\rule[-12pt]{0pt}{42pt}
\bpi(38,0)(-4,0)
\put(15,3){\circle*{4}}
\put(15,3){\line(0,1){11.66}}
\put(15,3){\line(5,-3){10}}
\put(15,3){\line(-5,-3){10}}
\put(15,21){\makebox(0,0){$\scs 1$}}
\put(1,-7){\makebox(0,0){$\scs 2$}}
\put(29,-7){\makebox(0,0){$\scs 3$}}
\epi,
\;\;\;\;
-\jb_1=
\rule[-15pt]{0pt}{29pt}
\bpi(14,0)
\put(6,-5){\line(0,1){12}}
\put(8,-5){\line(0,1){12}}
\put(7,7){\circle*{4}}
\put(7,-10){\makebox(0,0){$\scs 1$}}
\epi,
\;\;\;\;
-C
=
\rule[-2pt]{0pt}{14pt}
\bpi(10,0)
\put(5,3){\circle*{4}}
\epi.
\eeq
Propagators $G$ are indicated by lines connected at both ends.
The double line on $\jb$ indicates that the propagator is absorbed into
our new current $\jb$, so derivatives with respect to $G$ act only on
propagators not connected to a current, i.e.\ on single lines [see
however (\ref{jbjb})].
All space arguments that are not indicated by numbers are integrated over.

We can write (\ref{w0ssb2}) now as
\beq
\label{w0ssb2a}
W_0[C,\jb,G]=
\rule[-2pt]{0pt}{14pt}
\bpi(14,0)
\put(7,3){\circle*{4}}
\epi
+\frac{1}{2}
\rule[-2pt]{0pt}{14pt}
\bpi(27,0)
\put(7,3){\circle*{4}}
\put(7,2){\line(1,0){13}}
\put(7,4){\line(1,0){13}}
\put(20,3){\circle*{4}}
\epi
+\frac{1}{2}
\rule[-10pt]{0pt}{26pt}
\bpi(26,0)
\put(13,3){\circle{16}}
\epi,
\eeq
where by definition
\beq
\label{jbjb}
\rule[-4pt]{0pt}{14pt}
\bpi(27,0)
\put(7,3){\circle*{4}}
\put(9,2){\line(1,0){9}}
\put(9,4){\line(1,0){9}}
\put(20,3){\circle*{4}}
\epi
=
\int_{12}G_{12}^{-1}\jb_1\jb_2,
\eeq
(\ref{wissbid1e}) as
\bea
\label{wissbid1f}
\lefteqn{\rule[-18pt]{0pt}{42pt}
\bpi(57,0)(-17,0)
\put(19,3){\circle{32}}
\put(19,3){\makebox(1,0){$W_I$}}
\put(3,3){\circle{4}}
\put(-10,2){\line(1,0){11}}
\put(-10,4){\line(1,0){11}}
\put(-10,3){\circle*{4}}
\epi}
\nn\\
&=&
\frac{1}{2}
\rule[-11pt]{0pt}{35pt}
\bpi(38.24,0)(-4.12,0)
\put(15,3){\circle*{4}}
\put(14,3){\line(0,1){14}}
\put(16,3){\line(0,1){14}}
\put(15,17){\circle*{4}}
\put(15.5,2.13){\line(-5,-3){10.39}}
\put(14.5,3.87){\line(-5,-3){10.39}}
\put(2.88,-4){\circle*{4}}
\put(14.5,2.13){\line(5,-3){10.39}}
\put(15.5,3.87){\line(5,-3){10.39}}
\put(27.12,-4){\circle*{4}}
\epi
+\frac{1}{6}
\rule[-15.41pt]{0pt}{36.82pt}
\bpi(36.82,0)(-3.41,0)
\put(15,3){\circle*{4}}
\put(14.29,2.29){\line(-1,1){10}}
\put(15.71,3.71){\line(-1,1){10}}
\put(3.59,14.41){\circle*{4}}
\put(15.71,2.29){\line(-1,-1){10}}
\put(14.29,3.71){\line(-1,-1){10}}
\put(3.59,-8.41){\circle*{4}}
\put(14.29,3.71){\line(1,1){10}}
\put(15.71,2.29){\line(1,1){10}}
\put(26.41,14.41){\circle*{4}}
\put(15.71,3.71){\line(1,-1){10}}
\put(14.29,2.29){\line(1,-1){10}}
\put(26.41,-8.41){\circle*{4}}
\epi
+\frac{1}{2}
\rule[-10pt]{0pt}{21pt}
\bpi(41,0)
\put(28,3){\circle{16}}
\put(20,3){\circle*{4}}
\put(7,2){\line(1,0){11}}
\put(7,4){\line(1,0){11}}
\put(7,3){\circle*{4}}
\epi
+\frac{1}{2}
\rule[-15.41pt]{0pt}{36.82pt}
\bpi(39.41,0)(-3.41,0)
\put(15,3){\circle*{4}}
\put(14.29,2.29){\line(-1,1){10}}
\put(15.71,3.71){\line(-1,1){10}}
\put(3.59,14.41){\circle*{4}}
\put(15.71,2.29){\line(-1,-1){10}}
\put(14.29,3.71){\line(-1,-1){10}}
\put(3.59,-8.41){\circle*{4}}
\put(23,3){\circle{16}}
\epi
\nn\\
&&
+
\rule[-18pt]{0pt}{42pt}
\bpi(68.41,0)(-3.41,0)
\put(14.29,2.29){\line(-1,1){10}}
\put(15.71,3.71){\line(-1,1){10}}
\put(3.59,14.41){\circle*{4}}
\put(15.71,2.29){\line(-1,-1){10}}
\put(14.29,3.71){\line(-1,-1){10}}
\put(3.59,-8.41){\circle*{4}}
\put(15,3){\line(1,0){11}}
\put(15,3){\circle*{4}}
\put(28,3){\circle{4}}
\put(44,3){\circle{32}}
\put(44,3){\makebox(1,0){$W_I$}}
\epi
+\frac{1}{2}
\rule[-18pt]{0pt}{42pt}
\bpi(68.41,0)(-3.41,0)
\put(14.29,2.29){\line(-1,1){10}}
\put(15.71,3.71){\line(-1,1){10}}
\put(3.59,14.41){\circle*{4}}
\put(15.71,2.29){\line(-1,-1){10}}
\put(14.29,3.71){\line(-1,-1){10}}
\put(3.59,-8.41){\circle*{4}}
\put(3.59,3){\circle*{4}}
\put(3.59,2){\line(1,0){11.41}}
\put(3.59,4){\line(1,0){11.41}}
\put(15,3){\line(1,0){11}}
\put(15,3){\circle*{4}}
\put(28,3){\circle{4}}
\put(44,3){\circle{32}}
\put(44,3){\makebox(1,0){$W_I$}}
\epi
+\frac{1}{2}
\rule[-13pt]{0pt}{42pt}
\bpi(76,0)(2,0)
\put(9,3){\circle*{4}}
\put(41,3){\circle{4}}
\put(25,11){\circle{16}}
\put(9,2){\line(1,0){16}}
\put(9,4){\line(1,0){16}}
\put(25,3){\line(1,0){14}}
\put(25,3){\circle*{4}}
\put(57,3){\circle{32}}
\put(57,3){\makebox(1,0){$W_I$}}
\epi
+
\rule[-18pt]{0pt}{42pt}
\bpi(68.23,0)(-28.23,0)
\put(19,3){\circle{32}}
\put(19,3){\makebox(1,0){$W_I$}}
\put(5.77,12){\circle*{4}}
\put(5.77,-6){\circle*{4}}
\put(5.77,3){\oval(28,18)[l]}
\put(-8.23,3){\circle*{4}}
\put(-21.23,2){\line(1,0){13}}
\put(-21.23,4){\line(1,0){13}}
\put(-21.23,3){\circle*{4}}
\epi
+
\rule[-15.41pt]{0pt}{36.82pt}
\bpi(66.64,0)(-3.41,0)
\put(15,3){\circle*{4}}
\put(14.29,2.29){\line(-1,1){10}}
\put(15.71,3.71){\line(-1,1){10}}
\put(3.59,14.41){\circle*{4}}
\put(15.71,2.29){\line(-1,-1){10}}
\put(14.29,3.71){\line(-1,-1){10}}
\put(3.59,-8.41){\circle*{4}}
\put(42.23,3){\circle{32}}
\put(42.23,3){\makebox(1,0){$W_I$}}
\put(29,12){\circle*{4}}
\put(29,-6){\circle*{4}}
\put(29,3){\oval(28,18)[l]}
\epi
\nn\\
&&
+\frac{1}{3}
\rule[-18pt]{0pt}{42pt}
\bpi(68.23,0)(-28.23,0)
\put(19,3){\circle{32}}
\put(19,3){\makebox(1,0){$W_I$}}
\put(5.77,12){\circle*{4}}
\put(5.77,-6){\circle{4}}
\put(3.77,3){\oval(24,18)[l]}
\put(-8.23,3){\circle*{4}}
\put(-21.23,2){\line(1,0){13}}
\put(-21.23,4){\line(1,0){13}}
\put(-8.23,3){\line(1,0){11.23}}
\put(3,3){\circle*{4}}
\put(-21.23,3){\circle*{4}}
\epi
+\frac{1}{3}
\rule[-18pt]{0pt}{42pt}
\bpi(98.23,0)(-60,0)
\put(17.23,3){\circle{32}}
\put(17.23,3){\makebox(1,0){$W_I$}}
\put(4,12){\circle*{4}}
\put(4,-6){\circle*{4}}
\put(4,3){\oval(28,18)[l]}
\put(-11,3){\line(0,1){14}}
\put(-9,3){\line(0,1){14}}
\put(-10,17){\circle*{4}}
\put(-10,3){\circle*{4}}
\put(-23,3){\circle{4}}
\put(-10,3){\line(-1,0){11}}
\put(-39,3){\circle{32}}
\put(-39,3){\makebox(1,0){$W_I$}}
\epi
\eea
and (\ref{wissbid2e}) as
\bea
\label{wissbid2f}
\lefteqn{2
\rule[-18pt]{0pt}{42pt}
\bpi(53.23,0)(-13.23,0)
\put(19,3){\circle{32}}
\put(19,3){\makebox(1,0){$W_I$}}
\put(5.77,12){\circle*{4}}
\put(5.77,-6){\circle*{4}}
\put(5.77,3){\oval(28,18)[l]}
\epi
+
\rule[-18pt]{0pt}{42pt}
\bpi(57,0)(-17,0)
\put(19,3){\circle{32}}
\put(19,3){\makebox(1,0){$W_I$}}
\put(3,3){\circle{4}}
\put(-10,2){\line(1,0){11}}
\put(-10,4){\line(1,0){11}}
\put(-10,3){\circle*{4}}
\epi}
\nn\\
&=&
\frac{1}{2}
\rule[-11pt]{0pt}{35pt}
\bpi(38.24,0)(-4.12,0)
\put(15,3){\circle*{4}}
\put(14,3){\line(0,1){14}}
\put(16,3){\line(0,1){14}}
\put(15,17){\circle*{4}}
\put(15.5,2.13){\line(-5,-3){10.39}}
\put(14.5,3.87){\line(-5,-3){10.39}}
\put(2.88,-4){\circle*{4}}
\put(14.5,2.13){\line(5,-3){10.39}}
\put(15.5,3.87){\line(5,-3){10.39}}
\put(27.12,-4){\circle*{4}}
\epi
+\frac{1}{6}
\rule[-15.41pt]{0pt}{36.82pt}
\bpi(36.82,0)(-3.41,0)
\put(15,3){\circle*{4}}
\put(14.29,2.29){\line(-1,1){10}}
\put(15.71,3.71){\line(-1,1){10}}
\put(3.59,14.41){\circle*{4}}
\put(15.71,2.29){\line(-1,-1){10}}
\put(14.29,3.71){\line(-1,-1){10}}
\put(3.59,-8.41){\circle*{4}}
\put(14.29,3.71){\line(1,1){10}}
\put(15.71,2.29){\line(1,1){10}}
\put(26.41,14.41){\circle*{4}}
\put(15.71,3.71){\line(1,-1){10}}
\put(14.29,2.29){\line(1,-1){10}}
\put(26.41,-8.41){\circle*{4}}
\epi
+\frac{3}{2}
\rule[-10pt]{0pt}{21pt}
\bpi(41,0)
\put(28,3){\circle{16}}
\put(20,3){\circle*{4}}
\put(7,2){\line(1,0){13}}
\put(7,4){\line(1,0){13}}
\put(7,3){\circle*{4}}
\epi
+
\rule[-15.41pt]{0pt}{36.82pt}
\bpi(39.41,0)(-3.41,0)
\put(15,3){\circle*{4}}
\put(14.29,2.29){\line(-1,1){10}}
\put(15.71,3.71){\line(-1,1){10}}
\put(3.59,14.41){\circle*{4}}
\put(15.71,2.29){\line(-1,-1){10}}
\put(14.29,3.71){\line(-1,-1){10}}
\put(3.59,-8.41){\circle*{4}}
\put(23,3){\circle{16}}
\epi
+\frac{1}{2}
\rule[-10pt]{0pt}{26pt}
\bpi(42,0)
\put(21,3){\circle*{4}}
\put(13,3){\circle{16}}
\put(29,3){\circle{16}}
\epi
\nn\\
&&
+\frac{3}{2}
\rule[-18pt]{0pt}{42pt}
\bpi(68.41,0)(-3.41,0)
\put(14.29,2.29){\line(-1,1){10}}
\put(15.71,3.71){\line(-1,1){10}}
\put(3.59,14.41){\circle*{4}}
\put(15.71,2.29){\line(-1,-1){10}}
\put(14.29,3.71){\line(-1,-1){10}}
\put(3.59,-8.41){\circle*{4}}
\put(15,3){\line(1,0){11}}
\put(15,3){\circle*{4}}
\put(28,3){\circle{4}}
\put(44,3){\circle{32}}
\put(44,3){\makebox(1,0){$W_I$}}
\epi
+\frac{2}{3}
\rule[-18pt]{0pt}{42pt}
\bpi(68.41,0)(-3.41,0)
\put(14.29,2.29){\line(-1,1){10}}
\put(15.71,3.71){\line(-1,1){10}}
\put(3.59,14.41){\circle*{4}}
\put(15.71,2.29){\line(-1,-1){10}}
\put(14.29,3.71){\line(-1,-1){10}}
\put(3.59,-8.41){\circle*{4}}
\put(3.59,3){\circle*{4}}
\put(3.59,2){\line(1,0){11.41}}
\put(3.59,4){\line(1,0){11.41}}
\put(15,3){\line(1,0){11}}
\put(15,3){\circle*{4}}
\put(28,3){\circle{4}}
\put(44,3){\circle{32}}
\put(44,3){\makebox(1,0){$W_I$}}
\epi
+\frac{3}{2}
\rule[-18pt]{0pt}{42pt}
\bpi(71,0)
\put(13,3){\circle{16}}
\put(21,3){\circle*{4}}
\put(21,3){\line(1,0){11}}
\put(34,3){\circle{4}}
\put(50,3){\circle{32}}
\put(50,3){\makebox(1,0){$W_I$}}
\epi
+2
\rule[-13pt]{0pt}{42pt}
\bpi(76,0)(2,0)
\put(9,3){\circle*{4}}
\put(41,3){\circle{4}}
\put(25,11){\circle{16}}
\put(9,2){\line(1,0){16}}
\put(9,4){\line(1,0){16}}
\put(25,3){\line(1,0){14}}
\put(25,3){\circle*{4}}
\put(57,3){\circle{32}}
\put(57,3){\makebox(1,0){$W_I$}}
\epi
+3
\rule[-18pt]{0pt}{42pt}
\bpi(68.23,0)(-28.23,0)
\put(19,3){\circle{32}}
\put(19,3){\makebox(1,0){$W_I$}}
\put(5.77,12){\circle*{4}}
\put(5.77,-6){\circle*{4}}
\put(5.77,3){\oval(28,18)[l]}
\put(-8.23,3){\circle*{4}}
\put(-21.23,2){\line(1,0){13}}
\put(-21.23,4){\line(1,0){13}}
\put(-21.23,3){\circle*{4}}
\epi
\nn\\
&&
+2
\rule[-15.41pt]{0pt}{36.82pt}
\bpi(66.64,0)(-3.41,0)
\put(15,3){\circle*{4}}
\put(14.29,2.29){\line(-1,1){10}}
\put(15.71,3.71){\line(-1,1){10}}
\put(3.59,14.41){\circle*{4}}
\put(15.71,2.29){\line(-1,-1){10}}
\put(14.29,3.71){\line(-1,-1){10}}
\put(3.59,-8.41){\circle*{4}}
\put(42.23,3){\circle{32}}
\put(42.23,3){\makebox(1,0){$W_I$}}
\put(29,12){\circle*{4}}
\put(29,-6){\circle*{4}}
\put(29,3){\oval(28,18)[l]}
\epi
+2
\rule[-18pt]{0pt}{42pt}
\bpi(69.23,0)(-29.23,0)
\put(19,3){\circle{32}}
\put(19,3){\makebox(1,0){$W_I$}}
\put(5.77,12){\circle*{4}}
\put(5.77,-6){\circle*{4}}
\put(5.77,3){\oval(28,18)[l]}
\put(-8.23,3){\circle*{4}}
\put(-16.23,3){\circle{16}}
\epi
+
\rule[-18pt]{0pt}{42pt}
\bpi(55.23,0)(-15.23,0)
\put(19,3){\circle{32}}
\put(19,3){\makebox(1,0){$W_I$}}
\put(5.77,12){\circle*{4}}
\put(5.77,-6){\circle{4}}
\put(3.77,3){\oval(24,18)[l]}
\put(-8.23,3){\circle*{4}}
\put(-8.23,3){\line(1,0){11.23}}
\put(3,3){\circle*{4}}
\epi
+\frac{4}{3}
\rule[-18pt]{0pt}{42pt}
\bpi(68.23,0)(-28.23,0)
\put(19,3){\circle{32}}
\put(19,3){\makebox(1,0){$W_I$}}
\put(5.77,12){\circle*{4}}
\put(5.77,-6){\circle{4}}
\put(3.77,3){\oval(24,18)[l]}
\put(-8.23,3){\circle*{4}}
\put(-21.23,2){\line(1,0){13}}
\put(-21.23,4){\line(1,0){13}}
\put(-8.23,3){\line(1,0){11.23}}
\put(3,3){\circle*{4}}
\put(-21.23,3){\circle*{4}}
\epi
+\frac{2}{3}
\rule[-18pt]{0pt}{42pt}
\bpi(54.78,0)(-14.78,0)
\put(19,3){\circle{32}}
\put(19,3){\makebox(1,0){$W_I$}}
\put(5.22,11.4){\circle*{4}}
\put(10.87,17.1){\circle*{4}}
\put(5.22,-5.13){\circle*{4}}
\put(10.87,-10.78){\circle*{4}}
\put(10.87,3){\oval(41.3,27.88)[l]}
\put(5.22,3){\oval(30,16.53)[l]}
\put(-9.78,3){\circle*{4}}
\epi
\nn\\
&&
+
\rule[-18pt]{0pt}{42pt}
\bpi(98.23,0)(-60,0)
\put(17.23,3){\circle{32}}
\put(17.23,3){\makebox(1,0){$W_I$}}
\put(4,12){\circle*{4}}
\put(4,-6){\circle*{4}}
\put(4,3){\oval(28,18)[l]}
\put(-10,3){\circle*{4}}
\put(-23,3){\circle{4}}
\put(-10,3){\line(-1,0){11}}
\put(-39,3){\circle{32}}
\put(-39,3){\makebox(1,0){$W_I$}}
\epi
+\frac{4}{3}
\rule[-18pt]{0pt}{42pt}
\bpi(98.23,0)(-60,0)
\put(17.23,3){\circle{32}}
\put(17.23,3){\makebox(1,0){$W_I$}}
\put(4,12){\circle*{4}}
\put(4,-6){\circle*{4}}
\put(4,3){\oval(28,18)[l]}
\put(-11,3){\line(0,1){14}}
\put(-9,3){\line(0,1){14}}
\put(-10,17){\circle*{4}}
\put(-10,3){\circle*{4}}
\put(-23,3){\circle{4}}
\put(-10,3){\line(-1,0){11}}
\put(-39,3){\circle{32}}
\put(-39,3){\makebox(1,0){$W_I$}}
\epi
+\frac{2}{3}
\rule[-18pt]{0pt}{42pt}
\bpi(96.46,0)(-56.46,0)
\put(19,3){\circle{32}}
\put(19,3){\makebox(1,0){$W_I$}}
\put(5.77,12){\circle*{4}}
\put(5.77,-6){\circle*{4}}
\put(5.77,3){\oval(28,18)[l]}
\put(-8.23,3){\circle*{4}}
\put(-22.23,3){\oval(28,18)[r]}
\put(-22.23,12){\circle*{4}}
\put(-22.23,-6){\circle*{4}}
\put(-35.46,3){\circle{32}}
\put(-35.46,3){\makebox(1,0){$W_I$}}
\epi.
\eea

By construction, the simpler equation (\ref{wissbid1f}) involves only the
$J$-dependent terms and is therefore by itself not sufficient for an
investigation of the $J$-independent terms, for which (\ref{wissbid2f})
has to be used.

\subsection{Recursion Relations}
For later use note the following topological relations.
Let $n_4$ be the number of four-vertices, $n_3$ the number of
three-vertices, $n_1$ the number of $\jb$s, $n_G$ the number of free
propagators $G$ not connected to a $J$ and $L$ the number of loops in
a connected diagram $D$.
Then
\beq
\label{n1n3n4ngL}
3n_3+4n_4=n_1+2n_G,
\;\;\;\;
n_3+2n_4=2(L-1)+n_1
\eeq
and
\beq
\label{n1dngd}
\rule[-18pt]{0pt}{42pt}
\bpi(57,0)(-17,0)
\put(19,3){\circle{32}}
\put(19,3){\makebox(1,0){$D$}}
\put(3,3){\circle{4}}
\put(-10,2){\line(1,0){11}}
\put(-10,4){\line(1,0){11}}
\put(-10,3){\circle*{4}}
\epi
=
n_1
\rule[-18pt]{0pt}{42pt}
\bpi(42,0)
\put(21,3){\circle{32}}
\put(21,3){\makebox(1,0){$D$}}
\epi,
\;\;\;\;
\rule[-18pt]{0pt}{42pt}
\bpi(53.23,0)(-13.23,0)
\put(19,3){\circle{32}}
\put(19,3){\makebox(1,0){$D$}}
\put(5.77,12){\circle*{4}}
\put(5.77,-6){\circle*{4}}
\put(5.77,3){\oval(28,18)[l]}
\epi
=
n_G
\rule[-18pt]{0pt}{42pt}
\bpi(42,0)
\put(21,3){\circle{32}}
\put(21,3){\makebox(1,0){$D$}}
\epi.
\eeq

It is useful to consider a double expansion in the number $L$ of loops
and powers $n$ of $J$ or $\jb$,
\beq
\label{wdoubleex}
W\equiv
\rule[-18pt]{0pt}{42pt}
\bpi(42,0)
\put(21,3){\circle{32}}
\put(21,3){\makebox(0,0){$W$}}
\epi
=\sum_{L=0}^\infty\sum_{n=0}^\infty W^{(L,n)}\equiv
\sum_{L=0}^\infty\sum_{n=0}^\infty
\rule[-18pt]{0pt}{42pt}
\bpi(42,0)
\put(21,3){\circle{32}}
\put(21,3){\makebox(1,-2){$\ba{c}L\\n\ea$}}
\epi.
\eeq
Then the $L$-loop contribution to the connected $n$-point function
with vanishing source $J$ is given by
\beq
\label{wln}
G^{({\rm c})(L,n)}_{i_1,\ldots,i_n}
=\left.\frac{\de^n}{\de J_{i_1}\ldots\de J_{i_n}}W^{(L)}\right|_{J=0}
=\frac{\de^n}{\de J_{i_1}\ldots\de J_{i_n}}W^{(L,n)}.
\eeq

We have
\beq
\rule[-18pt]{0pt}{42pt}
\bpi(42,0)
\put(21,3){\circle{32}}
\put(21,3){\makebox(1,-2){$\ba{c}0\\1\ea$}}
\epi
=0
\eeq
and from (\ref{w0ssb2a})
\beq
\label{w000210}
\rule[-18pt]{0pt}{42pt}
\bpi(42,0)
\put(21,3){\circle{32}}
\put(21,3){\makebox(1,-2){$\ba{c}0\\0\ea$}}
\epi
=
\rule[-2pt]{0pt}{10pt}
\bpi(10,0)
\put(5,3){\circle*{4}}
\epi,
\;\;\;\;
\rule[-18pt]{0pt}{42pt}
\bpi(42,0)
\put(21,3){\circle{32}}
\put(21,3){\makebox(1,-2){$\ba{c}0\\2\ea$}}
\epi
=\frac{1}{2}
\rule[-2pt]{0pt}{10pt}
\bpi(26,0)
\put(5,3){\circle*{4}}
\put(21,3){\circle*{4}}
\put(5,2){\line(1,0){16}}
\put(5,4){\line(1,0){16}}
\epi,
\;\;\;\;
\rule[-18pt]{0pt}{42pt}
\bpi(42,0)
\put(21,3){\circle{32}}
\put(21,3){\makebox(1,-2){$\ba{c}1\\0\ea$}}
\epi
=\frac{1}{2}
\rule[-10pt]{0pt}{26pt}
\bpi(26,0)
\put(13,3){\circle{16}}
\epi.
\eeq
The other $W^{(L,n)}$ constitute $W_I$.

Using
\bea
\label{dwdjid}
\rule[-18pt]{0pt}{42pt}
\bpi(57,0)(-17,0)
\put(19,3){\circle{32}}
\put(19,3){\makebox(1,-2){$\ba{c}L\\n\ea$}}
\put(3,3){\circle{4}}
\put(-10,2){\line(1,0){11}}
\put(-10,4){\line(1,0){11}}
\put(-10,3){\circle*{4}}
\epi
&=&
n
\rule[-18pt]{0pt}{42pt}
\bpi(42,0)
\put(21,3){\circle{32}}
\put(21,3){\makebox(1,-2){$\ba{c}L\\n\ea$}}
\epi,
\eea
(\ref{wissbid1f}) can be split into
\beq
\label{w03}
\rule[-18pt]{0pt}{42pt}
\bpi(42,0)
\put(21,3){\circle{32}}
\put(21,3){\makebox(1,-2){$\ba{c}0\\3\ea$}}
\epi
=
\frac{1}{6}
\rule[-11pt]{0pt}{35pt}
\bpi(38.24,0)(-4.12,0)
\put(15,3){\circle*{4}}
\put(14,3){\line(0,1){14}}
\put(16,3){\line(0,1){14}}
\put(15,17){\circle*{4}}
\put(15.5,2.13){\line(-5,-3){10.39}}
\put(14.5,3.87){\line(-5,-3){10.39}}
\put(2.88,-4){\circle*{4}}
\put(14.5,2.13){\line(5,-3){10.39}}
\put(15.5,3.87){\line(5,-3){10.39}}
\put(27.12,-4){\circle*{4}}
\epi,
\eeq

\beq
\label{w04}
\rule[-18pt]{0pt}{42pt}
\bpi(42,0)
\put(21,3){\circle{32}}
\put(21,3){\makebox(1,-2){$\ba{c}0\\4\ea$}}
\epi
=
\frac{1}{24}
\rule[-15.41pt]{0pt}{36.82pt}
\bpi(36.82,0)(-3.41,0)
\put(15,3){\circle*{4}}
\put(14.29,2.29){\line(-1,1){10}}
\put(15.71,3.71){\line(-1,1){10}}
\put(3.59,14.41){\circle*{4}}
\put(15.71,2.29){\line(-1,-1){10}}
\put(14.29,3.71){\line(-1,-1){10}}
\put(3.59,-8.41){\circle*{4}}
\put(14.29,3.71){\line(1,1){10}}
\put(15.71,2.29){\line(1,1){10}}
\put(26.41,14.41){\circle*{4}}
\put(15.71,3.71){\line(1,-1){10}}
\put(14.29,2.29){\line(1,-1){10}}
\put(26.41,-8.41){\circle*{4}}
\epi
+\frac{1}{4}
\rule[-18pt]{0pt}{42pt}
\bpi(68.41,0)(-3.41,0)
\put(14.29,2.29){\line(-1,1){10}}
\put(15.71,3.71){\line(-1,1){10}}
\put(3.59,14.41){\circle*{4}}
\put(15.71,2.29){\line(-1,-1){10}}
\put(14.29,3.71){\line(-1,-1){10}}
\put(3.59,-8.41){\circle*{4}}
\put(15,3){\line(1,0){11}}
\put(15,3){\circle*{4}}
\put(28,3){\circle{4}}
\put(44,3){\circle{32}}
\put(44,3){\makebox(1,-2){$\ba{c}0\\3\ea$}}
\epi
+\frac{1}{4}
\rule[-18pt]{0pt}{42pt}
\bpi(68.23,0)(-28.23,0)
\put(19,3){\circle{32}}
\put(19,3){\makebox(1,-2){$\ba{c}0\\3\ea$}}
\put(5.77,12){\circle*{4}}
\put(5.77,-6){\circle*{4}}
\put(5.77,3){\oval(28,18)[l]}
\put(-8.23,3){\circle*{4}}
\put(-21.23,2){\line(1,0){13}}
\put(-21.23,4){\line(1,0){13}}
\put(-21.23,3){\circle*{4}}
\epi
=
\frac{1}{24}
\rule[-15.41pt]{0pt}{36.82pt}
\bpi(36.82,0)(-3.41,0)
\put(15,3){\circle*{4}}
\put(14.29,2.29){\line(-1,1){10}}
\put(15.71,3.71){\line(-1,1){10}}
\put(3.59,14.41){\circle*{4}}
\put(15.71,2.29){\line(-1,-1){10}}
\put(14.29,3.71){\line(-1,-1){10}}
\put(3.59,-8.41){\circle*{4}}
\put(14.29,3.71){\line(1,1){10}}
\put(15.71,2.29){\line(1,1){10}}
\put(26.41,14.41){\circle*{4}}
\put(15.71,3.71){\line(1,-1){10}}
\put(14.29,2.29){\line(1,-1){10}}
\put(26.41,-8.41){\circle*{4}}
\epi
+\frac{1}{8}
\rule[-15.41pt]{0pt}{36.82pt}
\bpi(49.82,0)(-3.41,0)
\put(15,3){\circle*{4}}
\put(15,3){\line(1,0){13}}
\put(28,3){\circle*{4}}
\put(14.29,2.29){\line(-1,1){10}}
\put(15.71,3.71){\line(-1,1){10}}
\put(3.59,14.41){\circle*{4}}
\put(15.71,2.29){\line(-1,-1){10}}
\put(14.29,3.71){\line(-1,-1){10}}
\put(3.59,-8.41){\circle*{4}}
\put(27.29,3.71){\line(1,1){10}}
\put(28.71,2.29){\line(1,1){10}}
\put(39.41,14.41){\circle*{4}}
\put(28.71,3.71){\line(1,-1){10}}
\put(27.29,2.29){\line(1,-1){10}}
\put(39.41,-8.41){\circle*{4}}
\epi,
\eeq

\beq
\label{w11}
\rule[-18pt]{0pt}{42pt}
\bpi(42,0)
\put(21,3){\circle{32}}
\put(21,3){\makebox(1,-2){$\ba{c}1\\1\ea$}}
\epi
=
\frac{1}{2}
\bpi(41,0)
\put(28,3){\circle{16}}
\put(20,3){\circle*{4}}
\put(7,2){\line(1,0){13}}
\put(7,4){\line(1,0){13}}
\put(7,3){\circle*{4}}
\epi,
\eeq

\beq
\label{w12}
\rule[-18pt]{0pt}{42pt}
\bpi(42,0)
\put(21,3){\circle{32}}
\put(21,3){\makebox(1,-2){$\ba{c}1\\2\ea$}}
\epi
=
\frac{1}{4}
\rule[-15.41pt]{0pt}{36.82pt}
\bpi(39.41,0)(-3.41,0)
\put(15,3){\circle*{4}}
\put(14.29,2.29){\line(-1,1){10}}
\put(15.71,3.71){\line(-1,1){10}}
\put(3.59,14.41){\circle*{4}}
\put(15.71,2.29){\line(-1,-1){10}}
\put(14.29,3.71){\line(-1,-1){10}}
\put(3.59,-8.41){\circle*{4}}
\put(23,3){\circle{16}}
\epi
+\frac{1}{2}
\rule[-18pt]{0pt}{42pt}
\bpi(68.41,0)(-3.41,0)
\put(14.29,2.29){\line(-1,1){10}}
\put(15.71,3.71){\line(-1,1){10}}
\put(3.59,14.41){\circle*{4}}
\put(15.71,2.29){\line(-1,-1){10}}
\put(14.29,3.71){\line(-1,-1){10}}
\put(3.59,-8.41){\circle*{4}}
\put(15,3){\line(1,0){11}}
\put(15,3){\circle*{4}}
\put(28,3){\circle{4}}
\put(44,3){\circle{32}}
\put(44,3){\makebox(1,-2){$\ba{c}1\\1\ea$}}
\epi
+\frac{1}{2}
\rule[-18pt]{0pt}{42pt}
\bpi(68.23,0)(-28.23,0)
\put(19,3){\circle{32}}
\put(19,3){\makebox(1,-2){$\ba{c}1\\1\ea$}}
\put(5.77,12){\circle*{4}}
\put(5.77,-6){\circle*{4}}
\put(5.77,3){\oval(28,18)[l]}
\put(-8.23,3){\circle*{4}}
\put(-21.23,2){\line(1,0){13}}
\put(-21.23,4){\line(1,0){13}}
\put(-21.23,3){\circle*{4}}
\epi
=
\frac{1}{4}
\rule[-15.41pt]{0pt}{36.82pt}
\bpi(39.41,0)(-3.41,0)
\put(15,3){\circle*{4}}
\put(14.29,2.29){\line(-1,1){10}}
\put(15.71,3.71){\line(-1,1){10}}
\put(3.59,14.41){\circle*{4}}
\put(15.71,2.29){\line(-1,-1){10}}
\put(14.29,3.71){\line(-1,-1){10}}
\put(3.59,-8.41){\circle*{4}}
\put(23,3){\circle{16}}
\epi
+\frac{1}{4}
\rule[-15.41pt]{0pt}{36.82pt}
\bpi(52.41,0)(-3.41,0)
\put(15,3){\circle*{4}}
\put(15,3){\line(1,0){13}}
\put(28,3){\circle*{4}}
\put(14.29,2.29){\line(-1,1){10}}
\put(15.71,3.71){\line(-1,1){10}}
\put(3.59,14.41){\circle*{4}}
\put(15.71,2.29){\line(-1,-1){10}}
\put(14.29,3.71){\line(-1,-1){10}}
\put(3.59,-8.41){\circle*{4}}
\put(36,3){\circle{16}}
\epi
+\frac{1}{4}
\rule[-10pt]{0pt}{26pt}
\bpi(56,0)
\put(7,3){\circle*{4}}
\put(7,2){\line(1,0){13}}
\put(7,4){\line(1,0){13}}
\put(20,3){\circle*{4}}
\put(28,3){\circle{16}}
\put(36,3){\circle*{4}}
\put(36,2){\line(1,0){13}}
\put(36,4){\line(1,0){13}}
\put(49,3){\circle*{4}}
\epi
\eeq
and the recursion relation
\bea
\label{wissbid1g}
n
\rule[-18pt]{0pt}{42pt}
\bpi(42,0)
\put(21,3){\circle{32}}
\put(21,3){\makebox(1,-2){$\ba{c}L\\n\ea$}}
\epi
&\doteq&
\rule[-18pt]{0pt}{42pt}
\bpi(68.41,0)(-3.41,0)
\put(14.29,2.29){\line(-1,1){10}}
\put(15.71,3.71){\line(-1,1){10}}
\put(3.59,14.41){\circle*{4}}
\put(15.71,2.29){\line(-1,-1){10}}
\put(14.29,3.71){\line(-1,-1){10}}
\put(3.59,-8.41){\circle*{4}}
\put(15,3){\line(1,0){11}}
\put(15,3){\circle*{4}}
\put(28,3){\circle{4}}
\put(44,3){\circle{32}}
\put(44,3){\makebox(1,-2){$\ba{c}L\\n{-}1\ea$}}
\epi
+\frac{1}{2}
\rule[-18pt]{0pt}{42pt}
\bpi(68.41,0)(-3.41,0)
\put(14.29,2.29){\line(-1,1){10}}
\put(15.71,3.71){\line(-1,1){10}}
\put(3.59,14.41){\circle*{4}}
\put(15.71,2.29){\line(-1,-1){10}}
\put(14.29,3.71){\line(-1,-1){10}}
\put(3.59,-8.41){\circle*{4}}
\put(3.59,3){\circle*{4}}
\put(3.59,2){\line(1,0){11.41}}
\put(3.59,4){\line(1,0){11.41}}
\put(15,3){\line(1,0){11}}
\put(15,3){\circle*{4}}
\put(28,3){\circle{4}}
\put(44,3){\circle{32}}
\put(44,3){\makebox(1,-2){$\ba{c}L\\n{-}2\ea$}}
\epi
+\frac{1}{2}
\rule[-13pt]{0pt}{42pt}
\bpi(76,0)(2,0)
\put(9,3){\circle*{4}}
\put(9,2){\line(1,0){16}}
\put(9,4){\line(1,0){16}}
\put(25,3){\circle*{4}}
\put(41,3){\circle{4}}
\put(25,11){\circle{16}}
\put(25,3){\line(1,0){14}}
\put(57,3){\circle{32}}
\put(57,3){\makebox(1,-2){$\ba{c}L{-}1\\n\ea$}}
\epi
+\rule[-18pt]{0pt}{42pt}
\bpi(68.23,0)(-28.23,0)
\put(19,3){\circle{32}}
\put(19,3){\makebox(1,-2){$\ba{c}L\\n{-}1\ea$}}
\put(5.77,12){\circle*{4}}
\put(5.77,-6){\circle*{4}}
\put(5.77,3){\oval(28,18)[l]}
\put(-8.23,3){\circle*{4}}
\put(-21.23,2){\line(1,0){13}}
\put(-21.23,4){\line(1,0){13}}
\put(-21.23,3){\circle*{4}}
\epi
+
\rule[-15.41pt]{0pt}{36.82pt}
\bpi(66.64,0)(-3.41,0)
\put(15,3){\circle*{4}}
\put(14.29,2.29){\line(-1,1){10}}
\put(15.71,3.71){\line(-1,1){10}}
\put(3.59,14.41){\circle*{4}}
\put(15.71,2.29){\line(-1,-1){10}}
\put(14.29,3.71){\line(-1,-1){10}}
\put(3.59,-8.41){\circle*{4}}
\put(42.23,3){\circle{32}}
\put(42.23,3){\makebox(1,-2){$\ba{c}L\\n{-}2\ea$}}
\put(29,12){\circle*{4}}
\put(29,-6){\circle*{4}}
\put(29,3){\oval(28,18)[l]}
\epi
\nn\\
&&
+\frac{1}{3}
\rule[-18pt]{0pt}{42pt}
\bpi(68.23,0)(-28.23,0)
\put(19,3){\circle{32}}
\put(19,3){\makebox(1,-2){$\ba{c}L{-}1\\n\ea$}}
\put(5.77,12){\circle*{4}}
\put(5.77,-6){\circle{4}}
\put(3.77,3){\oval(24,18)[l]}
\put(-8.23,3){\circle*{4}}
\put(-21.23,2){\line(1,0){13}}
\put(-21.23,4){\line(1,0){13}}
\put(-8.23,3){\line(1,0){11.23}}
\put(3,3){\circle*{4}}
\put(-21.23,3){\circle*{4}}
\epi
+\frac{1}{3}\sum_{l=0}^L\sum_{m=1}^n
\rule[-17pt]{0pt}{41pt}
\bpi(98.23,0)(-60,0)
\put(17.23,3){\circle{32}}
\put(17.23,3){\makebox(1,-2){$\ba{c}L{-}l\\n{-}m\ea$}}
\put(4,12){\circle*{4}}
\put(4,-6){\circle*{4}}
\put(4,3){\oval(28,18)[l]}
\put(-11,3){\line(0,1){13}}
\put(-9,3){\line(0,1){13}}
\put(-10,16){\circle*{4}}
\put(-10,3){\circle*{4}}
\put(-23,3){\circle{4}}
\put(-10,3){\line(-1,0){11}}
\put(-39,3){\circle{32}}
\put(-39,3){\makebox(1,-2){$\ba{c}l\\m\ea$}}
\epi,
\eea
where the dot on the equal sign means that the right hand side only
involves $W^{(i,j)}$ that are part of $W_I$, i.e.\ excluding
$(i,j)\in\{(0,0),(0,1),(0,2),(1,0)\}$ and negative $i$ or $j$.
Eq.\ (\ref{wissbid1g}) is valid for all $W^{(L,n)}$ which are
part of $W_I$ with the exception of $(L,n)\in\{(0,3),(0,4),(1,1),(1,2)\}$.

From (\ref{wissbid2f}) follow again equations leading with (\ref{n1dngd})
to (\ref{w03})-(\ref{w12}), but also
\bea
\label{w20a}
2
\rule[-18pt]{0pt}{42pt}
\bpi(53.23,0)(-13.23,0)
\put(19,3){\circle{32}}
\put(19,3){\makebox(1,-2){$\ba{c}2\\0\ea$}}
\put(5.77,12){\circle*{4}}
\put(5.77,-6){\circle*{4}}
\put(5.77,3){\oval(28,18)[l]}
\epi
&=&
\frac{1}{2}
\rule[-10pt]{0pt}{26pt}
\bpi(42,0)
\put(21,3){\circle*{4}}
\put(13,3){\circle{16}}
\put(29,3){\circle{16}}
\epi
+\frac{3}{2}
\rule[-18pt]{0pt}{42pt}
\bpi(71,0)(-31,0)
\put(19,3){\circle{32}}
\put(19,3){\makebox(1,-2){$\ba{c}1\\1\ea$}}
\put(3,3){\circle{4}}
\put(-10,3){\line(1,0){11}}
\put(-10,3){\circle*{4}}
\put(-18,3){\circle{16}}
\epi
+
\rule[-18pt]{0pt}{42pt}
\bpi(55.23,0)(-15.23,0)
\put(19,3){\circle{32}}
\put(19,3){\makebox(1,-2){$\ba{c}1\\1\ea$}}
\put(5.77,12){\circle*{4}}
\put(5.77,-6){\circle{4}}
\put(3.77,3){\oval(24,18)[l]}
\put(-8.23,3){\circle*{4}}
\put(-8.23,3){\line(1,0){11.23}}
\put(3,3){\circle*{4}}
\epi
\nn\\
&=&
\frac{1}{2}
\rule[-10pt]{0pt}{26pt}
\bpi(42,0)
\put(21,3){\circle*{4}}
\put(13,3){\circle{16}}
\put(29,3){\circle{16}}
\epi
+\frac{3}{4}
\rule[-10pt]{0pt}{26pt}
\bpi(60,0)
\put(15,3){\circle{16}}
\put(23,3){\circle*{4}}
\put(23,3){\line(1,0){16}}
\put(39,3){\circle*{4}}
\put(47,3){\circle{16}}
\epi
+\frac{1}{2}
\rule[-14pt]{0pt}{34pt}
\bpi(38,0)
\put(19,3){\circle{24}}
\put(7,3){\circle*{4}}
\put(31,3){\circle*{4}}
\put(7,3){\line(1,0){24}}
\epi,
\eea
which with (\ref{n1dngd}) becomes
\beq
\label{w20}
\rule[-18pt]{0pt}{42pt}
\bpi(42,0)
\put(21,3){\circle{32}}
\put(21,3){\makebox(1,-2){$\ba{c}2\\0\ea$}}
\epi
=
\frac{1}{8}
\rule[-10pt]{0pt}{26pt}
\bpi(60,0)
\put(15,3){\circle{16}}
\put(23,3){\circle*{4}}
\put(23,3){\line(1,0){16}}
\put(39,3){\circle*{4}}
\put(47,3){\circle{16}}
\epi
+\frac{1}{12}
\rule[-14pt]{0pt}{34pt}
\bpi(38,0)
\put(19,3){\circle{24}}
\put(7,3){\circle*{4}}
\put(31,3){\circle*{4}}
\put(7,3){\line(1,0){24}}
\epi
+\frac{1}{8}
\rule[-10pt]{0pt}{26pt}
\bpi(42,0)
\put(21,3){\circle*{4}}
\put(13,3){\circle{16}}
\put(29,3){\circle{16}}
\epi,
\eeq
and a recursion relation which we write down only for $n=0$,
since for $n>0$ the simpler relation (\ref{wissbid1g}) can be used:
\bea
\label{wissbid2g0}
\rule[-18pt]{0pt}{42pt}
\bpi(53.23,0)(-13.23,0)
\put(19,3){\circle{32}}
\put(19,3){\makebox(1,-2){$\ba{c}L\\0\ea$}}
\put(5.77,12){\circle*{4}}
\put(5.77,-6){\circle*{4}}
\put(5.77,3){\oval(28,18)[l]}
\epi
&=&
\frac{3}{4}
\rule[-18pt]{0pt}{42pt}
\bpi(71,0)(-31,0)
\put(19,3){\circle{32}}
\put(19,3){\makebox(1,-2){$\ba{c}L{-}1\\1\ea$}}
\put(3,3){\circle{4}}
\put(-10,3){\line(1,0){11}}
\put(-10,3){\circle*{4}}
\put(-18,3){\circle{16}}
\epi
+
\rule[-18pt]{0pt}{42pt}
\bpi(69.23,0)(-29.23,0)
\put(19,3){\circle{32}}
\put(19,3){\makebox(1,-2){$\ba{c}L{-}1\\0\ea$}}
\put(5.77,12){\circle*{4}}
\put(5.77,-6){\circle*{4}}
\put(5.77,3){\oval(28,18)[l]}
\put(-8.23,3){\circle*{4}}
\put(-16.23,3){\circle{16}}
\epi
+\frac{1}{2}
\rule[-18pt]{0pt}{42pt}
\bpi(55.23,0)(-15.23,0)
\put(19,3){\circle{32}}
\put(19,3){\makebox(1,-2){$\ba{c}L{-}1\\1\ea$}}
\put(5.77,12){\circle*{4}}
\put(5.77,-6){\circle{4}}
\put(3.77,3){\oval(24,18)[l]}
\put(-8.23,3){\circle*{4}}
\put(-8.23,3){\line(1,0){11.23}}
\put(3,3){\circle*{4}}
\epi
+\frac{1}{3}
\rule[-18pt]{0pt}{42pt}
\bpi(54.78,0)(-14.78,0)
\put(19,3){\circle{32}}
\put(19,3){\makebox(1,-2){$\ba{c}L{-}1\\0\ea$}}
\put(5.22,11.4){\circle*{4}}
\put(10.87,17.1){\circle*{4}}
\put(5.22,-5.13){\circle*{4}}
\put(10.87,-10.78){\circle*{4}}
\put(10.87,3){\oval(41.3,27.88)[l]}
\put(5.22,3){\oval(30,16.53)[l]}
\put(-9.78,3){\circle*{4}}
\epi
\nn\\
&&
+\frac{1}{2}\sum_{l=1}^{L-2}
\rule[-17pt]{0pt}{41pt}
\bpi(98.23,0)(-60,0)
\put(17.23,3){\circle{32}}
\put(17.23,3){\makebox(1,-2){$\ba{c}L{-}l\\0\ea$}}
\put(4,12){\circle*{4}}
\put(4,-6){\circle*{4}}
\put(4,3){\oval(28,18)[l]}
\put(-10,3){\circle*{4}}
\put(-23,3){\circle{4}}
\put(-10,3){\line(-1,0){11}}
\put(-39,3){\circle{32}}
\put(-39,3){\makebox(1,-1){$\ba{c}l\\1\ea$}}
\epi
+\frac{1}{3}\sum_{l=2}^{L-2}
\rule[-18pt]{0pt}{42pt}
\bpi(96.46,0)(-56.46,0)
\put(19,3){\circle{32}}
\put(19,3){\makebox(1,-2){$\ba{c}L{-}l\\0\ea$}}
\put(5.77,12){\circle*{4}}
\put(5.77,-6){\circle*{4}}
\put(5.77,3){\oval(28,18)[l]}
\put(-8.23,3){\circle*{4}}
\put(-22.23,3){\oval(28,18)[r]}
\put(-22.23,12){\circle*{4}}
\put(-22.23,-6){\circle*{4}}
\put(-35.46,3){\circle{32}}
\put(-35.46,3){\makebox(1,-2){$\ba{c}l\\0\ea$}}
\epi.
\eea
Eq.\ (\ref{wissbid2g0}) is valid for $L>2$.

Note that in (\ref{wissbid2g0})---but not in (\ref{wissbid1g})---the
right hand side involves graphs with more legs---namely one more---than
the left hand side.
This implies that for the generation of vacuum graphs, we have to consider
also one-point functions.
For all others it is enough to consider only diagrams with equal or
less numbers of legs.
Note further that if all lower loop orders contain only connected graphs,
then the recursion relations generate only connected graphs.
This establishes by induction that $W$ generates only connected graphs,
as shown before in \cite{Kleinert1,Kleinert2}.

As an example, we compute $W^{(3,0)}$ in appendix \ref{w30example}.
Combining (\ref{w000210}), (\ref{w20}) and the result (\ref{w30})
of appendix \ref{w30example}, we get
$W$ at $J=0$ in the three-loop approximation,
\bea
\label{w3loop}
W[J=0]
&=&
\rule[-2pt]{0pt}{10pt}
\bpi(10,0)
\put(5,3){\circle*{4}}
\epi
+\frac{1}{2}
\rule[-10pt]{0pt}{26pt}
\bpi(26,0)
\put(13,3){\circle{16}}
\epi
+\frac{1}{8}
\rule[-10pt]{0pt}{26pt}
\bpi(60,0)
\put(15,3){\circle{16}}
\put(23,3){\circle*{4}}
\put(23,3){\line(1,0){16}}
\put(39,3){\circle*{4}}
\put(47,3){\circle{16}}
\epi
+\frac{1}{12}
\rule[-14pt]{0pt}{34pt}
\bpi(38,0)
\put(19,3){\circle{24}}
\put(7,3){\circle*{4}}
\put(31,3){\circle*{4}}
\put(7,3){\line(1,0){24}}
\epi
+\frac{1}{8}
\rule[-10pt]{0pt}{26pt}
\bpi(46,0)
\put(23,3){\circle*{4}}
\put(15,3){\circle{16}}
\put(31,3){\circle{16}}
\epi
\nn\\
&&
+\frac{1}{16}
\rule[-10pt]{0pt}{26pt}
\bpi(84,0)
\put(13,3){\circle{16}}
\put(21,3){\circle*{4}}
\put(21,3){\line(1,0){13}}
\put(34,3){\circle*{4}}
\put(42,3){\circle{16}}
\put(50,3){\circle*{4}}
\put(50,3){\line(1,0){13}}
\put(63,3){\circle*{4}}
\put(71,3){\circle{16}}
\epi
+\frac{1}{48}
\rule[-19pt]{0pt}{53pt}
\bpi(57.18,0)(-13.59,0)
\put(15,3){\circle*{4}}
\put(15,3){\line(0,1){10}}
\put(15,13){\circle*{4}}
\put(15,21){\circle{16}}
\put(15,3){\line(-5,-3){10}}
\put(6.34,-2){\circle*{4}}
\put(-0.59,-6){\circle{16}}
\put(15,3){\line(5,-3){10}}
\put(23.66,-2){\circle*{4}}
\put(30.59,-6){\circle{16}}
\epi
+\frac{1}{8}
\rule[-16pt]{0pt}{38pt}
\bpi(63,0)
\put(13,3){\circle{16}}
\put(21,3){\circle*{4}}
\put(21,3){\line(1,0){13}}
\put(34,3){\circle*{4}}
\put(46,3){\circle{24}}
\put(46,-9){\circle*{4}}
\put(46,15){\circle*{4}}
\put(46,-9){\line(0,1){24}}
\epi
+\frac{1}{16}
\rule[-16pt]{0pt}{40pt}
\bpi(50,0)
\put(13,3){\circle{16}}
\put(37,3){\circle{16}}
\put(13,-5){\circle*{4}}
\put(13,11){\circle*{4}}
\put(37,-5){\circle*{4}}
\put(37,11){\circle*{4}}
\put(25,-3){\oval(24,16)[b]}
\put(25,11){\oval(24,16)[t]}
\epi
+\frac{1}{24}
\rule[-14pt]{0pt}{36pt}
\bpi(34.78,0)(-3.39,0)
\put(15,3){\circle{24}}
\put(15,3){\circle*{4}}
\put(15,15){\circle*{4}}
\put(4.61,-3){\circle*{4}}
\put(25.39,-3){\circle*{4}}
\put(15,3){\line(0,1){12}}
\put(15,3){\line(5,-3){10}}
\put(15,3){\line(-5,-3){10}}
\epi
\nn\\
&&
+\frac{1}{16}
\rule[-10pt]{0pt}{34pt}
\bpi(68,0)
\put(13,3){\circle{16}}
\put(21,3){\circle*{4}}
\put(21,3){\line(1,0){26}}
\put(34,3){\circle*{4}}
\put(34,11){\circle{16}}
\put(47,3){\circle*{4}}
\put(55,3){\circle{16}}
\epi
+\frac{1}{8}
\rule[-10pt]{0pt}{26pt}
\bpi(71,0)
\put(13,3){\circle{16}}
\put(21,3){\circle*{4}}
\put(21,3){\line(1,0){13}}
\put(34,3){\circle*{4}}
\put(42,3){\circle{16}}
\put(50,3){\circle*{4}}
\put(58,3){\circle{16}}
\epi
+\frac{1}{12}
\rule[-14pt]{0pt}{34pt}
\bpi(65,0)
\put(13,3){\circle{16}}
\put(21,3){\circle*{4}}
\put(21,3){\line(1,0){37}}
\put(34,3){\circle*{4}}
\put(46,3){\circle{24}}
\put(58,3){\circle*{4}}
\epi
+\frac{1}{8}
\rule[-16pt]{0pt}{38pt}
\bpi(50,0)
\put(13,3){\circle{16}}
\put(21,3){\circle*{4}}
\put(33,3){\circle{24}}
\put(33,-9){\circle*{4}}
\put(33,15){\circle*{4}}
\put(33,-9){\line(0,1){24}}
\epi
+\frac{1}{8}
\rule[-16pt]{0pt}{36pt}
\bpi(38,0)
\put(19,3){\circle{24}}
\put(7,-9){\oval(24,24)[rt]}
\put(31,-9){\oval(24,24)[lt]}
\put(19,-9){\circle*{4}}
\put(7,3){\circle*{4}}
\put(31,3){\circle*{4}}
\epi
\nn\\
&&
+\frac{1}{16}
\rule[-10pt]{0pt}{26pt}
\bpi(58,0)
\put(13,3){\circle{16}}
\put(21,3){\circle*{4}}
\put(29,3){\circle{16}}
\put(37,3){\circle*{4}}
\put(45,3){\circle{16}}
\epi
+\frac{1}{48}
\rule[-12pt]{0pt}{34pt}
\bpi(38,0)
\put(19,3){\circle{24}}
\put(19,3){\oval(24,8)}
\put(7,3){\circle*{4}}
\put(31,3){\circle*{4}}
\epi.
\eea
At this low loop order, it is still relatively easy to check that the
weights come out the same when using the combinatorial prescriptions
that come with the usual Feynman rules.
I have written a computer code implementing the recursion relations
for the connected graphs.
If we restrict ourselves to the symmetric case, it reproduces
the graphs and multiplicities (trivially related to the weights,
see \cite{KPKB}) of Tables I through III in \cite{KPKB} and also all relevant
entries in Tables V through VII there.

Except for the vacuum diagrams, we still have to use (\ref{wln}) to
convert the graphs representing $W$ into connected Greens functions.
For example, to compute the zero-loop contribution
$G^{({\rm c})(0,4)}_{1234}$ to the connected four-point function we
combine (\ref{wln}) and (\ref{w04}) to get
\beq
\label{gc1234}
G^{({\rm c})(0,4)}_{1234}
=
\frac{\de^4}{\de J_1\de J_2\de J_3\de J_4}
\left[
\frac{1}{24}
\rule[-15.41pt]{0pt}{36.82pt}
\bpi(36.82,0)(-3.41,0)
\put(15,3){\circle*{4}}
\put(14.29,2.29){\line(-1,1){10}}
\put(15.71,3.71){\line(-1,1){10}}
\put(3.59,14.41){\circle*{4}}
\put(15.71,2.29){\line(-1,-1){10}}
\put(14.29,3.71){\line(-1,-1){10}}
\put(3.59,-8.41){\circle*{4}}
\put(14.29,3.71){\line(1,1){10}}
\put(15.71,2.29){\line(1,1){10}}
\put(26.41,14.41){\circle*{4}}
\put(15.71,3.71){\line(1,-1){10}}
\put(14.29,2.29){\line(1,-1){10}}
\put(26.41,-8.41){\circle*{4}}
\epi
+\frac{1}{8}
\rule[-15.41pt]{0pt}{36.82pt}
\bpi(49.82,0)(-3.41,0)
\put(15,3){\circle*{4}}
\put(15,3){\line(1,0){13}}
\put(28,3){\circle*{4}}
\put(14.29,2.29){\line(-1,1){10}}
\put(15.71,3.71){\line(-1,1){10}}
\put(3.59,14.41){\circle*{4}}
\put(15.71,2.29){\line(-1,-1){10}}
\put(14.29,3.71){\line(-1,-1){10}}
\put(3.59,-8.41){\circle*{4}}
\put(27.29,3.71){\line(1,1){10}}
\put(28.71,2.29){\line(1,1){10}}
\put(39.41,14.41){\circle*{4}}
\put(28.71,3.71){\line(1,-1){10}}
\put(27.29,2.29){\line(1,-1){10}}
\put(39.41,-8.41){\circle*{4}}
\epi
\right]
=
\rule[-15pt]{0pt}{36pt}
\bpi(40,0)
\put(20,3){\circle*{4}}
\put(20,3){\line(-1,1){10}}
\put(5,13){\makebox(0,0){$\scs 1$}}
\put(20,3){\line(-1,-1){10}}
\put(5,-7){\makebox(0,0){$\scs 2$}}
\put(20,3){\line(1,1){10}}
\put(35,13){\makebox(0,0){$\scs 3$}}
\put(20,3){\line(1,-1){10}}
\put(35,-7){\makebox(0,0){$\scs 4$}}
\epi
+
\rule[-15pt]{0pt}{36pt}
\bpi(53,0)
\put(20,3){\circle*{4}}
\put(20,3){\line(1,0){13}}
\put(33,3){\circle*{4}}
\put(20,3){\line(-1,1){10}}
\put(5,13){\makebox(0,0){$\scs 1$}}
\put(20,3){\line(-1,-1){10}}
\put(5,-7){\makebox(0,0){$\scs 2$}}
\put(33,3){\line(1,1){10}}
\put(48,13){\makebox(0,0){$\scs 3$}}
\put(33,3){\line(1,-1){10}}
\put(48,-7){\makebox(0,0){$\scs 4$}}
\epi
+
\rule[-15pt]{0pt}{36pt}
\bpi(53,0)
\put(20,3){\circle*{4}}
\put(20,3){\line(1,0){13}}
\put(33,3){\circle*{4}}
\put(20,3){\line(-1,1){10}}
\put(5,13){\makebox(0,0){$\scs 1$}}
\put(20,3){\line(-1,-1){10}}
\put(5,-7){\makebox(0,0){$\scs 3$}}
\put(33,3){\line(1,1){10}}
\put(48,13){\makebox(0,0){$\scs 2$}}
\put(33,3){\line(1,-1){10}}
\put(48,-7){\makebox(0,0){$\scs 4$}}
\epi
+
\rule[-15pt]{0pt}{36pt}
\bpi(53,0)
\put(20,3){\circle*{4}}
\put(20,3){\line(1,0){13}}
\put(33,3){\circle*{4}}
\put(20,3){\line(-1,1){10}}
\put(5,13){\makebox(0,0){$\scs 1$}}
\put(20,3){\line(-1,-1){10}}
\put(5,-7){\makebox(0,0){$\scs 4$}}
\put(33,3){\line(1,1){10}}
\put(48,13){\makebox(0,0){$\scs 2$}}
\put(33,3){\line(1,-1){10}}
\put(48,-7){\makebox(0,0){$\scs 3$}}
\epi.
\eeq
That is, each diagram with $n$ external currents is multiplied by $n!$,
supplied by external arguments replacing the $J$s and then splits
into ``crossed'' graphs related by exchanging external arguments.
The external legs represent free correlation functions $G$.

\section{Effective Energy}
\label{effen}
Often in field theory, one is interested rather in the effective
energy or effective action $\Ga$ than the free energy $W$
and rather in the 1PI Feynman diagrams than the connected ones.

Therefore, we translate in the following the identities for $W$
into identities for $\Ga$ and derive recursion relations for
the 1PI Feynman diagrams representing the proper vertices.

\subsection{Relations between $W$ and $\Ga$}
Since the physical situation in which we are interested does not
necessarily correspond to $J=0$, let us for the purpose of performing
a Legendre transform introduce an additional source $\jh$ into the
definition of the partition function $Z$ and the negative free energy $W$,
\beq
\label{zwmod}
Z[\jh,C,J,G,K,L]=\exp(W[\jh,C,J,G,K,L])
=\int D\phi\exp\left(-E[\phi,C,J,G,K,L]+\int_1\jh_1\phi_1\right).
\eeq
Note that we trivially have the relations
\beq
Z[\jh,C,J,G,K,L]=Z[C,J-\jh,G,K,L]
\eeq
and
\beq
\label{wnewwold}
W[\jh,C,J,G,K,L]=W[C,J-\jh,G,K,L]
\eeq
between the quantities defined in (\ref{zwgeneral})
and those in (\ref{zwmod}).
With (\ref{w0ssb1}) follows then that
\bea
\label{w0ssb3}
W_0[\jh,C,J,G]
&\equiv&
W[\jh,C,J,G,K,L]|_{K,L=0}=W[C,J-\jh,G,K,L]|_{K,L=0}
=W_0[C,J-\jh,G]
\nn\\
&=&
-C+\frac{1}{2}\int_{12}G_{12}(J_1-\jh_1)(J_2-\jh_2)
-\frac{1}{2}\int_1(\ln G^{-1})_{11}.
\eea

Define the effective energy $\Ga$ by the Legendre transform
\beq
\Ga[\Phi,C,J,G,K,L]=-W[\jh,C,J,G,K,L]+\int_1\jh_1\Phi_1,
\eeq
where the new variable $\Phi$ is defined by
\beq
\label{chidef}
\Phi_1=\left(\frac{\de W}{\de\jh_1}\right)_{C,J,G,K,L},
\eeq
which implicitly defines $\jh$ as functional of $\Phi$.
As usual we have
\beq
\label{dgadchi}
\left(\frac{\de\Ga}{\de \Phi_1}\right)_{C,J,G,K,L}=\jh_1.
\eeq
Notice that as intended by introducing the extra source term and performing
the Legendre transform with respect to $\jh$ instead of $J$, we do not have
to set $J=0$ but only $\jh=0$ to have a proper effective energy giving
us the equation of state (or the equation of motion if we consider an
effective action) through its stationary points.

In the following, we assume $C$, $J$, $K$, $L$ to be fixed
and do not treat them as variables.
Let us use the notation
\beq
\left(\frac{\de^2F}{\de x_1\de x_2}\right)_{(y_1y_2)}
\equiv
\left(\frac{\de}{\de x_1}\right)_{y_1}
\left(\frac{\de}{\de x_2}\right)_{y_2}F.
\eeq

For deriving identities for $\Ga$ we first need some relations
between the functional derivatives of $W$ and $\Ga$.
With (\ref{chidef}) and (\ref{dgadchi}) we get
\beq
\label{p12}
P_{12}
\equiv
\left(\frac{\de^2W}{\de\jh_1\de\jh_2}\right)_G
=
\left(\frac{\de\Phi_2}{\de\jh_1}\right)_G
=
\left(\frac{\de\jh_1}{\de\Phi_2}\right)_G^{-1}
=
\left(\frac{\de^2\Ga}{\de\Phi_1\de\Phi_2}\right)_G^{-1}.
\eeq
For $\Phi=0$, $P$ is the usual propagator.
It will turn out useful to reexpress $P_{12}$ as
\bea
\label{p12a}
P_{12}
&=&
\left(\frac{\de^2\ln Z}{\de\jh_1\de\jh_2}\right)_G
=
\frac{1}{Z}\left(\frac{\de^2Z}{\de\jh_1\de\jh_2}\right)_G
-\frac{1}{Z^2}\left(\frac{\de Z}{\de\jh_1}\right)_G
\left(\frac{\de Z}{\de\jh_2}\right)_G
=
-\frac{2}{Z}\left(\frac{\de Z}{\de G_{12}^{-1}}\right)_{\jh}
-\frac{1}{Z^2}\left(\frac{\de Z}{\de\jh_1}\right)_G
\left(\frac{\de Z}{\de\jh_2}\right)_G
\nn\\
&=&
-2\left(\frac{\de W}{\de G_{12}^{-1}}\right)_{\jh}
-\left(\frac{\de W}{\de\jh_1}\right)_G
\left(\frac{\de W}{\de\jh_2}\right)_G
=
2\left(\frac{\de\Ga}{\de G_{12}^{-1}}\right)_\Phi-\Phi_1\Phi_2,
\eea
where we have used
\bea
\left(\frac{\de W}{\de G_{12}}\right)_{\jh}
&=&
\left(\frac{\de}{\de G_{12}}\right)_{\jh}
\left[\int_3\jh_3\Phi_3-\Ga\right]
=
\int_3\jh_3\left(\frac{\de\Phi_3}{\de G_{12}}\right)_{\jh}
-\left(\frac{\de\Ga}{\de G_{12}}\right)_{\jh}
\nn\\
&=&
\int_3\jh_3\left(\frac{\de\Phi_3}{\de G_{12}}\right)_{\jh}
-\left[\left(\frac{\de\Ga}{\de G_{12}}\right)_\Phi
+\int_3\left(\frac{\de\Ga}{\de\Phi_3}\right)_G
\left(\frac{\de\Phi_3}{\de G_{12}}\right)_{\jh}\right]
=
-\left(\frac{\de\Ga}{\de G_{12}}\right)_\Phi.
\eea
Further we have
\beq
\left(\frac{\de^2 W}{\de\jh_1\de G_{23}}\right)_{(G\jh)}
=
\left(\frac{\de\Phi_1}{\de G_{23}}\right)_{\jh}
=
-\int_4\left(\frac{\de\Phi_1}{\de\jh_4}\right)_G
\left(\frac{\de\jh_4}{\de G_{23}}\right)_\Phi
=
-\int_4P_{14}
\left(\frac{\de^2\Ga}{\de\Phi_4\de G_{23}}\right)_{(G\Phi)}
\eeq
and
\bea
\left(\frac{\de^2W}{\de G_{12}\de G_{34}}\right)_{\jh}
&=&
-\left(\frac{\de^2\Ga}{\de G_{12}\de G_{34}}\right)_{(\jh\Phi)}
=
-\left(\frac{\de^2\Ga}{\de G_{12}\de G_{34}}\right)_\Phi
-\int_5\left(\frac{\de^2\Ga}{\de\Phi_5\de G_{34}}\right)_{(G\Phi)}
\left(\frac{\de\Phi_5}{\de G_{12}}\right)_{\jh}
\nn\\
&=&
-\left(\frac{\de^2\Ga}{\de G_{12}\de G_{34}}\right)_\Phi
+\int_{56}\left(\frac{\de^2\Ga}{\de\Phi_5\de G_{34}}\right)_{(G\Phi)}
P_{56}\left(\frac{\de^2\Ga}{\de\Phi_6\de G_{12}}\right)_{(G\Phi)}.
\eea

\subsection{Identities for $\Ga_I$}
Making use of (\ref{wnewwold}) and the relations just derived,
(\ref{wssbidentity1}) and (\ref{wssbidentity2}) can be rewritten as
\beq
\label{gassbidentity1a}
0=-J_1+\frac{\de\Ga}{\de\Phi_1}-\int_2G_{12}^{-1}\Phi_2
-\int_{23}K_{123}\frac{\de\Ga}{\de G_{23}^{-1}}
-\frac{1}{3}\int_{234}L_{1234}\left(\int_5P_{25}
\frac{\de^2\Ga}{\de\Phi_5\de G_{34}^{-1}}
+\Phi_2\frac{\de\Ga}{\de G_{34}^{-1}}\right)
\eeq
and
\bea
\label{gassbidentity2a}
0
&=&
\de_{12}-J_1\Phi_2+\frac{\de\Ga}{\de\Phi_1}\Phi_2
-2\int_3G_{13}^{-1}\frac{\de\Ga}{\de G_{23}^{-1}}
-\int_{34}K_{134}
\left(\int_5P_{25}\frac{\de^2\Ga}{\de\Phi_5\de G_{34}^{-1}}
+\Phi_2\frac{\de\Ga}{\de G_{34}^{-1}}\right)
\nn\\
&&
+\frac{2}{3}\int_{345}L_{1345}\left(
\frac{\de^2\Ga}{\de G_{23}^{-1}\de G_{45}^{-1}}
-\int_{67}\frac{\de^2\Ga}{\de\Phi_6\de G_{45}^{-1}}
P_{67}\frac{\de^2\Ga}{\de\Phi_7\de G_{23}^{-1}}
-\frac{\de\Ga}{\de G_{23}^{-1}}
\frac{\de\Ga}{\de G_{45}^{-1}}\right).
\eea
We have omitted now indicating the variables that are kept fixed, since
everything is written in terms of the variables $\Phi$ and $G$.

Split $\Ga$ into a free and an interacting part,
\beq
\label{gaga0gai}
\Ga=\Ga_0+\Ga_I\equiv \Ga|_{K,L=0}+\Ga_I.
\eeq
Then,
\beq
\Ga_0[\Phi,C,J,G]
=-W_0[\jh,C,J,G]+\int_1\jh_1\Phi_1
=-W_0[C,J-\jh,G]+\int_1\jh_1\Phi_1
\eeq
with
\beq
\Phi_1
=\frac{\de W_0[\jh,C,J,G]}{\de\jh_1}
=\frac{\de W_0[C,J-\jh,G]}{\de\jh_1}
=\int_2G_{12}(\jh_2-J_2),
\eeq
i.e.\ 
\beq
\jh_1=J_1+\int_2G_{12}^{-1}\Phi_2.
\eeq
Using (\ref{w0ssb1}), we get
\beq
\label{ga0ssb}
\Ga_0[\Phi,C,J,G]
=
C+\frac{1}{2}\int_1(\ln G^{-1})_{11}
+\int_1J_1\Phi_1
+\frac{1}{2}\int_{12}G_{12}^{-1}\Phi_1\Phi_2.
\eeq

For $\Ga_0$, (\ref{gassbidentity1a}) and (\ref{gassbidentity2a}) reduce to
\beq
\label{ga0ssbidentity1}
0
=
-J_1+\frac{\de\Ga_0}{\de\Phi_1}
-\int_2G_{12}^{-1}\Phi_2
\eeq
and
\beq
\label{ga0ssbidentity2}
0
=
\de_{12}-J_1\Phi_2+\frac{\de\Ga_0}{\de\Phi_1}\Phi_2
-2\int_3G_{13}^{-1}\frac{\de\Ga_0}{\de G_{23}^{-1}},
\eeq
respectively, so that
\beq
\label{dga0ssbdphi1}
\frac{\de\Ga_0}{\de\Phi_1}=J_1+\int_2G_{12}^{-1}\Phi_2
\eeq
and
\beq
\label{dga0ssbdginv12}
\frac{\de\Ga_0}{\de G_{12}^{-1}}
=\frac{1}{2}\left(G_{12}+\Phi_1\Phi_2\right).
\eeq

In the following we also need
\beq
\label{d2ga0ssbdphi1dphi2}
\frac{\de^2\Ga_0}{\de\Phi_1\de\Phi_2}=G_{12}^{-1},
\eeq

\beq
\label{d2ga0ssbdginvdphi}
\frac{\de^2\Ga_0}{\de\Phi_1\de G_{23}^{-1}}
=\frac{1}{2}\left(\de_{12}\Phi_3+\de_{13}\Phi_2\right)
\eeq
and
\beq
\label{d2ga0dginv11dginv11}
\frac{\de^2\Ga_0}{\de G_{12}^{-1}\de G_{34}^{-1}}
=-\frac{1}{4}\left(G_{13}G_{24}+G_{14}G_{23}\right).
\eeq

Notice that with (\ref{p12a}), (\ref{dga0ssbdginv12}) and (\ref{ddginv})
we get
\beq
\label{p12b}
P_{12}=G_{12}+2\frac{\de\Ga_I}{\de G_{12}^{-1}}
=G_{12}-2\int_{34}G_{13}G_{24}\frac{\de\Ga_I}{\de G_{34}}.
\eeq

Subtracting (\ref{ga0ssbidentity1}) from (\ref{gassbidentity1a}),
multiplying with $\Phi_1$, integrating over $x_1$ and using
(\ref{dga0ssbdginv12}), (\ref{d2ga0ssbdginvdphi}) and (\ref{ddginv})
gives
\bea
\label{gaissbid1c}
0
&=&
\int_1\frac{\de\Ga_I}{\de\Phi_1}\Phi_1
\nn\\
&&
-\frac{1}{2}\int_{123}K_{123}\Phi_1\Phi_2\Phi_3
-\frac{1}{6}\int_{1234}L_{1234}\Phi_1\Phi_2\Phi_3\Phi_4
-\frac{1}{2}\int_{123}K_{123}\Phi_1G_{23}
-\frac{1}{2}\int_{1234}L_{1234}G_{12}\Phi_3\Phi_4
\nn\\
&&
+\int_{12345}K_{123}\Phi_1G_{24}G_{35}\frac{\de\Ga_I}{\de G_{45}}
+\int_{123456}L_{1234}\Phi_1\Phi_2G_{35}G_{46}\frac{\de\Ga_I}{\de G_{56}}
\nn\\
&&
+\frac{1}{3}\int_{1234567}L_{1234}\Phi_1G_{25}G_{36}G_{47}
\frac{\de^2\Ga_I}{\de\Phi_5\de G_{67}}
-\frac{2}{3}
\int_{123456789}L_{1234}\Phi_1G_{25}G_{67}G_{38}G_{49}
\frac{\de\Ga_I}{\de G_{56}}\frac{\de^2\Ga_I}{\de\Phi_7\de G_{89}}.
\eea

Subtracting (\ref{ga0ssbidentity2}) from (\ref{gassbidentity2a}),
setting $x_2=x_1$, integrating over $x_1$ and using
(\ref{dga0ssbdginv12}), (\ref{d2ga0ssbdginvdphi}),
(\ref{d2ga0dginv11dginv11}) and (\ref{ddginv}) gives
\bea
\label{gaissbid2c}
0
&=&
\int_1\frac{\de\Ga_I}{\de\Phi_1}\Phi_1
+2\int_{12}G_{12}\frac{\de\Ga_I}{\de G_{12}}
-\frac{1}{2}\int_{123}K_{123}\Phi_1\Phi_2\Phi_3
-\frac{1}{6}\int_{1234}L_{1234}\Phi_1\Phi_2\Phi_3\Phi_4
\nn\\
&&
-\frac{3}{2}\int_{123}K_{123}G_{12}\Phi_3
-\int_{1234}L_{1234}G_{12}\Phi_3\Phi_4
-\frac{1}{2}\int_{1234}L_{1234}G_{12}G_{34}
\nn\\
&&
+3\int_{12345}K_{123}\Phi_1G_{24}G_{35}\frac{\de\Ga_I}{\de G_{45}}
+\int_{123456}K_{123}G_{14}G_{25}G_{36}\frac{\de^2\Ga_I}{\de\Phi_4\de G_{56}}
\nn\\
&&
-2\int_{12345678}K_{123}G_{34}G_{56}G_{27}G_{38}\frac{\de\Ga_I}{\de G_{45}}
\frac{\de^2\Ga_I}{\de\Phi_6\de G_{78}}
\nn\\
&&
+2\int_{123456}L_{1234}\Phi_1\Phi_2G_{35}G_{46}\frac{\de\Ga_I}{\de G_{56}}
+2\int_{123456}L_{1234}G_{12}G_{35}G_{46}\frac{\de\Ga_I}{\de G_{56}}
\nn\\
&&
+\frac{4}{3}\int_{1234567}L_{1234}\Phi_1G_{25}G_{36}G_{47}
\frac{\de^2\Ga_I}{\de\Phi_5\de G_{67}}
+\frac{2}{3}\int_{12345678}L_{1234}G_{15}G_{26}G_{37}G_{48}
\frac{\de^2\Ga_I}{\de G_{56}\de G_{78}}
\nn\\
&&
-\frac{2}{3}\int_{12345678}L_{1234}G_{15}G_{26}G_{37}G_{48}
\frac{\de\Ga_I}{\de G_{56}}\frac{\de\Ga_I}{\de G_{78}}
-\frac{8}{3}\int_{12345689}L_{1234}
\Phi_1G_{25}G_{38}G_{49}G_{67}\frac{\de\Ga_I}{\de G_{56}}
\frac{\de^2\Ga_I}{\de\Phi_7\de G_{89}}
\nn\\
&&
-\frac{2}{3}\int_{1\bar{1}2\bar{2}3\bar{3}4\bar{4}56}L_{1234}
G_{1\bar{1}}G_{2\bar{2}}G_{3\bar{3}}G_{4\bar{4}}G_{56}
\frac{\de^2\Ga_I}{\de\Phi_5\de G_{\bar{1}\bar{2}}}
\frac{\de^2\Ga_I}{\de\Phi_6\de G_{\bar{3}\bar{4}}}
\nn\\
&&
+\frac{4}{3}\int_{1\bar{1}2\bar{2}3\bar{3}4\bar{4}5678}L_{1234}
G_{1\bar{1}}G_{2\bar{2}}G_{3\bar{3}}G_{4\bar{4}}G_{56}G_{78}
\frac{\de^2\Ga_I}{\de\Phi_5\de G_{\bar{1}\bar{2}}}
\frac{\de\Ga_I}{\de G_{67}}
\frac{\de^2\Ga_I}{\de\Phi_8\de G_{\bar{3}\bar{4}}}.
\eea

To represent (\ref{gaissbid1c}) and (\ref{gaissbid2c}) graphically,
write for the derivatives of  $\Ga_I$ with respect to $\Phi$ and $G$
\beq
-\frac{\de\Ga_I}{\de\Phi_1}=
\rule[-18pt]{0pt}{42pt}
\bpi(55,0)(-23,0)
\put(8,3){\circle{32}}
\put(8,3){\makebox(1,0){$\Ga_I$}}
\put(-8,3){\circle{4}}
\put(-18,1){$\scs 1$}
\epi,
\;\;\;\;
-\frac{\de\Ga_I}{\de G_{12}}=
\rule[-18pt]{0pt}{42pt}
\bpi(50,0)(-10,0)
\put(19,3){\circle{32}}
\put(19,3){\makebox(1,0){$\Ga_I$}}
\put(-0.5,13){\makebox(0,0){$\scs 1$}}
\put(-0.5,-7){\makebox(0,0){$\scs 2$}}
\put(5.77,12){\circle*{4}}
\put(5.77,-6){\circle*{4}}
\epi,
\;\;\;\;
-\frac{\de^2\Ga_I}{\de\Phi_3\de G_{12}}=
\rule[-18pt]{0pt}{42pt}
\bpi(47,0)(-7,0)
\put(19,3){\circle{32}}
\put(19,3){\makebox(1,0){$\Ga_I$}}
\put(6,21){\makebox(0,0){$\scs 1$}}
\put(0,15){\makebox(0,0){$\scs 2$}}
\put(2,-12){\makebox(0,0){$\scs 3$}}
\put(5.22,11.4){\circle*{4}}
\put(10.87,17.1){\circle*{4}}
\put(7.69,-8.31){\circle{4}}
\epi,
\;\;\;\;
-\frac{\de^2\Ga_I}{\de G_{12}\de G_{34}}=
\rule[-18pt]{0pt}{42pt}
\bpi(47,0)(-7,0)
\put(19,3){\circle{32}}
\put(19,3){\makebox(1,0){$\Ga_I$}}
\put(6,21){\makebox(0,0){$\scs 1$}}
\put(0,15){\makebox(0,0){$\scs 2$}}
\put(0,-9){\makebox(0,0){$\scs 3$}}
\put(6,-15){\makebox(0,0){$\scs 4$}}
\put(5.22,11.4){\circle*{4}}
\put(10.87,17.1){\circle*{4}}
\put(5.22,-5.13){\circle*{4}}
\put(10.87,-10.78){\circle*{4}}
\epi
\eeq
and use the vertices
\beq
\label{gavertices}
-L_{1234}=
\rule[-14pt]{0pt}{34pt}
\bpi(34,0)(-4,0)
\put(5,-5){\line(1,1){16}}
\put(5,11){\line(1,-1){16}}
\put(13,3){\circle*{4}}
\put(25,15){\makebox(0,0){$\scs 1$}}
\put(1,15){\makebox(0,0){$\scs 2$}}
\put(1,-9){\makebox(0,0){$\scs 3$}}
\put(25,-9){\makebox(0,0){$\scs 4$}}
\epi,
\;\;\;\;
\label{3vertex}
-K_{123}=
\rule[-12pt]{0pt}{42pt}
\bpi(38,0)(-4,0)
\put(15,3){\circle*{4}}
\put(15,3){\line(0,1){11.66}}
\put(15,3){\line(5,-3){10}}
\put(15,3){\line(-5,-3){10}}
\put(15,21){\makebox(0,0){$\scs 1$}}
\put(1,-7){\makebox(0,0){$\scs 2$}}
\put(29,-7){\makebox(0,0){$\scs 3$}}
\epi,
\;\;\;\;
\label{1vertex}
-J_1=
\rule[-15pt]{0pt}{29pt}
\bpi(14,0)
\put(6,-5){\line(0,1){12}}
\put(8,-5){\line(0,1){12}}
\put(7,7){\circle*{4}}
\put(7,-10){\makebox(0,0){$\scs 1$}}
\epi,
\;\;\;\;
\label{0vertex}
-C
=
\rule[-2pt]{0pt}{10pt}
\bpi(10,0)
\put(5,3){\circle*{4}}
\epi.
\eeq
Instances of $\Phi$ are indicated by
\beq
\Phi_1=
\rule[-15pt]{0pt}{29pt}
\bpi(14,0)
\put(6,-5){\line(0,1){10}}
\put(8,-5){\line(0,1){10}}
\put(7,7){\circle{4}}
\put(7,-10){\makebox(0,0){$\scs 1$}}
\epi.
\eeq
Free propagators $G$ are indicated by lines connected at both ends.
The double lines on $J$ and $\Phi$ indicate that there are no propagators
attached to them in the diagrams, so derivatives with respect to $G$
act only on single lines [see however (\ref{phiphi})].
All space arguments that are not indicated by numbers are integrated over.

Now (\ref{ga0ssb}) can be written as
\beq
\label{ga0ssba}
-\Ga_0=
\rule[-2pt]{0pt}{10pt}
\bpi(10,0)
\put(5,3){\circle*{4}}
\epi
+
\rule[-4pt]{0pt}{14pt}
\bpi(27,0)
\put(7,3){\circle{4}}
\put(9,2){\line(1,0){11}}
\put(9,4){\line(1,0){11}}
\put(20,3){\circle*{4}}
\epi
+
\rule[-4pt]{0pt}{14pt}
\bpi(27,0)
\put(7,3){\circle{4}}
\put(9,2){\line(1,0){9}}
\put(9,4){\line(1,0){9}}
\put(20,3){\circle{4}}
\epi
+
\rule[-10pt]{0pt}{26pt}
\bpi(26,0)
\put(13,3){\circle{16}}
\epi,
\eeq
where by definition
\beq
\label{phiphi}
\rule[-4pt]{0pt}{14pt}
\bpi(27,0)
\put(7,3){\circle{4}}
\put(9,2){\line(1,0){9}}
\put(9,4){\line(1,0){9}}
\put(20,3){\circle{4}}
\epi
=
-\int_{12}\Phi_1G_{12}^{-1}\Phi_2,
\eeq
and (\ref{p12b}) as
\beq
P_{12}=
\bpi(43,0)
\put(11,3){\line(1,0){20}}
\put(6,3){\makebox(1,0){$\scs 1$}}
\put(36,3){\makebox(0,0){$\scs 2$}}
\epi
+2
\rule[-18pt]{0pt}{42pt}
\bpi(80,0)(-21,0)
\put(19,3){\circle{32}}
\put(19,3){\makebox(1,0){$\Ga_I$}}
\put(-14,-10.78){\makebox(0,0){$\scs 1$}}
\put(10.87,-10.78){\circle*{4}}
\put(52,-10.78){\makebox(0,0){$\scs 2$}}
\put(27.13,-10.78){\circle*{4}}
\put(10.87,-10.78){\line(-1,0){20}}
\put(27.13,-10.78){\line(1,0){20}}
\epi.
\eeq

We can write (\ref{gaissbid1c}) as
\bea
\label{gaissbid1d}
\rule[-18pt]{0pt}{42pt}
\bpi(57,0)(-17,0)
\put(19,3){\circle{32}}
\put(19,3){\makebox(1,0){$\Ga_I$}}
\put(3,3){\circle{4}}
\put(-8,2){\line(1,0){9}}
\put(-8,4){\line(1,0){9}}
\put(-10,3){\circle{4}}
\epi
&=&
\frac{1}{2}
\rule[-11pt]{0pt}{35pt}
\bpi(38.24,0)(-4.12,0)
\put(15,3){\circle*{4}}
\put(14,3){\line(0,1){12}}
\put(16,3){\line(0,1){12}}
\put(15,17){\circle{4}}
\put(15.5,2.13){\line(-5,-3){10.39}}
\put(14.5,3.87){\line(-5,-3){10.39}}
\put(2.88,-4){\circle{4}}
\put(14.5,2.13){\line(5,-3){10.39}}
\put(15.5,3.87){\line(5,-3){10.39}}
\put(27.12,-4){\circle{4}}
\epi
+\frac{1}{6}
\rule[-15.41pt]{0pt}{36.82pt}
\bpi(36.82,0)(-3.41,0)
\put(15,3){\circle*{4}}
\put(14.29,2.29){\line(-1,1){10}}
\put(15.71,3.71){\line(-1,1){10}}
\put(3.59,14.41){\circle{4}}
\put(15.71,2.29){\line(-1,-1){10}}
\put(14.29,3.71){\line(-1,-1){10}}
\put(3.59,-8.41){\circle{4}}
\put(14.29,3.71){\line(1,1){10}}
\put(15.71,2.29){\line(1,1){10}}
\put(26.41,14.41){\circle{4}}
\put(15.71,3.71){\line(1,-1){10}}
\put(14.29,2.29){\line(1,-1){10}}
\put(26.41,-8.41){\circle{4}}
\epi
+\frac{1}{2}
\rule[-10pt]{0pt}{21pt}
\bpi(41,0)
\put(28,3){\circle{16}}
\put(20,3){\circle*{4}}
\put(9,2){\line(1,0){11}}
\put(9,4){\line(1,0){11}}
\put(7,3){\circle{4}}
\epi
+\frac{1}{2}
\rule[-15.41pt]{0pt}{36.82pt}
\bpi(39.41,0)(-3.41,0)
\put(15,3){\circle*{4}}
\put(14.29,2.29){\line(-1,1){10}}
\put(15.71,3.71){\line(-1,1){10}}
\put(3.59,14.41){\circle{4}}
\put(15.71,2.29){\line(-1,-1){10}}
\put(14.29,3.71){\line(-1,-1){10}}
\put(3.59,-8.41){\circle{4}}
\put(23,3){\circle{16}}
\epi
\nn\\
&&
+
\rule[-18pt]{0pt}{42pt}
\bpi(68.23,0)(-28.23,0)
\put(19,3){\circle{32}}
\put(19,3){\makebox(1,0){$\Ga_I$}}
\put(5.77,12){\circle*{4}}
\put(5.77,-6){\circle*{4}}
\put(5.77,3){\oval(28,18)[l]}
\put(-8.23,3){\circle*{4}}
\put(-19.23,2){\line(1,0){11}}
\put(-19.23,4){\line(1,0){11}}
\put(-21.23,3){\circle{4}}
\epi
+
\rule[-15.41pt]{0pt}{36.82pt}
\bpi(66.64,0)(-3.41,0)
\put(15,3){\circle*{4}}
\put(14.29,2.29){\line(-1,1){10}}
\put(15.71,3.71){\line(-1,1){10}}
\put(3.59,14.41){\circle{4}}
\put(15.71,2.29){\line(-1,-1){10}}
\put(14.29,3.71){\line(-1,-1){10}}
\put(3.59,-8.41){\circle{4}}
\put(42.23,3){\circle{32}}
\put(42.23,3){\makebox(1,0){$\Ga_I$}}
\put(29,12){\circle*{4}}
\put(29,-6){\circle*{4}}
\put(29,3){\oval(28,18)[l]}
\epi
+\frac{1}{3}
\rule[-18pt]{0pt}{42pt}
\bpi(68.23,0)(-28.23,0)
\put(19,3){\circle{32}}
\put(19,3){\makebox(1,0){$\Ga_I$}}
\put(5.77,12){\circle*{4}}
\put(5.77,-6){\circle{4}}
\put(3.77,3){\oval(24,18)[l]}
\put(-8.23,3){\circle*{4}}
\put(-19.23,2){\line(1,0){11}}
\put(-19.23,4){\line(1,0){11}}
\put(-8.23,3){\line(1,0){11.23}}
\put(3,3){\circle*{4}}
\put(-21.23,3){\circle{4}}
\epi
+\frac{2}{3}
\rule[-18pt]{0pt}{51pt}
\bpi(96,0)(-2,0)
\put(19,3){\circle{32}}
\put(19,3){\makebox(1,0){$\Ga_I$}}
\put(32.23,12){\circle*{4}}
\put(32.23,-6){\circle*{4}}
\put(46,12){\circle*{4}}
\put(32.23,12){\line(1,0){13.77}}
\put(32.23,-6){\line(1,0){25.54}}
\put(57.8,12){\oval(23.6,8)[l]}
\put(57.8,16){\line(1,0){5.87}}
\put(63.67,16){\circle*{4}}
\put(57.8,8){\circle*{4}}
\put(45,12){\line(0,1){12}}
\put(47,12){\line(0,1){12}}
\put(46,26){\circle{4}}
\put(73,3){\circle{32}}
\put(73,3){\makebox(1,0){$\Ga_I$}}
\put(59.77,-6){\circle{4}}
\epi
\eea
and (\ref{gaissbid2c}) as
\bea
\label{gaissbid2d}
\lefteqn{2
\rule[-18pt]{0pt}{42pt}
\bpi(53.23,0)(-13.23,0)
\put(19,3){\circle{32}}
\put(19,3){\makebox(1,0){$\Ga_I$}}
\put(5.77,12){\circle*{4}}
\put(5.77,-6){\circle*{4}}
\put(5.77,3){\oval(28,18)[l]}
\epi
+
\rule[-18pt]{0pt}{42pt}
\bpi(57,0)(-17,0)
\put(19,3){\circle{32}}
\put(19,3){\makebox(1,0){$\Ga_I$}}
\put(3,3){\circle{4}}
\put(-8,2){\line(1,0){9}}
\put(-8,4){\line(1,0){9}}
\put(-10,3){\circle{4}}
\epi}
\nn\\
&=&
\frac{1}{2}
\rule[-11pt]{0pt}{35pt}
\bpi(38.24,0)(-4.12,0)
\put(15,3){\circle*{4}}
\put(14,3){\line(0,1){12}}
\put(16,3){\line(0,1){12}}
\put(15,17){\circle{4}}
\put(15.5,2.13){\line(-5,-3){10.39}}
\put(14.5,3.87){\line(-5,-3){10.39}}
\put(2.88,-4){\circle{4}}
\put(14.5,2.13){\line(5,-3){10.39}}
\put(15.5,3.87){\line(5,-3){10.39}}
\put(27.12,-4){\circle{4}}
\epi
+\frac{1}{6}
\rule[-15.41pt]{0pt}{36.82pt}
\bpi(36.82,0)(-3.41,0)
\put(15,3){\circle*{4}}
\put(14.29,2.29){\line(-1,1){10}}
\put(15.71,3.71){\line(-1,1){10}}
\put(3.59,14.41){\circle{4}}
\put(15.71,2.29){\line(-1,-1){10}}
\put(14.29,3.71){\line(-1,-1){10}}
\put(3.59,-8.41){\circle{4}}
\put(14.29,3.71){\line(1,1){10}}
\put(15.71,2.29){\line(1,1){10}}
\put(26.41,14.41){\circle{4}}
\put(15.71,3.71){\line(1,-1){10}}
\put(14.29,2.29){\line(1,-1){10}}
\put(26.41,-8.41){\circle{4}}
\epi
+\frac{3}{2}
\rule[-10pt]{0pt}{21pt}
\bpi(41,0)
\put(28,3){\circle{16}}
\put(20,3){\circle*{4}}
\put(9,2){\line(1,0){11}}
\put(9,4){\line(1,0){11}}
\put(7,3){\circle{4}}
\epi
+
\rule[-15.41pt]{0pt}{36.82pt}
\bpi(39.41,0)(-3.41,0)
\put(15,3){\circle*{4}}
\put(14.29,2.29){\line(-1,1){10}}
\put(15.71,3.71){\line(-1,1){10}}
\put(3.59,14.41){\circle{4}}
\put(15.71,2.29){\line(-1,-1){10}}
\put(14.29,3.71){\line(-1,-1){10}}
\put(3.59,-8.41){\circle{4}}
\put(23,3){\circle{16}}
\epi
+\frac{1}{2}
\rule[-10pt]{0pt}{26pt}
\bpi(42,0)
\put(21,3){\circle*{4}}
\put(13,3){\circle{16}}
\put(29,3){\circle{16}}
\epi
\nn\\
&&
+3
\rule[-18pt]{0pt}{42pt}
\bpi(68.23,0)(-28.23,0)
\put(19,3){\circle{32}}
\put(19,3){\makebox(1,0){$\Ga_I$}}
\put(5.77,12){\circle*{4}}
\put(5.77,-6){\circle*{4}}
\put(5.77,3){\oval(28,18)[l]}
\put(-8.23,3){\circle*{4}}
\put(-19.23,2){\line(1,0){11}}
\put(-19.23,4){\line(1,0){11}}
\put(-21.23,3){\circle{4}}
\epi
+
\rule[-18pt]{0pt}{42pt}
\bpi(55.23,0)(-15.23,0)
\put(19,3){\circle{32}}
\put(19,3){\makebox(1,0){$\Ga_I$}}
\put(5.77,12){\circle*{4}}
\put(5.77,-6){\circle{4}}
\put(3.77,3){\oval(24,18)[l]}
\put(-8.23,3){\circle*{4}}
\put(-8.23,3){\line(1,0){11.23}}
\put(3,3){\circle*{4}}
\epi
+2
\rule[-18pt]{0pt}{42pt}
\bpi(96,0)(-2,0)
\put(19,3){\circle{32}}
\put(19,3){\makebox(1,0){$\Ga_I$}}
\put(32.23,12){\circle*{4}}
\put(32.23,-6){\circle*{4}}
\put(46,12){\circle*{4}}
\put(32.23,12){\line(1,0){13.77}}
\put(32.23,-6){\line(1,0){25.54}}
\put(57.8,12){\oval(23.6,8)[l]}
\put(57.8,16){\line(1,0){5.87}}
\put(63.67,16){\circle*{4}}
\put(57.8,8){\circle*{4}}
\put(73,3){\circle{32}}
\put(73,3){\makebox(1,0){$\Ga_I$}}
\put(59.77,-6){\circle{4}}
\epi
\nn\\
&&
+2
\rule[-15.41pt]{0pt}{36.82pt}
\bpi(66.64,0)(-3.41,0)
\put(15,3){\circle*{4}}
\put(14.29,2.29){\line(-1,1){10}}
\put(15.71,3.71){\line(-1,1){10}}
\put(3.59,14.41){\circle{4}}
\put(15.71,2.29){\line(-1,-1){10}}
\put(14.29,3.71){\line(-1,-1){10}}
\put(3.59,-8.41){\circle{4}}
\put(42.23,3){\circle{32}}
\put(42.23,3){\makebox(1,0){$\Ga_I$}}
\put(29,12){\circle*{4}}
\put(29,-6){\circle*{4}}
\put(29,3){\oval(28,18)[l]}
\epi
+2
\rule[-18pt]{0pt}{42pt}
\bpi(69.23,0)(-29.23,0)
\put(19,3){\circle{32}}
\put(19,3){\makebox(1,0){$\Ga_I$}}
\put(5.77,12){\circle*{4}}
\put(5.77,-6){\circle*{4}}
\put(5.77,3){\oval(28,18)[l]}
\put(-8.23,3){\circle*{4}}
\put(-16.23,3){\circle{16}}
\epi
+\frac{4}{3}
\rule[-18pt]{0pt}{42pt}
\bpi(68.23,0)(-28.23,0)
\put(19,3){\circle{32}}
\put(19,3){\makebox(1,0){$\Ga_I$}}
\put(5.77,12){\circle*{4}}
\put(5.77,-6){\circle{4}}
\put(3.77,3){\oval(24,18)[l]}
\put(-8.23,3){\circle*{4}}
\put(-19.23,2){\line(1,0){11}}
\put(-19.23,4){\line(1,0){11}}
\put(-8.23,3){\line(1,0){11.23}}
\put(3,3){\circle*{4}}
\put(-21.23,3){\circle{4}}
\epi
+\frac{2}{3}
\rule[-18pt]{0pt}{42pt}
\bpi(54.78,0)(-14.78,0)
\put(19,3){\circle{32}}
\put(19,3){\makebox(1,0){$\Ga_I$}}
\put(5.22,11.4){\circle*{4}}
\put(10.87,17.1){\circle*{4}}
\put(5.22,-5.13){\circle*{4}}
\put(10.87,-10.78){\circle*{4}}
\put(10.87,3){\oval(41.3,27.88)[l]}
\put(5.22,3){\oval(30,16.53)[l]}
\put(-9.78,3){\circle*{4}}
\epi
+\frac{2}{3}
\rule[-18pt]{0pt}{42pt}
\bpi(96.46,0)(-56.46,0)
\put(19,3){\circle{32}}
\put(19,3){\makebox(1,0){$\Ga_I$}}
\put(5.77,12){\circle*{4}}
\put(5.77,-6){\circle*{4}}
\put(5.77,3){\oval(28,18)[l]}
\put(-8.23,3){\circle*{4}}
\put(-22.23,3){\oval(28,18)[r]}
\put(-22.23,12){\circle*{4}}
\put(-22.23,-6){\circle*{4}}
\put(-35.46,3){\makebox(1,0){$\Ga_I$}}
\put(-35.46,3){\circle{32}}
\epi
\nn\\
&&
+\frac{8}{3}
\rule[-18pt]{0pt}{51pt}
\bpi(96,0)(-2,0)
\put(19,3){\circle{32}}
\put(19,3){\makebox(1,0){$\Ga_I$}}
\put(32.23,12){\circle*{4}}
\put(32.23,-6){\circle*{4}}
\put(46,12){\circle*{4}}
\put(32.23,12){\line(1,0){13.77}}
\put(32.23,-6){\line(1,0){25.54}}
\put(57.8,12){\oval(23.6,8)[l]}
\put(57.8,16){\line(1,0){5.87}}
\put(63.67,16){\circle*{4}}
\put(57.8,8){\circle*{4}}
\put(45,12){\line(0,1){12}}
\put(47,12){\line(0,1){12}}
\put(46,26){\circle{4}}
\put(73,3){\circle{32}}
\put(73,3){\makebox(1,0){$\Ga_I$}}
\put(59.77,-6){\circle{4}}
\epi
+\frac{2}{3}
\rule[-18pt]{0pt}{42pt}
\bpi(96,0)(-2,0)
\put(19,3){\circle{32}}
\put(19,3){\makebox(1,0){$\Ga_I$}}
\put(32.23,-6){\circle{4}}
\put(46,12){\circle*{4}}
\put(34.23,-6){\line(1,0){23.54}}
\put(34.2,12){\oval(23.6,8)[r]}
\put(34.2,16){\line(-1,0){5.87}}
\put(28.33,16){\circle*{4}}
\put(34.2,8){\circle*{4}}
\put(57.8,12){\oval(23.6,8)[l]}
\put(57.8,16){\line(1,0){5.87}}
\put(63.67,16){\circle*{4}}
\put(57.8,8){\circle*{4}}
\put(73,3){\circle{32}}
\put(73,3){\makebox(1,0){$\Ga_I$}}
\put(59.77,-6){\circle{4}}
\epi
+\frac{4}{3}
\rule[-40.23pt]{0pt}{64.23pt}
\bpi(114,0)(-2,0)
\put(19,3){\circle{32}}
\put(19,3){\makebox(1,0){$\Ga_I$}}
\put(32.23,-6){\circle{4}}
\put(55,12){\circle*{4}}
\put(34.23,-6){\line(1,0){11.77}}
\put(46,-6){\circle*{4}}
\put(34.2,12){\oval(41.6,8)[r]}
\put(34.2,16){\line(-1,0){5.87}}
\put(28.33,16){\circle*{4}}
\put(34.2,8){\circle*{4}}
\put(75.77,-6){\line(-1,0){11.77}}
\put(75.8,12){\oval(41.6,8)[l]}
\put(75.8,16){\line(1,0){5.87}}
\put(64,-6){\circle*{4}}
\put(81.67,16){\circle*{4}}
\put(75.8,8){\circle*{4}}
\put(55,-19.23){\circle{32}}
\put(55,-19.23){\makebox(1,0){$\Ga_I$}}
\put(91,3){\circle{32}}
\put(91,3){\makebox(1,0){$\Ga_I$}}
\put(77.77,-6){\circle{4}}
\epi.
\eea
Note that in the limit where $K=0$ and $\Phi=0$ this is identical
to (\ref{wisym}) in the limit $\De=0$ if there we replace
$W_I\rightarrow-\Ga_I$.

\subsection{Recursion Relations}
\label{gairecrel}
Consider a double expansion in the number of loops $L$ and
powers $n$ of $\Phi$,
\beq
\label{gadoubleex}
-\Ga=
\rule[-18pt]{0pt}{42pt}
\bpi(42,0)
\put(21,3){\circle{32}}
\put(21,3){\makebox(0,0){$\Ga$}}
\epi
=-\sum_{L=0}^\infty\sum_{n=0}^\infty \Ga^{(L,n)}=
\sum_{L=0}^\infty\sum_{n=0}^\infty
\rule[-18pt]{0pt}{42pt}
\bpi(42,0)
\put(21,3){\circle{32}}
\put(21,3){\makebox(1,-2){$\ba{c}L\\n\ea$}}
\epi
\eeq
[the double-indexed circles are not identical to those in (\ref{wdoubleex})].
Then the $L$-loop contribution to the proper $n$-point vertex with vanishing
external field is given by
\beq
\label{galn}
\Ga^{(L,n)}_{i_1,\ldots,i_n}
=\left.\frac{\de^n}{\de\Phi_{i_1}\ldots\de\Phi_{i_n}}\Ga^{(L)}\right|_{\Phi=0}
=\frac{\de^n}{\de\Phi_{i_1}\ldots\de\Phi_{i_n}}\Ga^{(L,n)}.
\eeq

We have from (\ref{ga0ssba})
\beq
\label{ga00010210}
\rule[-18pt]{0pt}{42pt}
\bpi(42,0)
\put(21,3){\circle{32}}
\put(21,3){\makebox(1,-2){$\ba{c}0\\0\ea$}}
\epi
=
\rule[-2pt]{0pt}{10pt}
\bpi(10,0)
\put(5,3){\circle*{4}}
\epi,
\;\;\;\;
\rule[-18pt]{0pt}{42pt}
\bpi(42,0)
\put(21,3){\circle{32}}
\put(21,3){\makebox(1,-2){$\ba{c}0\\1\ea$}}
\epi
=
\rule[-4pt]{0pt}{14pt}
\bpi(27,0)
\put(7,3){\circle{4}}
\put(9,2){\line(1,0){11}}
\put(9,4){\line(1,0){11}}
\put(20,3){\circle*{4}}
\epi,
\;\;\;\;
\rule[-18pt]{0pt}{42pt}
\bpi(42,0)
\put(21,3){\circle{32}}
\put(21,3){\makebox(1,-2){$\ba{c}0\\2\ea$}}
\epi
=\frac{1}{2}
\rule[-4pt]{0pt}{14pt}
\bpi(27,0)
\put(7,3){\circle{4}}
\put(9,2){\line(1,0){9}}
\put(9,4){\line(1,0){9}}
\put(20,3){\circle{4}}
\epi,
\;\;\;\;
\rule[-18pt]{0pt}{42pt}
\bpi(42,0)
\put(21,3){\circle{32}}
\put(21,3){\makebox(1,-2){$\ba{c}1\\0\ea$}}
\epi
=\frac{1}{2}
\rule[-10pt]{0pt}{26pt}
\bpi(26,0)
\put(13,3){\circle{16}}
\epi,
\eeq
The other $\Ga^{(L,n)}$ constitute $\Ga_I$.

Using
\bea
\label{dgadjid}
\rule[-18pt]{0pt}{42pt}
\bpi(57,0)(-17,0)
\put(19,3){\circle{32}}
\put(19,3){\makebox(1,-2){$\ba{c}L\\n\ea$}}
\put(3,3){\circle{4}}
\put(-8,2){\line(1,0){9}}
\put(-8,4){\line(1,0){9}}
\put(-10,3){\circle{4}}
\epi
&=&
n
\rule[-18pt]{0pt}{42pt}
\bpi(42,0)
\put(21,3){\circle{32}}
\put(21,3){\makebox(1,-2){$\ba{c}L\\n\ea$}}
\epi,
\eea
(\ref{gaissbid1d}) can be split into
\beq
\label{ga030411}
\rule[-18pt]{0pt}{42pt}
\bpi(42,0)
\put(21,3){\circle{32}}
\put(21,3){\makebox(1,-2){$\ba{c}0\\3\ea$}}
\epi
=
\frac{1}{6}
\rule[-11pt]{0pt}{35pt}
\bpi(38.24,0)(-4.12,0)
\put(15,3){\circle*{4}}
\put(14,3){\line(0,1){12}}
\put(16,3){\line(0,1){12}}
\put(15,17){\circle{4}}
\put(15.5,2.13){\line(-5,-3){10.39}}
\put(14.5,3.87){\line(-5,-3){10.39}}
\put(2.88,-4){\circle{4}}
\put(14.5,2.13){\line(5,-3){10.39}}
\put(15.5,3.87){\line(5,-3){10.39}}
\put(27.12,-4){\circle{4}}
\epi,
\;\;\;\;
\rule[-18pt]{0pt}{42pt}
\bpi(42,0)
\put(21,3){\circle{32}}
\put(21,3){\makebox(1,-2){$\ba{c}0\\4\ea$}}
\epi
=
\frac{1}{24}
\rule[-15.41pt]{0pt}{36.82pt}
\bpi(36.82,0)(-3.41,0)
\put(15,3){\circle*{4}}
\put(14.29,2.29){\line(-1,1){10}}
\put(15.71,3.71){\line(-1,1){10}}
\put(3.59,14.41){\circle{4}}
\put(15.71,2.29){\line(-1,-1){10}}
\put(14.29,3.71){\line(-1,-1){10}}
\put(3.59,-8.41){\circle{4}}
\put(14.29,3.71){\line(1,1){10}}
\put(15.71,2.29){\line(1,1){10}}
\put(26.41,14.41){\circle{4}}
\put(15.71,3.71){\line(1,-1){10}}
\put(14.29,2.29){\line(1,-1){10}}
\put(26.41,-8.41){\circle{4}}
\epi,
\;\;\;\;
\rule[-18pt]{0pt}{42pt}
\bpi(42,0)
\put(21,3){\circle{32}}
\put(21,3){\makebox(1,-2){$\ba{c}1\\1\ea$}}
\epi
=
\frac{1}{2}
\rule[-10pt]{0pt}{26pt}
\bpi(41,0)
\put(28,3){\circle{16}}
\put(20,3){\circle*{4}}
\put(9,2){\line(1,0){11}}
\put(9,4){\line(1,0){11}}
\put(7,3){\circle{4}}
\epi,
\eeq

\beq
\label{ga12}
\rule[-18pt]{0pt}{42pt}
\bpi(42,0)
\put(21,3){\circle{32}}
\put(21,3){\makebox(1,-2){$\ba{c}1\\2\ea$}}
\epi
=
\frac{1}{4}
\rule[-15.41pt]{0pt}{36.82pt}
\bpi(39.41,0)(-3.41,0)
\put(15,3){\circle*{4}}
\put(14.29,2.29){\line(-1,1){10}}
\put(15.71,3.71){\line(-1,1){10}}
\put(3.59,14.41){\circle{4}}
\put(15.71,2.29){\line(-1,-1){10}}
\put(14.29,3.71){\line(-1,-1){10}}
\put(3.59,-8.41){\circle{4}}
\put(23,3){\circle{16}}
\epi
+\frac{1}{2}
\rule[-15.41pt]{0pt}{36.82pt}
\bpi(66.64,0)(-3.41,0)
\put(15,3){\circle*{4}}
\put(14.29,2.29){\line(-1,1){10}}
\put(15.71,3.71){\line(-1,1){10}}
\put(3.59,14.41){\circle{4}}
\put(15.71,2.29){\line(-1,-1){10}}
\put(14.29,3.71){\line(-1,-1){10}}
\put(3.59,-8.41){\circle{4}}
\put(42.23,3){\circle{32}}
\put(42.23,3){\makebox(1,-2){$\ba{c}1\\1\ea$}}
\put(29,12){\circle*{4}}
\put(29,-6){\circle*{4}}
\put(29,3){\oval(28,18)[l]}
\epi
=
\frac{1}{4}
\rule[-15.41pt]{0pt}{36.82pt}
\bpi(39.41,0)(-3.41,0)
\put(15,3){\circle*{4}}
\put(14.29,2.29){\line(-1,1){10}}
\put(15.71,3.71){\line(-1,1){10}}
\put(3.59,14.41){\circle{4}}
\put(15.71,2.29){\line(-1,-1){10}}
\put(14.29,3.71){\line(-1,-1){10}}
\put(3.59,-8.41){\circle{4}}
\put(23,3){\circle{16}}
\epi
+\frac{1}{4}
\rule[-10pt]{0pt}{26pt}
\bpi(62,0)
\put(7,3){\circle{4}}
\put(9,2){\line(1,0){14}}
\put(9,4){\line(1,0){14}}
\put(23,3){\circle*{4}}
\put(31,3){\circle{16}}
\put(39,3){\circle*{4}}
\put(39,2){\line(1,0){14}}
\put(39,4){\line(1,0){14}}
\put(55,3){\circle{4}}
\epi
\eeq
and the recursion relation
\bea
\label{gaissbid1e}
n
\rule[-18pt]{0pt}{42pt}
\bpi(42,0)
\put(21,3){\circle{32}}
\put(21,3){\makebox(1,-2){$\ba{c}L\\n\ea$}}
\epi
&\doteq&
\rule[-18pt]{0pt}{42pt}
\bpi(68.23,0)(-28.23,0)
\put(19,3){\circle{32}}
\put(19,3){\makebox(1,-2){$\ba{c}L\\n{-}1\ea$}}
\put(5.77,12){\circle*{4}}
\put(5.77,-6){\circle*{4}}
\put(5.77,3){\oval(28,18)[l]}
\put(-8.23,3){\circle*{4}}
\put(-19.23,2){\line(1,0){11}}
\put(-19.23,4){\line(1,0){11}}
\put(-21.23,3){\circle{4}}
\epi
+
\rule[-15.41pt]{0pt}{36.82pt}
\bpi(66.64,0)(-3.41,0)
\put(15,3){\circle*{4}}
\put(14.29,2.29){\line(-1,1){10}}
\put(15.71,3.71){\line(-1,1){10}}
\put(3.59,14.41){\circle{4}}
\put(15.71,2.29){\line(-1,-1){10}}
\put(14.29,3.71){\line(-1,-1){10}}
\put(3.59,-8.41){\circle{4}}
\put(42.23,3){\circle{32}}
\put(42.23,3){\makebox(1,-2){$\ba{c}L\\n{-}2\ea$}}
\put(29,12){\circle*{4}}
\put(29,-6){\circle*{4}}
\put(29,3){\oval(28,18)[l]}
\epi
+\frac{1}{3}
\rule[-18pt]{0pt}{42pt}
\bpi(68.23,0)(-28.23,0)
\put(19,3){\circle{32}}
\put(19,3){\makebox(1,-2){$\ba{c}L{-}1\\n\ea$}}
\put(5.77,12){\circle*{4}}
\put(5.77,-6){\circle{4}}
\put(3.77,3){\oval(24,18)[l]}
\put(-8.23,3){\circle*{4}}
\put(-19.23,2){\line(1,0){11}}
\put(-19.23,4){\line(1,0){11}}
\put(-8.23,3){\line(1,0){11.23}}
\put(3,3){\circle*{4}}
\put(-21.23,3){\circle{4}}
\epi
+\frac{2}{3}\sum_{l=1}^{L-1}\sum_{m=0}^{n-1}
\rule[-18pt]{0pt}{51pt}
\bpi(96,0)(-2,0)
\put(19,3){\circle{32}}
\put(19,3){\makebox(1,-2){$\ba{c}l\\m\ea$}}
\put(32.23,12){\circle*{4}}
\put(32.23,-6){\circle*{4}}
\put(46,12){\circle*{4}}
\put(32.23,12){\line(1,0){13.77}}
\put(32.23,-6){\line(1,0){25.54}}
\put(57.8,12){\oval(23.6,8)[l]}
\put(57.8,16){\line(1,0){5.87}}
\put(63.67,16){\circle*{4}}
\put(57.8,8){\circle*{4}}
\put(45,12){\line(0,1){12}}
\put(47,12){\line(0,1){12}}
\put(46,26){\circle{4}}
\put(73,3){\circle{32}}
\put(73,3){\makebox(1,-2){$\ba{c}L{-}l\\n{-}m\ea$}}
\put(59.77,-6){\circle{4}}
\epi,
\nn\\
\eea
where the dot on the equal sign means that the right hand side only
involves $\Ga^{(i,j)}$ that are part of $\Ga_I$, i.e.\ excluding
$(i,j)\in\{(0,0),(0,1),(0,2),(1,0)\}$ and negative $i$ or $j$.
Eq.\ (\ref{gaissbid1e}) is valid for all $\Ga^{(L,n)}$ which are
part of $\Ga_I$ with the exception of $(L,n)\in\{(0,3),(0,4),(1,1),(1,2)\}$.

Note that from (\ref{gaissbid1e}) follows that
\beq
\rule[-18pt]{0pt}{42pt}
\bpi(42,0)
\put(21,3){\circle{32}}
\put(21,3){\makebox(1,-2){$\ba{c}0\\n\ea$}}
\epi
=0
\eeq
for $n>4$.

From (\ref{gaissbid2d}) follow again equations leading with
(\ref{n1dngd}) to (\ref{ga030411}) and (\ref{ga12}), but also
\beq
2
\rule[-18pt]{0pt}{42pt}
\bpi(53.23,0)(-13.23,0)
\put(19,3){\circle{32}}
\put(19,3){\makebox(1,-2){$\ba{c}2\\0\ea$}}
\put(5.77,12){\circle*{4}}
\put(5.77,-6){\circle*{4}}
\put(5.77,3){\oval(28,18)[l]}
\epi
=
\frac{1}{2}
\rule[-10pt]{0pt}{26pt}
\bpi(42,0)
\put(21,3){\circle*{4}}
\put(13,3){\circle{16}}
\put(29,3){\circle{16}}
\epi
+
\rule[-18pt]{0pt}{42pt}
\bpi(55.23,0)(-15.23,0)
\put(19,3){\circle{32}}
\put(19,3){\makebox(1,-2){$\ba{c}1\\1\ea$}}
\put(5.77,12){\circle*{4}}
\put(5.77,-6){\circle{4}}
\put(3.77,3){\oval(24,18)[l]}
\put(-8.23,3){\circle*{4}}
\put(-8.23,3){\line(1,0){11.23}}
\put(3,3){\circle*{4}}
\epi
=
\frac{1}{2}
\rule[-10pt]{0pt}{26pt}
\bpi(46,0)
\put(23,3){\circle*{4}}
\put(15,3){\circle{16}}
\put(31,3){\circle{16}}
\epi
+\frac{1}{2}
\rule[-14pt]{0pt}{34pt}
\bpi(38,0)
\put(19,3){\circle{24}}
\put(7,3){\circle*{4}}
\put(31,3){\circle*{4}}
\put(7,3){\line(1,0){24}}
\epi,
\eeq
which with (\ref{n1dngd}) becomes
\beq
\label{ga20}
\rule[-18pt]{0pt}{42pt}
\bpi(42,0)
\put(21,3){\circle{32}}
\put(21,3){\makebox(1,-2){$\ba{c}2\\0\ea$}}
\epi
=
\frac{1}{8}
\rule[-10pt]{0pt}{26pt}
\bpi(46,0)
\put(23,3){\circle*{4}}
\put(15,3){\circle{16}}
\put(31,3){\circle{16}}
\epi
+\frac{1}{12}
\rule[-14pt]{0pt}{34pt}
\bpi(38,0)
\put(19,3){\circle{24}}
\put(7,3){\circle*{4}}
\put(31,3){\circle*{4}}
\put(7,3){\line(1,0){24}}
\epi,
\eeq
and a recursion relation which we write down only for $n=0$,
since for $n>0$ the simpler relation (\ref{gaissbid1e}) can be used:
\bea
\label{gaissbid2vac}
\rule[-18pt]{0pt}{42pt}
\bpi(53.23,0)(-13.23,0)
\put(19,3){\circle{32}}
\put(19,3){\makebox(1,-2){$\ba{c}L\\0\ea$}}
\put(5.77,12){\circle*{4}}
\put(5.77,-6){\circle*{4}}
\put(5.77,3){\oval(28,18)[l]}
\epi
&=&
\frac{1}{2}
\rule[-18pt]{0pt}{42pt}
\bpi(55.23,0)(-15.23,0)
\put(19,3){\circle{32}}
\put(19,3){\makebox(1,-2){$\ba{c}L{-}1\\1\ea$}}
\put(5.77,12){\circle*{4}}
\put(5.77,-6){\circle{4}}
\put(3.77,3){\oval(24,18)[l]}
\put(-8.23,3){\circle*{4}}
\put(-8.23,3){\line(1,0){11.23}}
\put(3,3){\circle*{4}}
\epi
+
\rule[-18pt]{0pt}{42pt}
\bpi(69.23,0)(-29.23,0)
\put(19,3){\circle{32}}
\put(19,3){\makebox(1,-2){$\ba{c}L{-}1\\0\ea$}}
\put(5.77,12){\circle*{4}}
\put(5.77,-6){\circle*{4}}
\put(5.77,3){\oval(28,18)[l]}
\put(-8.23,3){\circle*{4}}
\put(-16.23,3){\circle{16}}
\epi
+\frac{1}{3}
\rule[-18pt]{0pt}{42pt}
\bpi(54.78,0)(-14.78,0)
\put(19,3){\circle{32}}
\put(19,3){\makebox(1,-2){$\ba{c}L{-}1\\0\ea$}}
\put(5.22,11.4){\circle*{4}}
\put(10.87,17.1){\circle*{4}}
\put(5.22,-5.13){\circle*{4}}
\put(10.87,-10.78){\circle*{4}}
\put(10.87,3){\oval(41.3,27.88)[l]}
\put(5.22,3){\oval(30,16.53)[l]}
\put(-9.78,3){\circle*{4}}
\epi
\nn\\
&&
+\sum_{l=1}^{L-2}
\rule[-18pt]{0pt}{42pt}
\bpi(96,0)(-2,0)
\put(19,3){\circle{32}}
\put(19,3){\makebox(1,-2){$\ba{c}L{-}l\\0\ea$}}
\put(32.23,12){\circle*{4}}
\put(32.23,-6){\circle*{4}}
\put(46,12){\circle*{4}}
\put(32.23,12){\line(1,0){13.77}}
\put(32.23,-6){\line(1,0){25.54}}
\put(57.8,12){\oval(23.6,8)[l]}
\put(57.8,16){\line(1,0){5.87}}
\put(63.67,16){\circle*{4}}
\put(57.8,8){\circle*{4}}
\put(73,3){\circle{32}}
\put(73,3){\makebox(1,-2){$\ba{c}l\\1\ea$}}
\put(59.77,-6){\circle{4}}
\epi
+\frac{1}{3}\sum_{l=2}^{L-2}
\rule[-18pt]{0pt}{42pt}
\bpi(96.46,0)(-56.46,0)
\put(19,3){\circle{32}}
\put(19,3){\makebox(1,-2){$\ba{c}L{-}l\\0\ea$}}
\put(5.77,12){\circle*{4}}
\put(5.77,-6){\circle*{4}}
\put(5.77,3){\oval(28,18)[l]}
\put(-8.23,3){\circle*{4}}
\put(-22.23,3){\oval(28,18)[r]}
\put(-22.23,12){\circle*{4}}
\put(-22.23,-6){\circle*{4}}
\put(-35.46,3){\circle{32}}
\put(-35.46,3){\makebox(1,-2){$\ba{c}l\\0\ea$}}
\epi
\nn\\
&&
+\frac{1}{3}\sum_{l=1}^{L-2}
\rule[-18pt]{0pt}{42pt}
\bpi(106,0)(-2,0)
\put(19,3){\circle{32}}
\put(19,3){\makebox(1,-2){$\ba{c}l\\1\ea$}}
\put(32.23,-6){\circle{4}}
\put(46,12){\circle*{4}}
\put(34.23,-6){\line(1,0){23.54}}
\put(34.2,12){\oval(23.6,8)[r]}
\put(34.2,16){\line(-1,0){5.87}}
\put(28.33,16){\circle*{4}}
\put(34.2,8){\circle*{4}}
\put(57.8,12){\oval(23.6,8)[l]}
\put(57.8,16){\line(1,0){5.87}}
\put(63.67,16){\circle*{4}}
\put(57.8,8){\circle*{4}}
\put(78,3){\oval(42,32)}
\put(78,3){\makebox(1,-2){$\ba{c}L{-}l{-}1\\1\ea$}}
\put(59.77,-6){\circle{4}}
\epi
+\frac{2}{3}\sum_{l_1=1}^{L-3}\sum_{l_2=1}^{L-l_1-2}
\rule[-40.23pt]{0pt}{64.23pt}
\bpi(134,0)(-2,0)
\put(19,3){\circle{32}}
\put(19,3){\makebox(1,-2){$\ba{c}l_1\\1\ea$}}
\put(32.23,-6){\circle{4}}
\put(65,12){\circle*{4}}
\put(34.23,-6){\line(1,0){11.77}}
\put(46,-6){\circle*{4}}
\put(34.2,12){\oval(61.6,8)[r]}
\put(34.2,16){\line(-1,0){5.87}}
\put(28.33,16){\circle*{4}}
\put(34.2,8){\circle*{4}}
\put(95.77,-6){\line(-1,0){11.77}}
\put(95.8,12){\oval(61.6,8)[l]}
\put(95.8,16){\line(1,0){5.87}}
\put(84,-6){\circle*{4}}
\put(101.67,16){\circle*{4}}
\put(95.8,8){\circle*{4}}
\put(65,-19.23){\oval(52,32)}
\put(65,-19.23){\makebox(1,-2){$\ba{c}L{-}l_1{-}l_2\\0\ea$}}
\put(111,3){\circle{32}}
\put(111,3){\makebox(1,-2){$\ba{c}l_2\\1\ea$}}
\put(97.77,-6){\circle{4}}
\epi.
\eea
Eq.\ (\ref{gaissbid2vac}) is valid for $L>2$.

Note that in (\ref{gaissbid2vac})---but not in (\ref{gaissbid1e})---the
right hand side involves graphs with more legs---namely one more---than
the left hand side.
This implies that for the generation of vacuum graphs, we have to consider
also one-point functions.
For all others it is enough to consider only diagrams with equal or
less numbers of legs.
Note further that if all lower loop orders contain only 1PI graphs
then the recursion relations also generate only 1PI graphs.
This establishes by induction that $\Ga$ generates only 1PI graphs,
as shown before in \cite{Kleinert1,Kleinert2}.

As an example, we compute $\Ga^{(3,0)}$ in appendix \ref{ga30example}.
Combining (\ref{ga00010210}), (\ref{ga20}) and the result (\ref{ga30})
of appendix \ref{ga30example}, we get the effective energy $\Ga$ at
$\Phi=0$ in the three-loop approximation,
\bea
\label{ga3loop}
-\Ga[\Phi=0]
&=&
\rule[-2pt]{0pt}{10pt}
\bpi(10,0)
\put(5,3){\circle*{4}}
\epi
+\frac{1}{2}
\rule[-10pt]{0pt}{26pt}
\bpi(26,0)
\put(13,3){\circle{16}}
\epi
+\frac{1}{8}
\rule[-10pt]{0pt}{26pt}
\bpi(46,0)
\put(23,3){\circle*{4}}
\put(15,3){\circle{16}}
\put(31,3){\circle{16}}
\epi
+\frac{1}{12}
\rule[-14pt]{0pt}{34pt}
\bpi(38,0)
\put(19,3){\circle{24}}
\put(7,3){\circle*{4}}
\put(31,3){\circle*{4}}
\put(7,3){\line(1,0){24}}
\epi
\nn\\
&&
+\frac{1}{16}
\rule[-10pt]{0pt}{26pt}
\bpi(58,0)
\put(13,3){\circle{16}}
\put(21,3){\circle*{4}}
\put(29,3){\circle{16}}
\put(37,3){\circle*{4}}
\put(45,3){\circle{16}}
\epi
+\frac{1}{48}
\rule[-12pt]{0pt}{34pt}
\bpi(38,0)
\put(19,3){\circle{24}}
\put(19,3){\oval(24,8)}
\put(7,3){\circle*{4}}
\put(31,3){\circle*{4}}
\epi
+\frac{1}{8}
\rule[-16pt]{0pt}{36pt}
\bpi(38,0)
\put(19,3){\circle{24}}
\put(7,-9){\oval(24,24)[rt]}
\put(31,-9){\oval(24,24)[lt]}
\put(19,-9){\circle*{4}}
\put(7,3){\circle*{4}}
\put(31,3){\circle*{4}}
\epi
+\frac{1}{8}
\rule[-16pt]{0pt}{38pt}
\bpi(50,0)
\put(17,3){\circle{24}}
\put(17,-9){\circle*{4}}
\put(17,-9){\line(0,1){24}}
\put(17,15){\circle*{4}}
\put(29,3){\circle*{4}}
\put(37,3){\circle{16}}
\epi
+\frac{1}{16}
\rule[-16pt]{0pt}{40pt}
\bpi(50,0)
\put(13,3){\circle{16}}
\put(37,3){\circle{16}}
\put(13,-5){\circle*{4}}
\put(13,11){\circle*{4}}
\put(37,-5){\circle*{4}}
\put(37,11){\circle*{4}}
\put(25,-3){\oval(24,16)[b]}
\put(25,11){\oval(24,16)[t]}
\epi
+\frac{1}{24}
\rule[-14pt]{0pt}{36pt}
\bpi(34.78,0)(-3.39,0)
\put(15,3){\circle{24}}
\put(15,3){\circle*{4}}
\put(15,15){\circle*{4}}
\put(4.61,-3){\circle*{4}}
\put(25.39,-3){\circle*{4}}
\put(15,3){\line(0,1){12}}
\put(15,3){\line(5,-3){10}}
\put(15,3){\line(-5,-3){10}}
\epi,
\eea
where propagator and vertices may contain a background-field dependence
as given e.g.\ by (\ref{cexample}) through (\ref{lexample}).
The corresponding effective potential $V$ in this model is then given
by $\Ga[\Phi=0,C,J,G,K,L]=\Omega V(\varphi)$, where $\Omega$ is the volume
of $d$-dimensional space.
That is, it can be computed from vacuum graphs with constant background
field $\varphi$.
Note that the right hand side of (\ref{ga3loop}) is the right hand side
of (\ref{w3loop}) with the one-particle reducible graphs omitted.

Except for the vacuum diagrams, we still have to use (\ref{galn}) to
convert the graphs representing $\Ga$ into proper vertices.
For example, to compute the one-loop contribution $\Ga^{(1,3)}_{123}$
to the $3$-point vertex, we first use (\ref{ga030411})-(\ref{gaissbid1e})
to get
\beq
\label{ga13}
\rule[-18pt]{0pt}{42pt}
\bpi(42,0)
\put(21,3){\circle{32}}
\put(21,3){\makebox(1,-2){$\ba{c}1\\3\ea$}}
\epi
=
\frac{1}{3}\left[
\rule[-18pt]{0pt}{42pt}
\bpi(68.23,0)(-28.23,0)
\put(19,3){\circle{32}}
\put(19,3){\makebox(1,0){$\ba{c}1\\2\ea$}}
\put(5.77,12){\circle*{4}}
\put(5.77,-6){\circle*{4}}
\put(5.77,3){\oval(28,18)[l]}
\put(-8.23,3){\circle*{4}}
\put(-19.23,2){\line(1,0){11}}
\put(-19.23,4){\line(1,0){11}}
\put(-21.23,3){\circle{4}}
\epi
+
\rule[-15.41pt]{0pt}{36.82pt}
\bpi(66.64,0)(-3.41,0)
\put(15,3){\circle*{4}}
\put(14.29,2.29){\line(-1,1){10}}
\put(15.71,3.71){\line(-1,1){10}}
\put(3.59,14.41){\circle{4}}
\put(15.71,2.29){\line(-1,-1){10}}
\put(14.29,3.71){\line(-1,-1){10}}
\put(3.59,-8.41){\circle{4}}
\put(42.23,3){\circle{32}}
\put(42.23,3){\makebox(1,0){$\ba{c}1\\1\ea$}}
\put(29,12){\circle*{4}}
\put(29,-6){\circle*{4}}
\put(29,3){\oval(28,18)[l]}
\epi
\right]
=
\frac{1}{4}
\rule[-15.41pt]{0pt}{36.82pt}
\bpi(54.41,0)(-3.41,0)
\put(23,3){\circle{16}}
\put(31,3){\circle*{4}}
\put(44,3){\circle{4}}
\put(31,2){\line(1,0){11}}
\put(31,4){\line(1,0){11}}
\put(15,3){\circle*{4}}
\put(14.29,2.29){\line(-1,1){10}}
\put(15.71,3.71){\line(-1,1){10}}
\put(3.59,14.41){\circle{4}}
\put(15.71,2.29){\line(-1,-1){10}}
\put(14.29,3.71){\line(-1,-1){10}}
\put(3.59,-8.41){\circle{4}}
\epi
+\frac{1}{6}
\rule[-15pt]{0pt}{47pt}
\bpi(47.1,0)(-11.05,0)
\put(15,3){\circle{16}}
\put(15,11){\circle*{4}}
\put(14,11){\line(0,1){12}}
\put(16,11){\line(0,1){12}}
\put(15,25){\circle{4}}
\put(21.93,-1){\circle*{4}}
\put(21.43,-1.87){\line(5,-3){10.39}}
\put(22.43,-0.13){\line(5,-3){10.39}}
\put(34.05,-8){\circle{4}}
\put(8.07,-1){\circle*{4}}
\put(8.57,-1.87){\line(-5,-3){10.39}}
\put(7.57,-0.13){\line(-5,-3){10.39}}
\put(-4.05,-8){\circle{4}}
\epi
\eeq
and then with (\ref{galn}) obtain
\bea
\label{ga123}
\Ga^{(1,3)}_{123}
&=&
-\frac{\de^3}{\de\Phi_1\de\Phi_2\de\Phi_3}
\left[
\frac{1}{4}
\rule[-15.41pt]{0pt}{36.82pt}
\bpi(54.41,0)(-3.41,0)
\put(23,3){\circle{16}}
\put(31,3){\circle*{4}}
\put(44,3){\circle{4}}
\put(31,2){\line(1,0){11}}
\put(31,4){\line(1,0){11}}
\put(15,3){\circle*{4}}
\put(14.29,2.29){\line(-1,1){10}}
\put(15.71,3.71){\line(-1,1){10}}
\put(3.59,14.41){\circle{4}}
\put(15.71,2.29){\line(-1,-1){10}}
\put(14.29,3.71){\line(-1,-1){10}}
\put(3.59,-8.41){\circle{4}}
\epi
+\frac{1}{6}
\rule[-15pt]{0pt}{47pt}
\bpi(47.1,0)(-11.05,0)
\put(15,3){\circle{16}}
\put(15,11){\circle*{4}}
\put(14,11){\line(0,1){12}}
\put(16,11){\line(0,1){12}}
\put(15,25){\circle{4}}
\put(21.93,-1){\circle*{4}}
\put(21.43,-1.87){\line(5,-3){10.39}}
\put(22.43,-0.13){\line(5,-3){10.39}}
\put(34.05,-8){\circle{4}}
\put(8.07,-1){\circle*{4}}
\put(8.57,-1.87){\line(-5,-3){10.39}}
\put(7.57,-0.13){\line(-5,-3){10.39}}
\put(-4.05,-8){\circle{4}}
\epi
\right]
\nn\\
&=&
-\frac{1}{2}\left[
\rule[-15.41pt]{0pt}{36.82pt}
\bpi(60.41,0)(-6.41,0)
\put(23,3){\circle{16}}
\put(31,3){\circle*{4}}
\put(47,3){\makebox(0,0){$\scs 3$}}
\put(31,2){\line(1,0){11}}
\put(31,4){\line(1,0){11}}
\put(15,3){\circle*{4}}
\put(14.29,2.29){\line(-1,1){10}}
\put(15.71,3.71){\line(-1,1){10}}
\put(0.59,14.41){\makebox(0,0){$\scs 1$}}
\put(15.71,2.29){\line(-1,-1){10}}
\put(14.29,3.71){\line(-1,-1){10}}
\put(0.59,-8.41){\makebox(0,0){$\scs 2$}}
\epi
+
\rule[-15.41pt]{0pt}{36.82pt}
\bpi(60.41,0)(-6.41,0)
\put(23,3){\circle{16}}
\put(31,3){\circle*{4}}
\put(47,3){\makebox(0,0){$\scs 2$}}
\put(31,2){\line(1,0){11}}
\put(31,4){\line(1,0){11}}
\put(15,3){\circle*{4}}
\put(14.29,2.29){\line(-1,1){10}}
\put(15.71,3.71){\line(-1,1){10}}
\put(0.59,14.41){\makebox(0,0){$\scs 3$}}
\put(15.71,2.29){\line(-1,-1){10}}
\put(14.29,3.71){\line(-1,-1){10}}
\put(0.59,-8.41){\makebox(0,0){$\scs 1$}}
\epi
+
\rule[-15.41pt]{0pt}{36.82pt}
\bpi(60.41,0)(-6.41,0)
\put(23,3){\circle{16}}
\put(31,3){\circle*{4}}
\put(47,3){\makebox(0,0){$\scs 1$}}
\put(31,2){\line(1,0){11}}
\put(31,4){\line(1,0){11}}
\put(15,3){\circle*{4}}
\put(14.29,2.29){\line(-1,1){10}}
\put(15.71,3.71){\line(-1,1){10}}
\put(0.59,14.41){\makebox(0,0){$\scs 2$}}
\put(15.71,2.29){\line(-1,-1){10}}
\put(14.29,3.71){\line(-1,-1){10}}
\put(0.59,-8.41){\makebox(0,0){$\scs 3$}}
\epi
\right]
-
\rule[-15pt]{0pt}{50pt}
\bpi(53.1,0)(-14.05,0)
\put(15,3){\circle{16}}
\put(15,11){\circle*{4}}
\put(14,11){\line(0,1){12}}
\put(16,11){\line(0,1){12}}
\put(15,28){\makebox(0,0){$\scs 1$}}
\put(21.93,-1){\circle*{4}}
\put(21.43,-1.87){\line(5,-3){10.39}}
\put(22.43,-0.13){\line(5,-3){10.39}}
\put(37.05,-8){\makebox(0,0){$\scs 3$}}
\put(8.07,-1){\circle*{4}}
\put(8.57,-1.87){\line(-5,-3){10.39}}
\put(7.57,-0.13){\line(-5,-3){10.39}}
\put(-7.05,-8){\makebox(0,0){$\scs 2$}}
\epi.
\eea
That is, each diagram with $n$ external fields $\Phi$ is multiplied by $-n!$,
supplied by external arguments replacing the $\Phi$s and then splits
into ``crossed'' graphs related by exchanging external arguments.
In contrast to the case of connected Greens functions, the external legs
carry only the external arguments and do not represent free correlation
functions $G$.

\subsection{Graphs for Renormalization}
For the purpose of perturbatively renormalizing standard $\phi^4$ theory,
we need the 1PI Feynman diagrams representing $\Ga^{(L,0)}$, $\Ga^{(L,2)}$
and $\Ga^{(L,4)}$ for the case $J=K=0$.
All $\Ga^{(L,n)}$ with odd $n$ are then identically zero.
The recursion relation for vacuum graphs with $L>2$ results from writing
(\ref{gaissbid2vac}) for $J=K=0$ and then making use of (\ref{n1n3n4ngL})
with $n_1=n_3=0$,
\beq
\label{gasymL0id}
\rule[-18pt]{0pt}{42pt}
\bpi(42,0)(-2,0)
\put(19,3){\circle{32}}
\put(19,3){\makebox(1,-2){$\ba{c}L\\0\ea$}}
\epi
=
\frac{1}{2(L-1)}\left[
\rule[-18pt]{0pt}{42pt}
\bpi(69.23,0)(-29.23,0)
\put(19,3){\circle{32}}
\put(19,3){\makebox(1,-2){$\ba{c}L{-}1\\0\ea$}}
\put(5.77,12){\circle*{4}}
\put(5.77,-6){\circle*{4}}
\put(5.77,3){\oval(28,18)[l]}
\put(-8.23,3){\circle*{4}}
\put(-16.23,3){\circle{16}}
\epi
+\frac{1}{3}
\rule[-18pt]{0pt}{42pt}
\bpi(54.78,0)(-14.78,0)
\put(19,3){\circle{32}}
\put(19,3){\makebox(1,-2){$\ba{c}L{-}1\\0\ea$}}
\put(5.22,11.4){\circle*{4}}
\put(10.87,17.1){\circle*{4}}
\put(5.22,-5.13){\circle*{4}}
\put(10.87,-10.78){\circle*{4}}
\put(10.87,3){\oval(41.3,27.88)[l]}
\put(5.22,3){\oval(30,16.53)[l]}
\put(-9.78,3){\circle*{4}}
\epi
+\frac{1}{3}\sum_{l=2}^{L-2}
\rule[-18pt]{0pt}{42pt}
\bpi(96.46,0)(-56.46,0)
\put(19,3){\circle{32}}
\put(19,3){\makebox(1,-2){$\ba{c}L{-}l\\0\ea$}}
\put(5.77,12){\circle*{4}}
\put(5.77,-6){\circle*{4}}
\put(5.77,3){\oval(28,18)[l]}
\put(-8.23,3){\circle*{4}}
\put(-22.23,3){\oval(28,18)[r]}
\put(-22.23,12){\circle*{4}}
\put(-22.23,-6){\circle*{4}}
\put(-35.46,3){\circle{32}}
\put(-35.46,3){\makebox(1,-2){$\ba{c}l\\0\ea$}}
\epi
\right].
\eeq
Notice that this is identical to (\ref{newrecrel}) for vanishing
two-point insertion.

For $n=2$ and $n=4$ we rewrite (\ref{gaissbid1e}) with $J=K=0$.
For $\Ga^{(L,2)}$ we get for $L>1$
\beq
\label{gasymL2id}
\rule[-18pt]{0pt}{42pt}
\bpi(42,0)
\put(21,3){\circle{32}}
\put(21,3){\makebox(1,-2){$\ba{c}L\\2\ea$}}
\epi
=
\frac{1}{2}
\rule[-15.41pt]{0pt}{36.82pt}
\bpi(66.64,0)(-3.41,0)
\put(15,3){\circle*{4}}
\put(14.29,2.29){\line(-1,1){10}}
\put(15.71,3.71){\line(-1,1){10}}
\put(3.59,14.41){\circle{4}}
\put(15.71,2.29){\line(-1,-1){10}}
\put(14.29,3.71){\line(-1,-1){10}}
\put(3.59,-8.41){\circle{4}}
\put(42.23,3){\circle{32}}
\put(42.23,3){\makebox(1,-2){$\ba{c}L\\0\ea$}}
\put(29,12){\circle*{4}}
\put(29,-6){\circle*{4}}
\put(29,3){\oval(28,18)[l]}
\epi
+\frac{1}{6}
\rule[-18pt]{0pt}{42pt}
\bpi(68.23,0)(-28.23,0)
\put(19,3){\circle{32}}
\put(19,3){\makebox(1,-2){$\ba{c}L{-}1\\2\ea$}}
\put(5.77,12){\circle*{4}}
\put(5.77,-6){\circle{4}}
\put(3.77,3){\oval(24,18)[l]}
\put(-8.23,3){\circle*{4}}
\put(-19.23,2){\line(1,0){11}}
\put(-19.23,4){\line(1,0){11}}
\put(-8.23,3){\line(1,0){11.23}}
\put(3,3){\circle*{4}}
\put(-21.23,3){\circle{4}}
\epi
+\frac{1}{3}\sum_{l=2}^{L-1}
\rule[-18pt]{0pt}{51pt}
\bpi(96,0)(-2,0)
\put(19,3){\circle{32}}
\put(19,3){\makebox(1,-2){$\ba{c}l\\0\ea$}}
\put(32.23,12){\circle*{4}}
\put(32.23,-6){\circle*{4}}
\put(46,12){\circle*{4}}
\put(32.23,12){\line(1,0){13.77}}
\put(32.23,-6){\line(1,0){25.54}}
\put(57.8,12){\oval(23.6,8)[l]}
\put(57.8,16){\line(1,0){5.87}}
\put(63.67,16){\circle*{4}}
\put(57.8,8){\circle*{4}}
\put(45,12){\line(0,1){12}}
\put(47,12){\line(0,1){12}}
\put(46,26){\circle{4}}
\put(73,3){\circle{32}}
\put(73,3){\makebox(1,-2){$\ba{c}L{-}l\\2\ea$}}
\put(59.77,-6){\circle{4}}
\epi
\eeq
while for $\Ga^{(L,4)}$ we get for $L>0$
\beq
\label{gasymL4id}
\rule[-18pt]{0pt}{42pt}
\bpi(42,0)
\put(21,3){\circle{32}}
\put(21,3){\makebox(1,-2){$\ba{c}L\\4\ea$}}
\epi
=
\frac{1}{4}
\rule[-15.41pt]{0pt}{36.82pt}
\bpi(66.64,0)(-3.41,0)
\put(15,3){\circle*{4}}
\put(14.29,2.29){\line(-1,1){10}}
\put(15.71,3.71){\line(-1,1){10}}
\put(3.59,14.41){\circle{4}}
\put(15.71,2.29){\line(-1,-1){10}}
\put(14.29,3.71){\line(-1,-1){10}}
\put(3.59,-8.41){\circle{4}}
\put(42.23,3){\circle{32}}
\put(42.23,3){\makebox(1,-2){$\ba{c}L\\2\ea$}}
\put(29,12){\circle*{4}}
\put(29,-6){\circle*{4}}
\put(29,3){\oval(28,18)[l]}
\epi
+\frac{1}{12}
\rule[-18pt]{0pt}{42pt}
\bpi(68.23,0)(-28.23,0)
\put(19,3){\circle{32}}
\put(19,3){\makebox(1,-2){$\ba{c}L{-}1\\4\ea$}}
\put(5.77,12){\circle*{4}}
\put(5.77,-6){\circle{4}}
\put(3.77,3){\oval(24,18)[l]}
\put(-8.23,3){\circle*{4}}
\put(-19.23,2){\line(1,0){11}}
\put(-19.23,4){\line(1,0){11}}
\put(-8.23,3){\line(1,0){11.23}}
\put(3,3){\circle*{4}}
\put(-21.23,3){\circle{4}}
\epi
+\frac{1}{6}\sum_{l=2}^{L-1}
\rule[-18pt]{0pt}{51pt}
\bpi(96,0)(-2,0)
\put(19,3){\circle{32}}
\put(19,3){\makebox(1,-2){$\ba{c}l\\0\ea$}}
\put(32.23,12){\circle*{4}}
\put(32.23,-6){\circle*{4}}
\put(46,12){\circle*{4}}
\put(32.23,12){\line(1,0){13.77}}
\put(32.23,-6){\line(1,0){25.54}}
\put(57.8,12){\oval(23.6,8)[l]}
\put(57.8,16){\line(1,0){5.87}}
\put(63.67,16){\circle*{4}}
\put(57.8,8){\circle*{4}}
\put(45,12){\line(0,1){12}}
\put(47,12){\line(0,1){12}}
\put(46,26){\circle{4}}
\put(73,3){\circle{32}}
\put(73,3){\makebox(1,-2){$\ba{c}L{-}l\\4\ea$}}
\put(59.77,-6){\circle{4}}
\epi
+\frac{1}{6}\sum_{l=1}^{L-1}
\rule[-18pt]{0pt}{51pt}
\bpi(96,0)(-2,0)
\put(19,3){\circle{32}}
\put(19,3){\makebox(1,-2){$\ba{c}l\\2\ea$}}
\put(32.23,12){\circle*{4}}
\put(32.23,-6){\circle*{4}}
\put(46,12){\circle*{4}}
\put(32.23,12){\line(1,0){13.77}}
\put(32.23,-6){\line(1,0){25.54}}
\put(57.8,12){\oval(23.6,8)[l]}
\put(57.8,16){\line(1,0){5.87}}
\put(63.67,16){\circle*{4}}
\put(57.8,8){\circle*{4}}
\put(45,12){\line(0,1){12}}
\put(47,12){\line(0,1){12}}
\put(46,26){\circle{4}}
\put(73,3){\circle{32}}
\put(73,3){\makebox(1,-2){$\ba{c}L{-}l\\2\ea$}}
\put(59.77,-6){\circle{4}}
\epi.
\eeq

Since now we have written down only the recursion relations without
starting with the identities for $\Ga_I$ again, we use for the
low-order terms not covered by (\ref{gasymL0id})-(\ref{gasymL4id})
just the results (\ref{ga00010210}), (\ref{ga030411}), (\ref{ga12}),
(\ref{ga20}) of Section \ref{gairecrel} with $J=K=0$,
\bea
\rule[-18pt]{0pt}{42pt}
\bpi(42,0)
\put(21,3){\circle{32}}
\put(21,3){\makebox(1,-2){$\ba{c}0\\0\ea$}}
\epi
=
\rule[-2pt]{0pt}{10pt}
\bpi(10,0)
\put(5,3){\circle*{4}}
\epi,
\;\;\;\;
\rule[-18pt]{0pt}{42pt}
\bpi(42,0)
\put(21,3){\circle{32}}
\put(21,3){\makebox(1,-2){$\ba{c}1\\0\ea$}}
\epi
=
\frac{1}{2}
\rule[-10pt]{0pt}{16pt}
\bpi(26,0)
\put(13,3){\circle{16}}
\epi,
\;\;\;\;
\rule[-18pt]{0pt}{42pt}
\bpi(42,0)
\put(21,3){\circle{32}}
\put(21,3){\makebox(1,-2){$\ba{c}2\\0\ea$}}
\epi
=
\frac{1}{8}
\rule[-10pt]{0pt}{26pt}
\bpi(46,0)
\put(23,3){\circle*{4}}
\put(15,3){\circle{16}}
\put(31,3){\circle{16}}
\epi,
\nn\\
\label{gasym001020021204}
\\
\rule[-18pt]{0pt}{42pt}
\bpi(42,0)
\put(21,3){\circle{32}}
\put(21,3){\makebox(1,-2){$\ba{c}0\\2\ea$}}
\epi
=\frac{1}{2}
\rule[-4pt]{0pt}{14pt}
\bpi(27,0)
\put(7,3){\circle{4}}
\put(9,2){\line(1,0){9}}
\put(9,4){\line(1,0){9}}
\put(20,3){\circle{4}}
\epi,
\;\;\;\;
\rule[-18pt]{0pt}{42pt}
\bpi(42,0)
\put(21,3){\circle{32}}
\put(21,3){\makebox(1,-2){$\ba{c}1\\2\ea$}}
\epi
=
\frac{1}{4}
\rule[-15.41pt]{0pt}{36.82pt}
\bpi(39.41,0)(-3.41,0)
\put(15,3){\circle*{4}}
\put(14.29,2.29){\line(-1,1){10}}
\put(15.71,3.71){\line(-1,1){10}}
\put(3.59,14.41){\circle{4}}
\put(15.71,2.29){\line(-1,-1){10}}
\put(14.29,3.71){\line(-1,-1){10}}
\put(3.59,-8.41){\circle{4}}
\put(23,3){\circle{16}}
\epi,
\;\;\;\;
\rule[-18pt]{0pt}{42pt}
\bpi(42,0)
\put(21,3){\circle{32}}
\put(21,3){\makebox(1,-2){$\ba{c}0\\4\ea$}}
\epi
=
\frac{1}{24}
\rule[-15.41pt]{0pt}{36.82pt}
\bpi(36.82,0)(-3.41,0)
\put(15,3){\circle*{4}}
\put(14.29,2.29){\line(-1,1){10}}
\put(15.71,3.71){\line(-1,1){10}}
\put(3.59,14.41){\circle{4}}
\put(15.71,2.29){\line(-1,-1){10}}
\put(14.29,3.71){\line(-1,-1){10}}
\put(3.59,-8.41){\circle{4}}
\put(14.29,3.71){\line(1,1){10}}
\put(15.71,2.29){\line(1,1){10}}
\put(26.41,14.41){\circle{4}}
\put(15.71,3.71){\line(1,-1){10}}
\put(14.29,2.29){\line(1,-1){10}}
\put(26.41,-8.41){\circle{4}}
\epi.
\nn
\eea

It is now easy to use (\ref{gasymL0id})-(\ref{gasymL4id}) to obtain e.g.\
(compare to the 1PI graphs in the tables in \cite{KPKB}; for the vacuum
graphs, compare also with Table \ref{allgraphs} in this work)
\beq
\label{gasym30}
\rule[-18pt]{0pt}{42pt}
\bpi(42,0)(-2,0)
\put(19,3){\circle{32}}
\put(19,3){\makebox(1,-2){$\ba{c}3\\0\ea$}}
\epi
=
\frac{1}{4}
\rule[-18pt]{0pt}{42pt}
\bpi(69.23,0)(-29.23,0)
\put(19,3){\circle{32}}
\put(19,3){\makebox(1,-2){$\ba{c}2\\0\ea$}}
\put(5.77,12){\circle*{4}}
\put(5.77,-6){\circle*{4}}
\put(5.77,3){\oval(28,18)[l]}
\put(-8.23,3){\circle*{4}}
\put(-16.23,3){\circle{16}}
\epi
+\frac{1}{12}
\rule[-18pt]{0pt}{42pt}
\bpi(54.78,0)(-14.78,0)
\put(19,3){\circle{32}}
\put(19,3){\makebox(1,-2){$\ba{c}2\\0\ea$}}
\put(5.22,11.4){\circle*{4}}
\put(10.87,17.1){\circle*{4}}
\put(5.22,-5.13){\circle*{4}}
\put(10.87,-10.78){\circle*{4}}
\put(10.87,3){\oval(41.3,27.88)[l]}
\put(5.22,3){\oval(30,16.53)[l]}
\put(-9.78,3){\circle*{4}}
\epi
=
\frac{1}{16}
\rule[-10pt]{0pt}{26pt}
\bpi(58,0)
\put(13,3){\circle{16}}
\put(21,3){\circle*{4}}
\put(29,3){\circle{16}}
\put(37,3){\circle*{4}}
\put(45,3){\circle{16}}
\epi
+\frac{1}{48}
\rule[-12pt]{0pt}{34pt}
\bpi(38,0)
\put(19,3){\circle{24}}
\put(19,3){\oval(24,8)}
\put(7,3){\circle*{4}}
\put(31,3){\circle*{4}}
\epi,
\eeq

\bea
\label{gasym40}
\rule[-18pt]{0pt}{42pt}
\bpi(42,0)(-2,0)
\put(19,3){\circle{32}}
\put(19,3){\makebox(1,-2){$\ba{c}4\\0\ea$}}
\epi
&=&
\frac{1}{6}
\rule[-18pt]{0pt}{42pt}
\bpi(69.23,0)(-29.23,0)
\put(19,3){\circle{32}}
\put(19,3){\makebox(1,-2){$\ba{c}3\\0\ea$}}
\put(5.77,12){\circle*{4}}
\put(5.77,-6){\circle*{4}}
\put(5.77,3){\oval(28,18)[l]}
\put(-8.23,3){\circle*{4}}
\put(-16.23,3){\circle{16}}
\epi
+\frac{1}{18}
\rule[-18pt]{0pt}{42pt}
\bpi(54.78,0)(-14.78,0)
\put(19,3){\circle{32}}
\put(19,3){\makebox(1,-2){$\ba{c}3\\0\ea$}}
\put(5.22,11.4){\circle*{4}}
\put(10.87,17.1){\circle*{4}}
\put(5.22,-5.13){\circle*{4}}
\put(10.87,-10.78){\circle*{4}}
\put(10.87,3){\oval(41.3,27.88)[l]}
\put(5.22,3){\oval(30,16.53)[l]}
\put(-9.78,3){\circle*{4}}
\epi
+\frac{1}{18}
\rule[-18pt]{0pt}{42pt}
\bpi(96.46,0)(-56.46,0)
\put(19,3){\circle{32}}
\put(19,3){\makebox(1,-2){$\ba{c}2\\0\ea$}}
\put(5.77,12){\circle*{4}}
\put(5.77,-6){\circle*{4}}
\put(5.77,3){\oval(28,18)[l]}
\put(-8.23,3){\circle*{4}}
\put(-22.23,3){\oval(28,18)[r]}
\put(-22.23,12){\circle*{4}}
\put(-22.23,-6){\circle*{4}}
\put(-35.46,3){\makebox(1,-2){$\ba{c}2\\0\ea$}}
\put(-35.46,3){\circle{32}}
\epi
\nn\\
&=&
\frac{1}{48}
\rule[-14pt]{0pt}{34pt}
\bpi(34,12)
\put(17,3){\circle{24}}
\put(6.6,9){\line(1,0){20.8}}
\put(6.6,9){\line(3,-5){10.4}}
\put(27.4,9){\line(-3,-5){10.4}}
\put(6.6,9){\circle*{4}}
\put(27.4,9){\circle*{4}}
\put(17,-9){\circle*{4}}
\epi
+\frac{1}{24}
\rule[-14pt]{0pt}{50pt}
\bpi(34,12)
\put(17,3){\circle{24}}
\put(17,3){\oval(24,8)}
\put(17,23){\circle{16}}
\put(5,3){\circle*{4}}
\put(29,3){\circle*{4}}
\put(17,15){\circle*{4}}
\epi
+\frac{1}{32}
\rule[-10pt]{0pt}{26pt}
\bpi(74,12)
\put(13,3){\circle{16}}
\put(29,3){\circle{16}}
\put(45,3){\circle{16}}
\put(61,3){\circle{16}}
\put(21,3){\circle*{4}}
\put(37,3){\circle*{4}}
\put(53,3){\circle*{4}}
\epi
+\frac{1}{48}
\rule[-18pt]{0pt}{50pt}
\bpi(53.7,12)
\put(26.85,3){\circle{16}}
\put(26.85,19){\circle{16}}
\put(13,-5){\circle{16}}
\put(40.7,-5){\circle{16}}
\put(26.85,11){\circle*{4}}
\put(19.95,-1){\circle*{4}}
\put(33.75,-1){\circle*{4}}
\epi,
\eea

\beq
\label{gasym22}
\rule[-18pt]{0pt}{42pt}
\bpi(42,0)
\put(21,3){\circle{32}}
\put(21,3){\makebox(1,-2){$\ba{c}2\\2\ea$}}
\epi
=
\frac{1}{2}
\rule[-15.41pt]{0pt}{36.82pt}
\bpi(66.64,0)(-3.41,0)
\put(15,3){\circle*{4}}
\put(14.29,2.29){\line(-1,1){10}}
\put(15.71,3.71){\line(-1,1){10}}
\put(3.59,14.41){\circle{4}}
\put(15.71,2.29){\line(-1,-1){10}}
\put(14.29,3.71){\line(-1,-1){10}}
\put(3.59,-8.41){\circle{4}}
\put(42.23,3){\circle{32}}
\put(42.23,3){\makebox(1,0){$\ba{c}2\\0\ea$}}
\put(29,12){\circle*{4}}
\put(29,-6){\circle*{4}}
\put(29,3){\oval(28,18)[l]}
\epi
+\frac{1}{6}
\rule[-18pt]{0pt}{42pt}
\bpi(68.23,0)(-28.23,0)
\put(19,3){\circle{32}}
\put(19,3){\makebox(1,0){$\ba{c}1\\2\ea$}}
\put(5.77,12){\circle*{4}}
\put(5.77,-6){\circle{4}}
\put(3.77,3){\oval(24,18)[l]}
\put(-8.23,3){\circle*{4}}
\put(-19.23,2){\line(1,0){11}}
\put(-19.23,4){\line(1,0){11}}
\put(-8.23,3){\line(1,0){11.23}}
\put(3,3){\circle*{4}}
\put(-21.23,3){\circle{4}}
\epi
=
\frac{1}{8}
\rule[-14pt]{0pt}{34pt}
\bpi(55.41,0)(-3.41,0)
\put(15,3){\circle*{4}}
\put(14.29,2.29){\line(-1,1){10}}
\put(15.71,3.71){\line(-1,1){10}}
\put(3.59,14.41){\circle{4}}
\put(15.71,2.29){\line(-1,-1){10}}
\put(14.29,3.71){\line(-1,-1){10}}
\put(3.59,-8.41){\circle{4}}
\put(23,3){\circle{16}}
\put(31,3){\circle*{4}}
\put(39,3){\circle{16}}
\epi
+\frac{1}{12}
\rule[-14pt]{0pt}{34pt}
\bpi(64,0)
\put(9,2){\line(1,0){11}}
\put(9,4){\line(1,0){11}}
\put(20,3){\line(1,0){24}}
\put(44,2){\line(1,0){11}}
\put(44,4){\line(1,0){11}}
\put(7,3){\circle{4}}
\put(20,3){\circle*{4}}
\put(32,3){\circle{24}}
\put(44,3){\circle*{4}}
\put(57,3){\circle{4}}
\epi,
\eeq

\bea
\label{gasym32}
\rule[-18pt]{0pt}{42pt}
\bpi(42,0)(-2,0)
\put(19,3){\circle{32}}
\put(19,3){\makebox(1,-2){$\ba{c}3\\2\ea$}}
\epi
&=&
\frac{1}{2}
\rule[-15.41pt]{0pt}{36.82pt}
\bpi(66.64,0)(-3.41,0)
\put(15,3){\circle*{4}}
\put(14.29,2.29){\line(-1,1){10}}
\put(15.71,3.71){\line(-1,1){10}}
\put(3.59,14.41){\circle{4}}
\put(15.71,2.29){\line(-1,-1){10}}
\put(14.29,3.71){\line(-1,-1){10}}
\put(3.59,-8.41){\circle{4}}
\put(42.23,3){\circle{32}}
\put(42.23,3){\makebox(1,-2){$\ba{c}3\\0\ea$}}
\put(29,12){\circle*{4}}
\put(29,-6){\circle*{4}}
\put(29,3){\oval(28,18)[l]}
\epi
+\frac{1}{6}
\rule[-18pt]{0pt}{42pt}
\bpi(68.23,0)(-28.23,0)
\put(19,3){\circle{32}}
\put(19,3){\makebox(1,-2){$\ba{c}2\\2\ea$}}
\put(5.77,12){\circle*{4}}
\put(5.77,-6){\circle{4}}
\put(3.77,3){\oval(24,18)[l]}
\put(-8.23,3){\circle*{4}}
\put(-19.23,2){\line(1,0){11}}
\put(-19.23,4){\line(1,0){11}}
\put(-8.23,3){\line(1,0){11.23}}
\put(3,3){\circle*{4}}
\put(-21.23,3){\circle{4}}
\epi
+\frac{1}{3}
\rule[-18pt]{0pt}{51pt}
\bpi(96,0)(-2,0)
\put(19,3){\circle{32}}
\put(19,3){\makebox(1,-2){$\ba{c}2\\0\ea$}}
\put(32.23,12){\circle*{4}}
\put(32.23,-6){\circle*{4}}
\put(46,12){\circle*{4}}
\put(32.23,12){\line(1,0){13.77}}
\put(32.23,-6){\line(1,0){25.54}}
\put(57.8,12){\oval(23.6,8)[l]}
\put(57.8,16){\line(1,0){5.87}}
\put(63.67,16){\circle*{4}}
\put(57.8,8){\circle*{4}}
\put(45,12){\line(0,1){12}}
\put(47,12){\line(0,1){12}}
\put(46,26){\circle{4}}
\put(73,3){\circle{32}}
\put(73,3){\makebox(1,-2){$\ba{c}1\\2\ea$}}
\put(59.77,-6){\circle{4}}
\epi
\nn\\
&=&
\frac{1}{16}
\rule[-15.41pt]{0pt}{36.82pt}
\bpi(71.41,0)(-3.41,0)
\put(15,3){\circle*{4}}
\put(14.29,2.29){\line(-1,1){10}}
\put(15.71,3.71){\line(-1,1){10}}
\put(3.59,14.41){\circle{4}}
\put(15.71,2.29){\line(-1,-1){10}}
\put(14.29,3.71){\line(-1,-1){10}}
\put(3.59,-8.41){\circle{4}}
\put(23,3){\circle{16}}
\put(31,3){\circle*{4}}
\put(39,3){\circle{16}}
\put(47,3){\circle*{4}}
\put(55,3){\circle{16}}
\epi
+\frac{1}{16}
\rule[-23.41pt]{0pt}{39.41pt}
\bpi(58,0)
\put(13,3){\circle{16}}
\put(21,3){\circle*{4}}
\put(29,3){\circle{16}}
\put(37,3){\circle*{4}}
\put(45,3){\circle{16}}
\put(29,-5){\circle*{4}}
\put(29.71,-5.71){\line(-1,-1){10}}
\put(28.29,-4.29){\line(-1,-1){10}}
\put(17.59,-16.41){\circle{4}}
\put(28.29,-5.71){\line(1,-1){10}}
\put(29.71,-4.29){\line(1,-1){10}}
\put(40.41,-16.41){\circle{4}}
\epi
+\frac{1}{24}
\rule[-16pt]{0pt}{22pt}
\bpi(48.41,0)(-3.41,0)
\put(15,3){\circle*{4}}
\put(14.29,2.29){\line(-1,1){10}}
\put(15.71,3.71){\line(-1,1){10}}
\put(3.59,14.41){\circle{4}}
\put(15.71,2.29){\line(-1,-1){10}}
\put(14.29,3.71){\line(-1,-1){10}}
\put(3.59,-8.41){\circle{4}}
\put(27,3){\circle{24}}
\put(27,3){\oval(8,24)}
\put(27,-9){\circle*{4}}
\put(27,15){\circle*{4}}
\epi
+\frac{1}{8}
\rule[-14pt]{0pt}{50pt}
\bpi(64,0)
\put(9,2){\line(1,0){11}}
\put(9,4){\line(1,0){11}}
\put(7,3){\circle{4}}
\put(20,3){\line(1,0){24}}
\put(20,3){\circle*{4}}
\put(32,3){\circle{24}}
\put(32,15){\circle*{4}}
\put(32,23){\circle{16}}
\put(44,3){\circle*{4}}
\put(44,2){\line(1,0){11}}
\put(44,4){\line(1,0){11}}
\put(57,3){\circle{4}}
\epi
+\frac{1}{8}
\rule[-16pt]{0pt}{36pt}
\bpi(64,0)
\put(7,3){\circle{4}}
\put(9,2){\line(1,0){11}}
\put(9,4){\line(1,0){11}}
\put(20,3){\circle*{4}}
\put(32,3){\circle{24}}
\put(20,-9){\oval(24,24)[tr]}
\put(44,-9){\oval(24,24)[tl]}
\put(32,-9){\circle*{4}}
\put(44,3){\circle*{4}}
\put(44,2){\line(1,0){11}}
\put(44,4){\line(1,0){11}}
\put(57,3){\circle{4}}
\epi,
\;\;\;\;
\eea

\beq
\label{gasym14}
\rule[-18pt]{0pt}{42pt}
\bpi(42,0)
\put(21,3){\circle{32}}
\put(21,3){\makebox(1,-2){$\ba{c}1\\4\ea$}}
\epi
=
\frac{1}{4}
\rule[-15.41pt]{0pt}{36.82pt}
\bpi(66.64,0)(-3.41,0)
\put(15,3){\circle*{4}}
\put(14.29,2.29){\line(-1,1){10}}
\put(15.71,3.71){\line(-1,1){10}}
\put(3.59,14.41){\circle{4}}
\put(15.71,2.29){\line(-1,-1){10}}
\put(14.29,3.71){\line(-1,-1){10}}
\put(3.59,-8.41){\circle{4}}
\put(42.23,3){\circle{32}}
\put(42.23,3){\makebox(1,-2){$\ba{c}1\\2\ea$}}
\put(29,12){\circle*{4}}
\put(29,-6){\circle*{4}}
\put(29,3){\oval(28,18)[l]}
\epi
+\frac{1}{12}
\rule[-18pt]{0pt}{42pt}
\bpi(68.23,0)(-28.23,0)
\put(19,3){\circle{32}}
\put(19,3){\makebox(1,-2){$\ba{c}0\\4\ea$}}
\put(5.77,12){\circle*{4}}
\put(5.77,-6){\circle{4}}
\put(3.77,3){\oval(24,18)[l]}
\put(-8.23,3){\circle*{4}}
\put(-19.23,2){\line(1,0){11}}
\put(-19.23,4){\line(1,0){11}}
\put(-8.23,3){\line(1,0){11.23}}
\put(3,3){\circle*{4}}
\put(-21.23,3){\circle{4}}
\epi
=
\frac{1}{16}
\rule[-15.41pt]{0pt}{36.82pt}
\bpi(52.82,0)(-3.41,0)
\put(23,3){\circle{16}}
\put(15,3){\circle*{4}}
\put(31,3){\circle*{4}}
\put(14.29,2.29){\line(-1,1){10}}
\put(15.71,3.71){\line(-1,1){10}}
\put(3.59,14.41){\circle{4}}
\put(15.71,2.29){\line(-1,-1){10}}
\put(14.29,3.71){\line(-1,-1){10}}
\put(3.59,-8.41){\circle{4}}
\put(31.71,2.29){\line(1,1){10}}
\put(30.29,3.71){\line(1,1){10}}
\put(42.41,14.41){\circle{4}}
\put(30.29,2.29){\line(1,-1){10}}
\put(31.71,3.71){\line(1,-1){10}}
\put(42.41,-8.41){\circle{4}}
\epi,
\eeq

\bea
\label{gasym24}
\rule[-18pt]{0pt}{42pt}
\bpi(42,0)
\put(21,3){\circle{32}}
\put(21,3){\makebox(1,-2){$\ba{c}2\\4\ea$}}
\epi
&=&
\frac{1}{4}
\rule[-15.41pt]{0pt}{36.82pt}
\bpi(66.64,0)(-3.41,0)
\put(15,3){\circle*{4}}
\put(14.29,2.29){\line(-1,1){10}}
\put(15.71,3.71){\line(-1,1){10}}
\put(3.59,14.41){\circle{4}}
\put(15.71,2.29){\line(-1,-1){10}}
\put(14.29,3.71){\line(-1,-1){10}}
\put(3.59,-8.41){\circle{4}}
\put(42.23,3){\circle{32}}
\put(42.23,3){\makebox(1,-2){$\ba{c}2\\2\ea$}}
\put(29,12){\circle*{4}}
\put(29,-6){\circle*{4}}
\put(29,3){\oval(28,18)[l]}
\epi
+\frac{1}{12}
\rule[-18pt]{0pt}{42pt}
\bpi(68.23,0)(-28.23,0)
\put(19,3){\circle{32}}
\put(19,3){\makebox(1,-2){$\ba{c}1\\4\ea$}}
\put(5.77,12){\circle*{4}}
\put(5.77,-6){\circle{4}}
\put(3.77,3){\oval(24,18)[l]}
\put(-8.23,3){\circle*{4}}
\put(-19.23,2){\line(1,0){11}}
\put(-19.23,4){\line(1,0){11}}
\put(-8.23,3){\line(1,0){11.23}}
\put(3,3){\circle*{4}}
\put(-21.23,3){\circle{4}}
\epi
+\frac{1}{6}
\rule[-18pt]{0pt}{51pt}
\bpi(96,0)(-2,0)
\put(19,3){\circle{32}}
\put(19,3){\makebox(1,-2){$\ba{c}1\\2\ea$}}
\put(32.23,12){\circle*{4}}
\put(32.23,-6){\circle*{4}}
\put(46,12){\circle*{4}}
\put(32.23,12){\line(1,0){13.77}}
\put(32.23,-6){\line(1,0){25.54}}
\put(57.8,12){\oval(23.6,8)[l]}
\put(57.8,16){\line(1,0){5.87}}
\put(63.67,16){\circle*{4}}
\put(57.8,8){\circle*{4}}
\put(45,12){\line(0,1){12}}
\put(47,12){\line(0,1){12}}
\put(46,26){\circle{4}}
\put(73,3){\circle{32}}
\put(73,3){\makebox(1,-2){$\ba{c}1\\2\ea$}}
\put(59.77,-6){\circle{4}}
\epi
\nn\\
&=&
\frac{1}{32}
\rule[-15.41pt]{0pt}{36.82pt}
\bpi(68.82,0)(-3.41,0)
\put(15,3){\circle*{4}}
\put(23,3){\circle{16}}
\put(31,3){\circle*{4}}
\put(39,3){\circle{16}}
\put(47,3){\circle*{4}}
\put(14.29,2.29){\line(-1,1){10}}
\put(15.71,3.71){\line(-1,1){10}}
\put(3.59,14.41){\circle{4}}
\put(15.71,2.29){\line(-1,-1){10}}
\put(14.29,3.71){\line(-1,-1){10}}
\put(3.59,-8.41){\circle{4}}
\put(47.71,2.29){\line(1,1){10}}
\put(46.29,3.71){\line(1,1){10}}
\put(58.41,14.41){\circle{4}}
\put(46.29,2.29){\line(1,-1){10}}
\put(47.71,3.71){\line(1,-1){10}}
\put(58.41,-8.41){\circle{4}}
\epi
+\frac{1}{16}
\rule[-15.41pt]{0pt}{47.41pt}
\bpi(52.82,0)(-3.41,0)
\put(15,3){\circle*{4}}
\put(23,3){\circle{16}}
\put(23,11){\circle*{4}}
\put(23,19){\circle{16}}
\put(31,3){\circle*{4}}
\put(14.29,2.29){\line(-1,1){10}}
\put(15.71,3.71){\line(-1,1){10}}
\put(3.59,14.41){\circle{4}}
\put(15.71,2.29){\line(-1,-1){10}}
\put(14.29,3.71){\line(-1,-1){10}}
\put(3.59,-8.41){\circle{4}}
\put(31.71,2.29){\line(1,1){10}}
\put(30.29,3.71){\line(1,1){10}}
\put(42.41,14.41){\circle{4}}
\put(30.29,2.29){\line(1,-1){10}}
\put(31.71,3.71){\line(1,-1){10}}
\put(42.41,-8.41){\circle{4}}
\epi
+\frac{1}{8}
\rule[-14pt]{0pt}{47.41pt}
\bpi(64,0)
\put(31.29,14.29){\line(-1,1){10}}
\put(32.71,15.71){\line(-1,1){10}}
\put(20.59,26.41){\circle{4}}
\put(31.29,15.71){\line(1,1){10}}
\put(32.71,14.29){\line(1,1){10}}
\put(43.41,26.41){\circle{4}}
\put(9,2){\line(1,0){11}}
\put(9,4){\line(1,0){11}}
\put(7,3){\circle{4}}
\put(20,3){\line(1,0){24}}
\put(20,3){\circle*{4}}
\put(32,3){\circle{24}}
\put(32,15){\circle*{4}}
\put(44,3){\circle*{4}}
\put(44,2){\line(1,0){11}}
\put(44,4){\line(1,0){11}}
\put(57,3){\circle{4}}
\epi.
\eea
In this way, all the graphs needed for the renormalization of
$\phi^4$ theory can be obtained (for a five-loop treatment see
\cite{5loop,Buch}).
There is no need to go beyond $L$ loop order to determine all 1PI
zero-, two- and four-point graphs through $L$ loops.
I have written a computer code implementing the recursion relations
for the 1PI graphs.
If we restrict ourselves to the symmetric case, it reproduces the 1PI
graphs and their multiplicities (trivially related to the weights,
see \cite{KPKB}) of Tables I through III in \cite{KPKB} and also all
relevant entries in Tables V through VII there.

\subsection{Absorption of Tadpoles}
\label{multiloop}
Here we discuss the absorption of tadpoles, i.e.\ $\Phi$-independent
subdiagrams of the form
\beq
\label{snails}
\rule[-19pt]{0pt}{53pt}
\bpi(42,0)(-2,-10)
\put(19,3){\circle{32}}
\put(3,3){\line(1,1){16}}
\put(4.42,-3.58){\line(1,1){21.17}}
\put(7.68,-8.31){\line(1,1){22.63}}
\put(12.42,-11.58){\line(1,1){21.17}}
\put(19,-13){\line(1,1){16}}
\put(19,-10.23){\oval(18,28)[b]}
\put(19,-24.23){\circle*{4}}
\put(4,-24.23){\line(1,0){30}}
\epi
\eeq
into the propagator for diagrams representing the proper vertices
$\Ga^{(L,n)}$ with $n>0$ in the theory.
For standard $\phi^4$ theory, this amounts to an absorption of
momentum-independent propagator corrections into the mass.
This drastically reduces the amount of remaining diagrams and
therefore simplifies the bookkeeping for higher-loop calculations
\cite{KuMu}.

Let us first indicate the changes to be introduced into the treatment
of the asymmetric case to arrive at recursion relations for the
$\Ga^{(L,n)}$  in the presence of a two-point insertion as defined in
Section \ref{sym}.
Since (\ref{wssbidentity1}) and (\ref{wssbidentity2}) receive the
additional terms
\beq
\int_2\De_{12}\frac{\de W}{\de J_2}
\eeq
and
\beq
2\int_3\De_{13}\frac{\de W}{\de G_{23}^{-1}}
\eeq
on their respective right hand sides, the changes on the right hand sides
of (\ref{gassbidentity1a}) and (\ref{gassbidentity2a}) is the addition of
the terms
\beq
\label{twopointdelta}
-\int_2\De_{12}\Phi_2
\eeq
and
\beq
-2\int_3\De_{13}\frac{\de\Ga}{\de G_{23}^{-1}}
=
-\int_3\De_{13}\left[\left(G_{23}+\Phi_2\Phi_3\right)
+2\frac{\de\Ga_I}{\de G_{23}^{-1}}\right],
\eeq
respectively, where we have used (\ref{dga0ssbdginv12}).
This leads to the addition of
\beq
-\int_{12}\De_{12}\Phi_1\Phi_2
\eeq
and
\beq
-\int_{12}\De_{12}\Phi_1\Phi_2-\int_{12}G_{12}\De_{12}
+2\int_{12}\De_{12}G_{13}G_{24}\frac{\de\Ga_I}{\de G_{34}}
\eeq
to the right hand sides of (\ref{gaissbid1c}) and (\ref{gaissbid2c}),
respectively.
Then, in a notation which by now should be obvious, the right hand sides of
(\ref{gaissbid1d}) and (\ref{gaissbid2d}) receive the addition of
\beq
\rule[-7pt]{0pt}{20pt}
\bpi(50,0)
\put(7,3){\circle{4}}
\put(9,2){\line(1,0){11}}
\put(9,4){\line(1,0){11}}
\put(30,2){\line(1,0){11}}
\put(30,4){\line(1,0){11}}
\put(43,3){\circle{4}}
\put(20,-2){\line(1,0){10}}
\put(20,-2){\line(0,1){10}}
\put(20,8){\line(1,0){10}}
\put(30,-2){\line(0,1){10}}
\put(25,3){\makebox(1,0){$\scs\De$}}
\epi
\eeq
and
\beq
\label{cvgaissbid2d}
\rule[-7pt]{0pt}{20pt}
\bpi(50,0)
\put(7,3){\circle{4}}
\put(9,2){\line(1,0){11}}
\put(9,4){\line(1,0){11}}
\put(30,2){\line(1,0){11}}
\put(30,4){\line(1,0){11}}
\put(43,3){\circle{4}}
\put(20,-2){\line(1,0){10}}
\put(20,-2){\line(0,1){10}}
\put(20,8){\line(1,0){10}}
\put(30,-2){\line(0,1){10}}
\put(25,3){\makebox(1,0){$\scs\De$}}
\epi
+
\rule[-11pt]{0pt}{28pt}
\bpi(33,0)
\put(19,8){\oval(18,8)[t]}
\put(19,-2){\oval(18,8)[b]}
\put(28,-2){\line(0,1){10}}
\put(5,-2){\line(1,0){10}}
\put(5,-2){\line(0,1){10}}
\put(5,8){\line(1,0){10}}
\put(15,-2){\line(0,1){10}}
\put(10,3){\makebox(1,0){$\scs\De$}}
\epi
+2
\rule[-18pt]{0pt}{42pt}
\bpi(58.23,0)(-18.23,0)
\put(19,3){\circle{32}}
\put(19,3){\makebox(2,0){$\Ga_I$}}
\put(5.77,12){\circle*{4}}
\put(5.77,-6){\circle*{4}}
\put(5.77,8){\oval(28,8)[tl]}
\put(5.77,-2){\oval(28,8)[bl]}
\put(-13.23,-2){\line(1,0){10}}
\put(-13.23,-2){\line(0,1){10}}
\put(-13.23,8){\line(1,0){10}}
\put(-3.23,-2){\line(0,1){10}}
\put(-8.23,3){\makebox(1,0){$\scs\De$}}
\epi,
\eeq
respectively. 
Note that for $K=\Phi=0$ and $\Ga_I\rightarrow-W_I$, the second resulting
equation is identical to (\ref{wisym}).

Eqs.\ (\ref{gadoubleex}) through (\ref{ga030411}) remain unchanged, while
the right hand side of (\ref{ga12}) receives the addition of
\beq
\label{cvga12}
\frac{1}{2}
\rule[-7pt]{0pt}{20pt}
\bpi(50,0)
\put(7,3){\circle{4}}
\put(9,2){\line(1,0){11}}
\put(9,4){\line(1,0){11}}
\put(30,2){\line(1,0){11}}
\put(30,4){\line(1,0){11}}
\put(43,3){\circle{4}}
\put(20,-2){\line(1,0){10}}
\put(20,-2){\line(0,1){10}}
\put(20,8){\line(1,0){10}}
\put(30,-2){\line(0,1){10}}
\put(25,3){\makebox(1,0){$\scs 1$}}
\epi.
\eeq
For $n\neq2$, (\ref{gaissbid1e}) remains unchanged.
For $n=2$ with $L>1$, the right hand side of (\ref{gaissbid1e}) receives
the addition of
\beq
\rule[-7pt]{0pt}{20pt}
\bpi(50,0)
\put(7,3){\circle{4}}
\put(9,2){\line(1,0){11}}
\put(9,4){\line(1,0){11}}
\put(30,2){\line(1,0){11}}
\put(30,4){\line(1,0){11}}
\put(43,3){\circle{4}}
\put(20,-2){\line(1,0){10}}
\put(20,-2){\line(0,1){10}}
\put(20,8){\line(1,0){10}}
\put(30,-2){\line(0,1){10}}
\put(25,3){\makebox(1,0){$\scs L$}}
\epi.
\eeq
Finally, the right hand sides of (\ref{ga20}) and (\ref{gaissbid2vac})
receive the additions of
\beq
\frac{1}{2}
\rule[-11pt]{0pt}{28pt}
\bpi(33,0)
\put(19,8){\oval(18,8)[t]}
\put(19,-2){\oval(18,8)[b]}
\put(28,-2){\line(0,1){10}}
\put(5,-2){\line(1,0){10}}
\put(5,-2){\line(0,1){10}}
\put(5,8){\line(1,0){10}}
\put(15,-2){\line(0,1){10}}
\put(10,3){\makebox(1,0){$\scs 1$}}
\epi
\eeq
and
\beq
\frac{1}{2}
\rule[-11pt]{0pt}{28pt}
\bpi(36,0)(-3,0)
\put(19,8){\oval(18,8)[t]}
\put(19,-2){\oval(18,8)[b]}
\put(28,-2){\line(0,1){10}}
\put(0,-2){\line(1,0){18}}
\put(0,-2){\line(0,1){10}}
\put(0,8){\line(1,0){18}}
\put(18,-2){\line(0,1){10}}
\put(9,3){\makebox(0,0){$\scs L-1$}}
\epi
+\sum_{l=1}^{L-2}
\rule[-18pt]{0pt}{42pt}
\bpi(61.23,0)(-21.23,0)
\put(19,3){\circle{32}}
\put(19,3){\makebox(1,-2){$\ba{c}L{-}l\\0\ea$}}
\put(5.77,12){\circle*{4}}
\put(5.77,-6){\circle*{4}}
\put(5.77,8){\oval(32,8)[tl]}
\put(5.77,-2){\oval(32,8)[bl]}
\put(-16.23,-2){\line(1,0){10}}
\put(-16.23,-2){\line(0,1){10}}
\put(-16.23,8){\line(1,0){10}}
\put(-6.23,-2){\line(0,1){10}}
\put(-11.23,3){\makebox(0,0){$\scs l$}}
\epi,
\eeq
respectively.

It is not hard to see then that with and only with the choices
\beq
\label{tp1}
\bpi(46,0)(2,0)
\put(7,3){\line(1,0){13}}
\put(30,3){\line(1,0){13}}
\put(20,-2){\line(1,0){10}}
\put(20,-2){\line(0,1){10}}
\put(20,8){\line(1,0){10}}
\put(30,-2){\line(0,1){10}}
\put(25,3){\makebox(1,0){$\scs 1$}}
\epi
=
-\frac{1}{2}
\rule[-12pt]{0pt}{36pt}
\bpi(30,0)
\put(15,3){\circle*{4}}
\put(15,11){\circle{16}}
\put(15,3){\line(-1,-1){10}}
\put(15,3){\line(1,-1){10}}
\epi
\eeq
at the one-loop level and
\beq
\label{tpl}
\bpi(46,0)(2,0)
\put(7,3){\line(1,0){13}}
\put(30,3){\line(1,0){13}}
\put(20,-2){\line(1,0){10}}
\put(20,-2){\line(0,1){10}}
\put(20,8){\line(1,0){10}}
\put(30,-2){\line(0,1){10}}
\put(25,3){\makebox(1,0){$\scs L$}}
\epi
=-
\rule[-29pt]{0pt}{63pt}
\bpi(42,0)(-2,-10)
\put(19,3){\circle{32}}
\put(19,3){\makebox(1,-2){$\ba{c}L\\0\ea$}}
\put(10,-10.23){\circle*{4}}
\put(28,-10.23){\circle*{4}}
\put(19,-10.23){\oval(18,28)[b]}
\put(19,-24.23){\circle*{4}}
\put(19,-24.23){\line(-1,-1){10}}
\put(19,-24.23){\line(1,-1){10}}
\epi
\eeq
for $L>1$, all tadpole corrections to propagators in 1PI $n$-point functions
with $n>0$ will be canceled.
This drastically reduces the number of diagrams to be considered at higher
loop orders.
Notice that with (\ref{p12b}), the insertions (\ref{tp1}) and (\ref{tpl})
can be summarized by writing
\beq
\label{propinloop}
\De_{12}=\left.-\frac{1}{2}\int_{34}L_{1234}P_{34}\right|_{\Phi=0},
\eeq
i.e.\ by inserting the full propagator into a one-loop tadpole [compare to
(\ref{p12}) and the comment following it].
It turns out that this cancellation is also true for
\beq
\rule[-18pt]{0pt}{42pt}
\bpi(53.23,0)(-13.23,0)
\put(19,3){\circle{32}}
\put(19,3){\makebox(1,-2){$\ba{c}L\\0\ea$}}
\put(5.77,12){\circle*{4}}
\put(5.77,-6){\circle*{4}}
\put(5.77,3){\oval(28,18)[l]}
\epi,
\eeq
but not for
\beq
\rule[-18pt]{0pt}{42pt}
\bpi(42,0)
\put(21,3){\circle{32}}
\put(21,3){\makebox(1,-2){$\ba{c}L\\0\ea$}}
\epi,
\eeq
i.e.\ not for the vacuum graphs with their proper weights.
A simple diagrammatical explanation for this failure is that the
combinatorics do not work out since as a matter of principle it is
undefined which part of a vacuum diagram with a cutvertex (a vertex which
connects two otherwise unconnected parts of a diagram) is the tadpole
and which part is the rest of the diagram.
A reflection of this problem was already encountered in Section
\ref{oneloop}, where the two-loop diagram (\ref{twoloopvac}) survived
our one-loop resummation.

Let us emphasize that the values (\ref{tp1}) and (\ref{tpl}) for the
two-point insertions have to be used {\em after} evaluating the recursion
relations.

Let us now establish the connection between our resummation above and
the one used in \cite{KuMu}.
In that work, a distinction is established between $\Phi$-independent
subdiagrams of the form (\ref{snails}), called ``snail diagrams'' there,
and $\Phi$-independent subdiagrams of the forms
\beq
\label{tadpoles}
\rule[-29pt]{0pt}{63pt}
\bpi(42,0)(-2,-10)
\put(19,3){\circle{32}}
\put(3,3){\line(1,1){16}}
\put(4.42,-3.58){\line(1,1){21.17}}
\put(7.68,-8.31){\line(1,1){22.63}}
\put(12.42,-11.58){\line(1,1){21.17}}
\put(19,-13){\line(1,1){16}}
\put(19,-24.23){\line(0,1){11.23}}
\put(19,-24.23){\circle*{4}}
\put(9,-24.23){\line(1,0){20}}
\epi,~~~~~~
\rule[-39pt]{0pt}{73pt}
\bpi(42,0)(-2,-10)
\put(19,3){\circle{32}}
\put(3,3){\line(1,1){16}}
\put(4.42,-3.58){\line(1,1){21.17}}
\put(7.68,-8.31){\line(1,1){22.63}}
\put(12.42,-11.58){\line(1,1){21.17}}
\put(19,-13){\line(1,1){16}}
\put(19,-34.23){\line(0,1){21.23}}
\put(19,-24.23){\circle*{4}}
\put(9,-24.23){\line(1,0){20}}
\epi,
\eeq
called ``tadpole diagrams'' there.
Ref.\ \cite{KuMu} uses the usual Schwinger-Dyson equations to adjust
the triple coupling and mass such that there are no more graphs of
the $n$-point functions with $n>1$ to consider for the effective action
(equivalent to the effective energy in our treatment) that contain any
``snail'' or ``tadpole'' subdiagrams.

One notices that graphs of the form (\ref{tadpoles}) are absent altogether
in our treatment of the effective energy, which contains only truly
1PI diagrams in contrast to a weaker definition of one-particle
irreducibility used in \cite{KuMu}, which allows also for
$\Phi$-independent subdiagrams of the form (\ref{tadpoles}).
This absence can be traced to the fact that we work with a general
background field $\varphi$ in (\ref{ephi}).
For the computation of scattering processes, $\varphi$ has to be adjusted
to the radiatively corrected vacuum expectation value $v$ of $\phi$, i.e.\
the true minimum of the effective potential, whose shift from the
tree-level value $v_0$ can perturbatively be computed as a sum
$v=v_0+{\rm corrections}$.
If we expanded $v$ subsequently in our graphs, the corrections
would lead to exactly the diagrams containing ``tadpole diagrams'' as
subdiagrams used as a starting point in \cite{KuMu}.
In other words, the formalism we use already takes care of the
resummation of all tadpoles (\ref{tadpoles}) in the effective action,
so that the triple coupling and the mass experience an appropriate
correction when computing the corrections to $v_0$ and setting $\varphi=v$.
This has nothing to do with our recursion relations, but could have been
used by the authors of \cite{KuMu} from the start as well.

For the other class of subdiagrams, the ``snail diagrams'' (\ref{snails}),
our result (\ref{propinloop}) agrees with the result of \cite{KuMu} that
the sum of all such subdiagrams amounts to a full propagator in a one-loop
``snail diagram'' and that therefore an appropriate split of the mass term
in standard $\phi^4$ theory will achieve a cancellation of all such
``snail diagrams.''

\section{Discussion}
\label{summary}
In this work we have derived efficient recursion relations to generate
connected and 1PI Feynman diagrams for $\phi^4$ theory both with and
without $\phi\rightarrow-\phi$ symmetry.
Although we used also external sources $J$ and field expectations $\Phi$
as functional variables, we were able to keep the recursion relations
simple by using as much as possible the free propagator $G$ as a functional
variable.

Taking $W$ as functional of both $G$ and $J$ and $\Ga$ as functional
of both $G$ and $\Phi$ allowed us to combine the advantages of both the
``current approach'' and the ``kernel approach'' \cite{KPKB}:
By considering diagrams with arguments $J$ and $\Phi$ on the external
legs we avoided having to deal with ``crossed'' diagrams which are related
by exchanging external arguments on their legs.
This helps keep the number of diagrams at intermediate steps low.
Only when we finally want to convert the coefficient functions of $W$
and $\Ga$ (in an expansion in powers of $J$ and $\Phi$, respectively)
into Greens functions as in (\ref{gc1234}) or (\ref{ga123}) do we have
to consider ``crossed'' diagrams.

The applications of the recursion relations lie potentially in both
statistical and particle physics.
Together with a powerful numeric integration method, the relations could
be used to push the computation of critical exponents in three dimensions
to higher loop orders, see e.g.\ \cite{Buch,MuNi,crit}.

Similar recursion relations can be set up for theories with other field
contents as well.
They are a convenient starting point for the investigation of
resummations of classes of Feynman diagrams.
Simple one-loop and multi-loop tadpole resummation examples were given
in Sections \ref{oneloop} and \ref{multiloop}, respectively.
Since the identities from which the recursion relations are derived
are non-perturbative, they might also be useful for other expansions
than the ones organized by the number of loops or powers of coupling
constants.
Another field for future investigations is the systematic solution
of recursion relations for Legendre transforms of higher order than the
effective energy \cite{Kleinert1,Kleinert2,holegendre}.
Also, the exploitation of derivatives with respect to tensors representing
interactions as in \cite{BKP} seems promising to further simplify
identities and recursion relations.

\section*{Acknowledgments}
It is a pleasure to thank H.~Kleinert and A.~Pelster for many useful
discussions and for a careful reading of the manuscript.

This work was supported by the Deutsche Forschungsgemeinschaft (DFG).

\appendix

\section{Derivatives with Respect to Symmetric $G$ and $G^{-1}$}
\label{symg}
The basic properties of derivatives with respect to an unconstrained
tensor $H_{12}$ and its inverse $H^{-1}_{12}$ are
\beq
\label{hcomhinvcom}
\left[\frac{\de}{\de H_{12}},\frac{\de}{\de H_{34}}\right]=
\left[\frac{\de}{\de H^{-1}_{12}},\frac{\de}{\de H^{-1}_{34}}\right]=0
\eeq
and
\beq
\label{dhdh}
\frac{\de H_{12}}{\de H_{34}}=
\frac{\de H^{-1}_{12}}{\de H^{-1}_{34}}=\de_{13}\de_{24},
\eeq
where, according to our conventions, the labels could mean discrete
as well as continuos variables and the $\de$'s are an according
combination of Kronecker $\de$'s and Dirac $\de$ functions.
From 
\beq
0=\frac{\de}{\de H_{34}}\de_{12}
=\frac{\de}{\de H_{34}}\int_5 H^{-1}_{15}H_{52}
=\int_5\frac{\de H^{-1}_{15}}{\de H_{34}}H_{52}
+\int_5H^{-1}_{15}\frac{\de H_{52}}{\de H_{34}}
=\int_5\frac{\de H^{-1}_{15}}{\de H_{34}}H_{52}
+H^{-1}_{13}\de_{24}
\eeq
we get
\beq
\label{dhinvdh}
\frac{\de H^{-1}_{12}}{\de H_{34}}=-H^{-1}_{13}H^{-1}_{42}
\eeq
and therefore
\beq
\label{ddh}
\frac{\de}{\de H_{12}}
=\int_{34}\frac{\de H^{-1}_{34}}{\de H_{12}}\frac{\de}{\de H^{-1}_{34}}
=-\int_{34}H^{-1}_{31}H^{-1}_{24}\frac{\de}{\de H^{-1}_{34}}.
\eeq
By exchanging $H$ and $H^{-1}$ in the derivation of (\ref{dhinvdh})
and (\ref{ddh}) we get
\beq
\label{dhdhinv}
\frac{\de H_{12}}{\de H^{-1}_{34}}=-H_{13}H_{42}
\eeq
and
\beq
\label{ddhinv}
\frac{\de}{\de H^{-1}_{12}}
=\int_{34}\frac{\de H_{34}}{\de H^{-1}_{12}}\frac{\de}{\de H_{34}}
=-\int_{34}H_{31}H_{24}\frac{\de}{\de H_{34}}.
\eeq

When considering symmetric tensors $G$ and $G^{-1}$, we have to define
what we mean by derivatives with respect to them.
While (\ref{dhdh}), (\ref{dhinvdh}) and (\ref{dhdhinv})
obviously need appropriate symmetrizations, we would like to keep
(\ref{hcomhinvcom}), (\ref{ddh}) and (\ref{ddhinv}) untouched.

Let us for the following considerations keep $H$ unconstrained and define
$G$ to be its symmetric part,
\beq
G_{12}\equiv\frac{1}{2}\left(H_{12}+H_{21}\right).
\eeq
Define the derivative with respect to $G$ by
\beq
\label{ddgdef}
\frac{\de}{\de G_{12}}\equiv
\frac{1}{2}\left(\frac{\de}{\de H_{12}}+\frac{\de}{\de H_{21}}\right),
\eeq
so that with (\ref{hcomhinvcom}) immediately follows
\beq
\label{gcom}
\left[\frac{\de}{\de G_{12}},\frac{\de}{\de G_{34}}\right]=0.
\eeq
Then, if $\de/\de G_{12}$ acts on a functional that depends on 
$H$ only through $G$, it acts exactly to remove an appearance of
$G$ in a symmetric way,
\beq
\label{dgdg}
\frac{\de G_{12}}{\de G_{34}}
=\frac{1}{4}
\left(\frac{\de}{\de H_{34}}+\frac{\de}{\de H_{43}}\right)
\left(H_{12}+H_{21}\right)
=\frac{1}{2}\left(\de_{13}\de_{24}+\de_{14}\de_{23}\right).
\eeq

We also need derivatives with respect to $G^{-1}$, which is also
symmetric in its indices.
Since in general the symmetrized version of $H^{-1}$ is not identical
to $G^{-1}$, it turns out to be inconvenient to define derivatives
with respect to $G^{-1}$ by just replacing $G$ and $H$ by $G^{-1}$
and $H^{-1}$  in (\ref{ddgdef}), respectively.
Define instead
\beq
\label{ddginv}
\frac{\de}{\de G^{-1}_{12}}\equiv
-\int_{34}G_{13}G_{24}\frac{\de}{\de G_{34}},
\eeq
which trivially implies
\beq
\label{ddg}
\frac{\de}{\de G_{12}}\equiv
-\int_{34}G^{-1}_{13}G^{-1}_{24}
\frac{\de}{\de G^{-1}_{34}}.
\eeq
Using (\ref{gcom}) and (\ref{dgdg}) it is easy to check that
\beq
\label{ginvcom}
\left[\frac{\de}{\de G^{-1}_{12}},
\frac{\de}{\de G^{-1}_{34}}\right]=0.
\eeq

Using that
\beq
0=\frac{\de}{\de G_{34}}\de_{12}
=\frac{\de}{\de G_{34}}\int_5G^{-1}_{15}G_{52}
=\int_5\frac{\de G^{-1}_{15}}{\de G_{34}}G_{52}
+\int_5G^{-1}_{15}\frac{\de G_{52}}{\de G_{34}}
=\int_5\frac{\de G^{-1}_{15}}{\de G_{34}}G_{52}
+\frac{1}{2}\left(G^{-1}_{13}\de_{24}
+G^{-1}_{14}\de_{23}\right)
\eeq
and therefore
\beq
\label{dginvdg}
\frac{\de G^{-1}_{12}}{\de G_{34}}
=
-\frac{1}{2}\left(G^{-1}_{13}G^{-1}_{24}
+G^{-1}_{14}G^{-1}_{23}\right),
\eeq
we get
\beq
\label{dginvdginv}
\frac{\de G^{-1}_{12}}{\de G^{-1}_{34}}
=
-\int_{56}G_{35}G_{46}
\frac{\de G^{-1}_{12}}{\de G_{56}}
=
\frac{1}{2}\int_{56}G_{35}G_{46}
\left(G^{-1}_{15}G^{-1}_{26}
+G^{-1}_{16}G^{-1}_{25}\right)
=\frac{1}{2}\left(\de_{13}\de_{24}+\de_{14}\de_{23}\right)
\eeq
and therefore, repeating the steps that lead to (\ref{dginvdg})
with the roles of $G$ and $G^{-1}$ exchanged,
\beq
\label{dgdginv}
\frac{\de G_{12}}{\de G^{-1}_{34}}
=
-\frac{1}{2}\left(G_{13}G_{24}
+G_{14}G_{23}\right).
\eeq

The upshot of these considerations is that we can work with symmetric
$G$ and $G^{-1}$ in the first place if we use the equations (\ref{gcom})
and (\ref{ddginv})-(\ref{ginvcom}), as well as the
symmetrized relations (\ref{dgdg}) and (\ref{dginvdg})-(\ref{dgdginv}).

\section{Elimination of $(\de^2W_I/\de\jb_1\de\jb_2)_G$}
\label{d2wdjbdjb}
In the course of changing variables from $J$ to $\jb$ in Section
\ref{chofvar}, double derivatives with respect to $\jb$ appear.
However, we want to replace this kind of terms with derivatives
with respect to $G$ to keep the resulting recursion relations as
simple as possible.

From the definition (\ref{zwgeneral}) of $Z$ and $W$ we have
\beq
-2\left(\frac{\de Z}{\de G_{12}^{-1}}\right)_J
=\left(\frac{\de^2Z}{\de J_1\de J_2}\right)_G
\eeq
and therefore
\beq
\label{wgwjj}
-2\left(\frac{\de W}{\de G_{12}^{-1}}\right)_J
=\left(\frac{\de^2W}{\de J_1\de J_2}\right)_G
+\left(\frac{\de W}{\de J_1}\right)_G
\left(\frac{\de W}{\de J_2}\right)_G.
\eeq
From 
\beq
\left(\frac{\de}{\de J_1}\right)_G
=\int_2\left(\frac{\de\jb_2}{\de J_1}\right)_G
\left(\frac{\de}{\de\jb_2}\right)_G
=\int_2G_{12}\left(\frac{\de}{\de\jb_2}\right)_G
\eeq
we get
\beq
\left(\frac{\de W}{\de J_1}\right)_G
=\int_2G_{12}\left(\frac{\de W}{\de\jb_2}\right)_G
\eeq
and
\beq
\left(\frac{\de^2W}{\de J_1\de J_2}\right)_G
=\int_{34}G_{13}G_{24}\left(\frac{\de^2W}{\de\jb_3\de\jb_4}\right)_G.
\eeq
Also,
\bea
\label{wjjwjbjb}
\left(\frac{\de W}{\de G_{12}}\right)_J
&=&
\left(\frac{\de W}{\de G_{12}}\right)_{\jb}
+\int_3\left(\frac{\de\jb_3}{\de G_{12}}\right)_J
\left(\frac{\de W}{\de\jb_3}\right)_G
=
\left(\frac{\de W}{\de G_{12}}\right)_{\jb}
+\frac{1}{2}\int_3(\de_{13}J_2+\de_{23}J_1)
\left(\frac{\de W}{\de\jb_3}\right)_G
\nn\\
&=&
\left(\frac{\de W}{\de G_{12}}\right)_{\jb}
+\frac{1}{2}\left(\frac{\de W}{\de\jb_1}\right)_GJ_2
+\frac{1}{2}\left(\frac{\de W}{\de\jb_2}\right)_GJ_1
\nn\\
&=&
\left(\frac{\de W}{\de G_{12}}\right)_{\jb}
+\frac{1}{2}\left(\frac{\de W}{\de\jb_1}\right)_G
\int_3G_{23}^{-1}\jb_3
+\frac{1}{2}\left(\frac{\de W}{\de\jb_2}\right)_G
\int_3G_{13}^{-1}\jb_3.
\eea
Combining (\ref{wgwjj}) through (\ref{wjjwjbjb}) yields
\beq
\label{wjbjbwjbwjb}
\left(\frac{\de^2W}{\de\jb_1\de\jb_2}\right)_G
+\left(\frac{\de W}{\de\jb_1}\right)_G
\left(\frac{\de W}{\de\jb_2}\right)_G
=
2\left(\frac{\de W}{\de G_{12}}\right)_{\jb}
+\frac{1}{2}
\left(\frac{\de W}{\de\jb_1}\right)_G\int_3G_{23}^{-1}\jb_3
+\frac{1}{2}
\left(\frac{\de W}{\de\jb_2}\right)_G\int_3G_{13}^{-1}\jb_3.
\eeq
From (\ref{w0ssb2}) we have
\beq
\left(\frac{\de W_0}{\de G_{12}}\right)_{\jb}
=\frac{1}{2}G_{12}^{-1}-\frac{1}{2}\int_{34}G_{13}^{-1}G_{24}^{-1}\jb_3\jb_4,
\eeq
\beq
\left(\frac{\de W_0}{\de\jb_1}\right)_G=\int_2G_{12}^{-1}\jb_2,
\eeq
\beq
\left(\frac{\de^2W_0}{\de\jb_1\de\jb_2}\right)_G=G_{12}^{-1}
\eeq
and combining this with (\ref{ww0wi}) and (\ref{wjbjbwjbwjb}) finally
gives
\beq
\left(\frac{\de^2W_I}{\de\jb_1\de\jb_2}\right)_G
+\left(\frac{\de W_I}{\de\jb_1}\right)_G
\left(\frac{\de W_I}{\de\jb_2}\right)_G
=
2\left(\frac{\de W_I}{\de G_{12}}\right)_{\jb}.
\eeq

\section{Graphs for $W^{(3,0)}$}
\label{w30example}
To demonstrate the use of the recursion relations for the Feynman
diagrams constituting $W$, we compute here $W^{(3,0)}$.
From (\ref{wissbid2g0}) and (\ref{wissbid1g}) we get
\beq
\rule[-18pt]{0pt}{42pt}
\bpi(53.23,0)(-13.23,0)
\put(19,3){\circle{32}}
\put(19,3){\makebox(1,-2){$\ba{c}3\\0\ea$}}
\put(5.77,12){\circle*{4}}
\put(5.77,-6){\circle*{4}}
\put(5.77,3){\oval(28,18)[l]}
\epi
=
\frac{3}{4}
\rule[-18pt]{0pt}{42pt}
\bpi(71,0)(-31,0)
\put(19,3){\circle{32}}
\put(19,3){\makebox(1,-2){$\ba{c}2\\1\ea$}}
\put(3,3){\circle{4}}
\put(-10,3){\line(1,0){11}}
\put(-10,3){\circle*{4}}
\put(-18,3){\circle{16}}
\epi
+
\rule[-18pt]{0pt}{42pt}
\bpi(69.23,0)(-29.23,0)
\put(19,3){\circle{32}}
\put(19,3){\makebox(1,-2){$\ba{c}2\\0\ea$}}
\put(5.77,12){\circle*{4}}
\put(5.77,-6){\circle*{4}}
\put(5.77,3){\oval(28,18)[l]}
\put(-8.23,3){\circle*{4}}
\put(-16.23,3){\circle{16}}
\epi
+\frac{1}{2}
\rule[-18pt]{0pt}{42pt}
\bpi(55.23,0)(-15.23,0)
\put(19,3){\circle{32}}
\put(19,3){\makebox(1,-2){$\ba{c}2\\1\ea$}}
\put(5.77,12){\circle*{4}}
\put(5.77,-6){\circle{4}}
\put(3.77,3){\oval(24,18)[l]}
\put(-8.23,3){\circle*{4}}
\put(-8.23,3){\line(1,0){11.23}}
\put(3,3){\circle*{4}}
\epi
+\frac{1}{3}
\rule[-18pt]{0pt}{42pt}
\bpi(54.78,0)(-14.78,0)
\put(19,3){\circle{32}}
\put(19,3){\makebox(1,-2){$\ba{c}2\\0\ea$}}
\put(5.22,11.4){\circle*{4}}
\put(10.87,17.1){\circle*{4}}
\put(5.22,-5.13){\circle*{4}}
\put(10.87,-10.78){\circle*{4}}
\put(10.87,3){\oval(41.3,27.88)[l]}
\put(5.22,3){\oval(30,16.53)[l]}
\put(-9.78,3){\circle*{4}}
\epi
+\frac{1}{2}
\rule[-17pt]{0pt}{41pt}
\bpi(98.23,0)(-60,0)
\put(17.23,3){\circle{32}}
\put(17.23,3){\makebox(1,-2){$\ba{c}2\\0\ea$}}
\put(4,12){\circle*{4}}
\put(4,-6){\circle*{4}}
\put(4,3){\oval(28,18)[l]}
\put(-10,3){\circle*{4}}
\put(-23,3){\circle{4}}
\put(-10,3){\line(-1,0){11}}
\put(-39,3){\circle{32}}
\put(-39,3){\makebox(1,-2){$\ba{c}1\\1\ea$}}
\epi
\eeq
and
\beq
\rule[-18pt]{0pt}{42pt}
\bpi(42,0)
\put(21,3){\circle{32}}
\put(21,3){\makebox(1,-2){$\ba{c}2\\1\ea$}}
\epi
=
\frac{1}{2}
\rule[-13pt]{0pt}{42pt}
\bpi(76,0)(2,0)
\put(9,3){\circle*{4}}
\put(9,2){\line(1,0){16}}
\put(9,4){\line(1,0){16}}
\put(25,3){\circle*{4}}
\put(41,3){\circle{4}}
\put(25,11){\circle{16}}
\put(25,3){\line(1,0){14}}
\put(57,3){\circle{32}}
\put(57,3){\makebox(1,-2){$\ba{c}1\\1\ea$}}
\epi
+\rule[-18pt]{0pt}{42pt}
\bpi(68.23,0)(-28.23,0)
\put(19,3){\circle{32}}
\put(19,3){\makebox(1,-2){$\ba{c}2\\0\ea$}}
\put(5.77,12){\circle*{4}}
\put(5.77,-6){\circle*{4}}
\put(5.77,3){\oval(28,18)[l]}
\put(-8.23,3){\circle*{4}}
\put(-21.23,2){\line(1,0){13}}
\put(-21.23,4){\line(1,0){13}}
\put(-21.23,3){\circle*{4}}
\epi
+\frac{1}{3}
\rule[-18pt]{0pt}{42pt}
\bpi(68.23,0)(-28.23,0)
\put(19,3){\circle{32}}
\put(19,3){\makebox(1,-2){$\ba{c}1\\1\ea$}}
\put(5.77,12){\circle*{4}}
\put(5.77,-6){\circle{4}}
\put(3.77,3){\oval(24,18)[l]}
\put(-8.23,3){\circle*{4}}
\put(-21.23,2){\line(1,0){13}}
\put(-21.23,4){\line(1,0){13}}
\put(-8.23,3){\line(1,0){11.23}}
\put(3,3){\circle*{4}}
\put(-21.23,3){\circle*{4}}
\epi,
\eeq
respectively.
With (\ref{w11}) and (\ref{w20}) we get
\beq
\frac{1}{2}
\rule[-13pt]{0pt}{42pt}
\bpi(76,0)(2,0)
\put(9,3){\circle*{4}}
\put(25,3){\circle*{4}}
\put(41,3){\circle{4}}
\put(25,11){\circle{16}}
\put(9,2){\line(1,0){16}}
\put(9,4){\line(1,0){16}}
\put(25,3){\line(1,0){14}}
\put(57,3){\circle{32}}
\put(57,3){\makebox(1,-2){$\ba{c}1\\1\ea$}}
\epi
=
\frac{1}{4}
\rule[-10pt]{0pt}{33pt}
\bpi(55,0)
\put(13,3){\circle{16}}
\put(21,3){\circle*{4}}
\put(21,3){\line(1,0){13}}
\put(34,3){\circle*{4}}
\put(33,3){\line(0,1){13}}
\put(35,3){\line(0,1){13}}
\put(34,16){\circle*{4}}
\put(42,3){\circle{16}}
\epi,
\eeq

\beq
\rule[-18pt]{0pt}{42pt}
\bpi(68.23,0)(-28.23,0)
\put(19,3){\circle{32}}
\put(19,3){\makebox(1,-2){$\ba{c}2\\0\ea$}}
\put(5.77,12){\circle*{4}}
\put(5.77,-6){\circle*{4}}
\put(5.77,3){\oval(28,18)[l]}
\put(-8.23,3){\circle*{4}}
\put(-21.23,2){\line(1,0){13}}
\put(-21.23,4){\line(1,0){13}}
\put(-21.23,3){\circle*{4}}
\epi
=
\frac{1}{4}
\rule[-10pt]{0pt}{26pt}
\bpi(70,0)
\put(7,3){\circle*{4}}
\put(7,2){\line(1,0){13}}
\put(7,4){\line(1,0){13}}
\put(20,3){\circle*{4}}
\put(28,3){\circle{16}}
\put(36,3){\circle*{4}}
\put(36,3){\line(1,0){13}}
\put(49,3){\circle*{4}}
\put(57,3){\circle{16}}
\epi
+\frac{1}{8}
\rule[-10pt]{0pt}{33pt}
\bpi(68,0)
\put(13,3){\circle{16}}
\put(21,3){\circle*{4}}
\put(21,3){\line(1,0){26}}
\put(34,3){\circle*{4}}
\put(33,3){\line(0,1){13}}
\put(35,3){\line(0,1){13}}
\put(34,16){\circle*{4}}
\put(47,3){\circle*{4}}
\put(55,3){\circle{16}}
\epi
+\frac{1}{4}
\rule[-16pt]{0pt}{38pt}
\bpi(49,0)
\put(7,3){\circle*{4}}
\put(7,2){\line(1,0){13}}
\put(7,4){\line(1,0){13}}
\put(20,3){\circle*{4}}
\put(32,3){\circle{24}}
\put(32,-9){\circle*{4}}
\put(32,15){\circle*{4}}
\put(32,-9){\line(0,1){24}}
\epi
+\frac{1}{4}
\rule[-10pt]{0pt}{26pt}
\bpi(57,0)
\put(7,3){\circle*{4}}
\put(7,2){\line(1,0){13}}
\put(7,4){\line(1,0){13}}
\put(20,3){\circle*{4}}
\put(28,3){\circle{16}}
\put(36,3){\circle*{4}}
\put(44,3){\circle{16}}
\epi
\eeq
and
\beq
\frac{1}{3}
\rule[-18pt]{0pt}{42pt}
\bpi(68.23,0)(-28.23,0)
\put(19,3){\circle{32}}
\put(19,3){\makebox(1,-2){$\ba{c}1\\1\ea$}}
\put(5.77,12){\circle*{4}}
\put(5.77,-6){\circle{4}}
\put(3.77,3){\oval(24,18)[l]}
\put(-8.23,3){\circle*{4}}
\put(-21.23,2){\line(1,0){13}}
\put(-21.23,4){\line(1,0){13}}
\put(-8.23,3){\line(1,0){11.23}}
\put(3,3){\circle*{4}}
\put(-21.23,3){\circle*{4}}
\epi
=
\frac{1}{6}
\rule[-14pt]{0pt}{34pt}
\bpi(51,0)
\put(7,3){\circle*{4}}
\put(7,2){\line(1,0){13}}
\put(7,4){\line(1,0){13}}
\put(20,3){\line(1,0){24}}
\put(20,3){\circle*{4}}
\put(32,3){\circle{24}}
\put(44,3){\circle*{4}}
\epi
\eeq
and thus
\bea
\label{w21}
\rule[-18pt]{0pt}{42pt}
\bpi(42,0)
\put(21,3){\circle{32}}
\put(21,3){\makebox(1,-2){$\ba{c}2\\1\ea$}}
\epi
&=&
\frac{1}{4}
\rule[-10pt]{0pt}{26pt}
\bpi(70,0)
\put(7,3){\circle*{4}}
\put(7,2){\line(1,0){13}}
\put(7,4){\line(1,0){13}}
\put(20,3){\circle*{4}}
\put(28,3){\circle{16}}
\put(36,3){\circle*{4}}
\put(36,3){\line(1,0){13}}
\put(49,3){\circle*{4}}
\put(57,3){\circle{16}}
\epi
+\frac{1}{8}
\rule[-10pt]{0pt}{33pt}
\bpi(68,0)
\put(13,3){\circle{16}}
\put(21,3){\circle*{4}}
\put(21,3){\line(1,0){26}}
\put(34,3){\circle*{4}}
\put(33,3){\line(0,1){13}}
\put(35,3){\line(0,1){13}}
\put(34,16){\circle*{4}}
\put(47,3){\circle*{4}}
\put(55,3){\circle{16}}
\epi
+\frac{1}{4}
\rule[-16pt]{0pt}{38pt}
\bpi(49,0)
\put(7,3){\circle*{4}}
\put(7,2){\line(1,0){13}}
\put(7,4){\line(1,0){13}}
\put(20,3){\circle*{4}}
\put(32,3){\circle{24}}
\put(32,-9){\circle*{4}}
\put(32,15){\circle*{4}}
\put(32,-9){\line(0,1){24}}
\epi
\nn\\
&&
+\frac{1}{4}
\rule[-10pt]{0pt}{33pt}
\bpi(55,0)
\put(13,3){\circle{16}}
\put(21,3){\circle*{4}}
\put(21,3){\line(1,0){13}}
\put(34,3){\circle*{4}}
\put(33,3){\line(0,1){13}}
\put(35,3){\line(0,1){13}}
\put(34,16){\circle*{4}}
\put(42,3){\circle{16}}
\epi
+\frac{1}{4}
\rule[-10pt]{0pt}{26pt}
\bpi(57,0)
\put(7,3){\circle*{4}}
\put(7,2){\line(1,0){13}}
\put(7,4){\line(1,0){13}}
\put(20,3){\circle*{4}}
\put(28,3){\circle{16}}
\put(36,3){\circle*{4}}
\put(44,3){\circle{16}}
\epi
+\frac{1}{6}
\rule[-14pt]{0pt}{34pt}
\bpi(51,0)
\put(7,3){\circle*{4}}
\put(7,2){\line(1,0){13}}
\put(7,4){\line(1,0){13}}
\put(20,3){\line(1,0){24}}
\put(20,3){\circle*{4}}
\put(32,3){\circle{24}}
\put(44,3){\circle*{4}}
\epi.
\eea
With (\ref{w11}), (\ref{w20}) and (\ref{w21}) we have
\bea
\frac{3}{2}
\rule[-18pt]{0pt}{42pt}
\bpi(71,0)(-31,0)
\put(19,3){\circle{32}}
\put(19,3){\makebox(1,-2){$\ba{c}2\\1\ea$}}
\put(3,3){\circle{4}}
\put(-10,3){\line(1,0){11}}
\put(-10,3){\circle*{4}}
\put(-18,3){\circle{16}}
\epi
&=&
\frac{3}{8}
\rule[-10pt]{0pt}{26pt}
\bpi(84,0)
\put(13,3){\circle{16}}
\put(21,3){\circle*{4}}
\put(21,3){\line(1,0){13}}
\put(34,3){\circle*{4}}
\put(42,3){\circle{16}}
\put(50,3){\circle*{4}}
\put(50,3){\line(1,0){13}}
\put(63,3){\circle*{4}}
\put(71,3){\circle{16}}
\epi
+
\frac{3}{16}
\rule[-19pt]{0pt}{53pt}
\bpi(57.18,0)(-13.59,0)
\put(15,3){\circle*{4}}
\put(15,3){\line(0,1){10}}
\put(15,13){\circle*{4}}
\put(15,21){\circle{16}}
\put(15,3){\line(-5,-3){10}}
\put(6.34,-2){\circle*{4}}
\put(-0.59,-6){\circle{16}}
\put(15,3){\line(5,-3){10}}
\put(23.66,-2){\circle*{4}}
\put(30.59,-6){\circle{16}}
\epi
+\frac{3}{8}
\rule[-16pt]{0pt}{38pt}
\bpi(63,0)
\put(13,3){\circle{16}}
\put(21,3){\circle*{4}}
\put(21,3){\line(1,0){13}}
\put(34,3){\circle*{4}}
\put(46,3){\circle{24}}
\put(46,-9){\circle*{4}}
\put(46,15){\circle*{4}}
\put(46,-9){\line(0,1){24}}
\epi
\nn\\
&&
+\frac{3}{8}
\rule[-10pt]{0pt}{34pt}
\bpi(68,0)
\put(13,3){\circle{16}}
\put(21,3){\circle*{4}}
\put(21,3){\line(1,0){26}}
\put(34,3){\circle*{4}}
\put(34,11){\circle{16}}
\put(47,3){\circle*{4}}
\put(55,3){\circle{16}}
\epi
+\frac{3}{8}
\rule[-10pt]{0pt}{26pt}
\bpi(71,0)
\put(13,3){\circle{16}}
\put(21,3){\circle*{4}}
\put(21,3){\line(1,0){13}}
\put(34,3){\circle*{4}}
\put(42,3){\circle{16}}
\put(50,3){\circle*{4}}
\put(58,3){\circle{16}}
\epi
+\frac{1}{4}
\rule[-12pt]{0pt}{30pt}
\bpi(65,0)
\put(13,3){\circle{16}}
\put(21,3){\circle*{4}}
\put(21,3){\line(1,0){37}}
\put(34,3){\circle*{4}}
\put(46,3){\circle{24}}
\put(58,3){\circle*{4}}
\epi,
\eea

\bea
\rule[-18pt]{0pt}{42pt}
\bpi(55.23,0)(-15.23,0)
\put(19,3){\circle{32}}
\put(19,3){\makebox(1,-2){$\ba{c}2\\1\ea$}}
\put(5.77,12){\circle*{4}}
\put(5.77,-6){\circle{4}}
\put(3.77,3){\oval(24,18)[l]}
\put(-8.23,3){\circle*{4}}
\put(-8.23,3){\line(1,0){11.23}}
\put(3,3){\circle*{4}}
\epi
&=&
\frac{1}{4}
\rule[-10pt]{0pt}{26pt}
\bpi(84,0)
\put(13,3){\circle{16}}
\put(21,3){\circle*{4}}
\put(21,3){\line(1,0){13}}
\put(34,3){\circle*{4}}
\put(42,3){\circle{16}}
\put(50,3){\circle*{4}}
\put(50,3){\line(1,0){13}}
\put(63,3){\circle*{4}}
\put(71,3){\circle{16}}
\epi
+
\rule[-16pt]{0pt}{38pt}
\bpi(63,0)
\put(13,3){\circle{16}}
\put(21,3){\circle*{4}}
\put(21,3){\line(1,0){13}}
\put(34,3){\circle*{4}}
\put(46,3){\circle{24}}
\put(46,-9){\circle*{4}}
\put(46,15){\circle*{4}}
\put(46,-9){\line(0,1){24}}
\epi
+\frac{3}{4}
\rule[-16pt]{0pt}{40pt}
\bpi(50,0)
\put(13,3){\circle{16}}
\put(37,3){\circle{16}}
\put(13,-5){\circle*{4}}
\put(13,11){\circle*{4}}
\put(37,-5){\circle*{4}}
\put(37,11){\circle*{4}}
\put(25,-3){\oval(24,16)[b]}
\put(25,11){\oval(24,16)[t]}
\epi
+\frac{1}{2}
\rule[-14pt]{0pt}{36pt}
\bpi(34.78,0)(-3.39,0)
\put(15,3){\circle{24}}
\put(15,3){\circle*{4}}
\put(15,15){\circle*{4}}
\put(4.61,-3){\circle*{4}}
\put(25.39,-3){\circle*{4}}
\put(15,3){\line(0,1){12}}
\put(15,3){\line(5,-3){10}}
\put(15,3){\line(-5,-3){10}}
\epi
\nn\\
&&
+\frac{1}{4}
\rule[-10pt]{0pt}{26pt}
\bpi(71,0)
\put(13,3){\circle{16}}
\put(21,3){\circle*{4}}
\put(21,3){\line(1,0){13}}
\put(34,3){\circle*{4}}
\put(42,3){\circle{16}}
\put(50,3){\circle*{4}}
\put(58,3){\circle{16}}
\epi
+\frac{1}{4}
\rule[-14pt]{0pt}{34pt}
\bpi(65,0)
\put(13,3){\circle{16}}
\put(21,3){\circle*{4}}
\put(21,3){\line(1,0){37}}
\put(34,3){\circle*{4}}
\put(46,3){\circle{24}}
\put(58,3){\circle*{4}}
\epi
+\frac{3}{4}
\rule[-16pt]{0pt}{38pt}
\bpi(50,0)
\put(13,3){\circle{16}}
\put(21,3){\circle*{4}}
\put(33,3){\circle{24}}
\put(33,-9){\circle*{4}}
\put(33,15){\circle*{4}}
\put(33,-9){\line(0,1){24}}
\epi
+\frac{3}{4}
\rule[-16pt]{0pt}{36pt}
\bpi(38,0)
\put(19,3){\circle{24}}
\put(7,-9){\oval(24,24)[rt]}
\put(31,-9){\oval(24,24)[lt]}
\put(19,-9){\circle*{4}}
\put(7,3){\circle*{4}}
\put(31,3){\circle*{4}}
\epi,
\eea

\beq
\rule[-17pt]{0pt}{41pt}
\bpi(98.23,0)(-60,0)
\put(17.23,3){\circle{32}}
\put(17.23,3){\makebox(1,-2){$\ba{c}2\\0\ea$}}
\put(4,12){\circle*{4}}
\put(4,-6){\circle*{4}}
\put(4,3){\oval(28,18)[l]}
\put(-10,3){\circle*{4}}
\put(-23,3){\circle{4}}
\put(-10,3){\line(-1,0){11}}
\put(-39,3){\circle{32}}
\put(-39,3){\makebox(1,-2){$\ba{c}1\\1\ea$}}
\epi
=
\frac{1}{8}
\rule[-10pt]{0pt}{26pt}
\bpi(84,0)
\put(13,3){\circle{16}}
\put(21,3){\circle*{4}}
\put(21,3){\line(1,0){13}}
\put(34,3){\circle*{4}}
\put(42,3){\circle{16}}
\put(50,3){\circle*{4}}
\put(50,3){\line(1,0){13}}
\put(63,3){\circle*{4}}
\put(71,3){\circle{16}}
\epi
+
\frac{1}{16}
\rule[-19pt]{0pt}{53pt}
\bpi(57.18,0)(-13.59,0)
\put(15,3){\circle*{4}}
\put(15,3){\line(0,1){10}}
\put(15,13){\circle*{4}}
\put(15,21){\circle{16}}
\put(15,3){\line(-5,-3){10}}
\put(6.34,-2){\circle*{4}}
\put(-0.59,-6){\circle{16}}
\put(15,3){\line(5,-3){10}}
\put(23.66,-2){\circle*{4}}
\put(30.59,-6){\circle{16}}
\epi
+\frac{1}{8}
\rule[-16pt]{0pt}{38pt}
\bpi(63,0)
\put(13,3){\circle{16}}
\put(21,3){\circle*{4}}
\put(21,3){\line(1,0){13}}
\put(34,3){\circle*{4}}
\put(46,3){\circle{24}}
\put(46,-9){\circle*{4}}
\put(46,15){\circle*{4}}
\put(46,-9){\line(0,1){24}}
\epi
+\frac{1}{8}
\rule[-10pt]{0pt}{26pt}
\bpi(71,0)
\put(13,3){\circle{16}}
\put(21,3){\circle*{4}}
\put(21,3){\line(1,0){13}}
\put(34,3){\circle*{4}}
\put(42,3){\circle{16}}
\put(50,3){\circle*{4}}
\put(58,3){\circle{16}}
\epi,
\eeq

\beq
2
\rule[-18pt]{0pt}{42pt}
\bpi(69.23,0)(-29.23,0)
\put(19,3){\circle{32}}
\put(19,3){\makebox(1,-2){$\ba{c}2\\0\ea$}}
\put(5.77,12){\circle*{4}}
\put(5.77,-6){\circle*{4}}
\put(5.77,3){\oval(28,18)[l]}
\put(-8.23,3){\circle*{4}}
\put(-16.23,3){\circle{16}}
\epi
=
\frac{1}{4}
\rule[-10pt]{0pt}{34pt}
\bpi(68,0)
\put(13,3){\circle{16}}
\put(21,3){\circle*{4}}
\put(21,3){\line(1,0){26}}
\put(34,3){\circle*{4}}
\put(34,11){\circle{16}}
\put(47,3){\circle*{4}}
\put(55,3){\circle{16}}
\epi
+\frac{1}{2}
\rule[-10pt]{0pt}{26pt}
\bpi(71,0)
\put(13,3){\circle{16}}
\put(21,3){\circle*{4}}
\put(21,3){\line(1,0){13}}
\put(34,3){\circle*{4}}
\put(42,3){\circle{16}}
\put(50,3){\circle*{4}}
\put(58,3){\circle{16}}
\epi
+\rule[-16pt]{0pt}{38pt}
\bpi(50,0)
\put(13,3){\circle{16}}
\put(21,3){\circle*{4}}
\put(33,3){\circle{24}}
\put(33,-9){\circle*{4}}
\put(33,15){\circle*{4}}
\put(33,-9){\line(0,1){24}}
\epi
+\frac{1}{2}
\rule[-10pt]{0pt}{26pt}
\bpi(58,0)
\put(13,3){\circle{16}}
\put(21,3){\circle*{4}}
\put(29,3){\circle{16}}
\put(37,3){\circle*{4}}
\put(45,3){\circle{16}}
\epi,
\eeq

\beq
\frac{2}{3}
\rule[-18pt]{0pt}{42pt}
\bpi(54.78,0)(-14.78,0)
\put(19,3){\circle{32}}
\put(19,3){\makebox(1,-2){$\ba{c}2\\0\ea$}}
\put(5.22,11.4){\circle*{4}}
\put(10.87,17.1){\circle*{4}}
\put(5.22,-5.13){\circle*{4}}
\put(10.87,-10.78){\circle*{4}}
\put(10.87,3){\oval(41.3,27.88)[l]}
\put(5.22,3){\oval(30,16.53)[l]}
\put(-9.78,3){\circle*{4}}
\epi
=
\frac{1}{3}
\rule[-14pt]{0pt}{34pt}
\bpi(65,0)
\put(13,3){\circle{16}}
\put(21,3){\circle*{4}}
\put(21,3){\line(1,0){37}}
\put(34,3){\circle*{4}}
\put(46,3){\circle{24}}
\put(58,3){\circle*{4}}
\epi
+\frac{1}{2}
\rule[-16pt]{0pt}{36pt}
\bpi(38,0)
\put(19,3){\circle{24}}
\put(7,-9){\oval(24,24)[rt]}
\put(31,-9){\oval(24,24)[lt]}
\put(19,-9){\circle*{4}}
\put(7,3){\circle*{4}}
\put(31,3){\circle*{4}}
\epi
+\frac{1}{6}
\rule[-12pt]{0pt}{34pt}
\bpi(38,0)
\put(19,3){\circle{24}}
\put(19,3){\oval(24,8)}
\put(7,3){\circle*{4}}
\put(31,3){\circle*{4}}
\epi,
\eeq
such that
\bea
2
\rule[-18pt]{0pt}{42pt}
\bpi(53.23,0)(-13.23,0)
\put(19,3){\circle{32}}
\put(19,3){\makebox(1,-2){$\ba{c}3\\0\ea$}}
\put(5.77,12){\circle*{4}}
\put(5.77,-6){\circle*{4}}
\put(5.77,3){\oval(28,18)[l]}
\epi
&=&
\frac{3}{4}
\rule[-10pt]{0pt}{26pt}
\bpi(84,0)
\put(13,3){\circle{16}}
\put(21,3){\circle*{4}}
\put(21,3){\line(1,0){13}}
\put(34,3){\circle*{4}}
\put(42,3){\circle{16}}
\put(50,3){\circle*{4}}
\put(50,3){\line(1,0){13}}
\put(63,3){\circle*{4}}
\put(71,3){\circle{16}}
\epi
+\frac{1}{4}
\rule[-19pt]{0pt}{53pt}
\bpi(57.18,0)(-13.59,0)
\put(15,3){\circle*{4}}
\put(15,3){\line(0,1){10}}
\put(15,13){\circle*{4}}
\put(15,21){\circle{16}}
\put(15,3){\line(-5,-3){10}}
\put(6.34,-2){\circle*{4}}
\put(-0.59,-6){\circle{16}}
\put(15,3){\line(5,-3){10}}
\put(23.66,-2){\circle*{4}}
\put(30.59,-6){\circle{16}}
\epi
+\frac{3}{2}
\rule[-16pt]{0pt}{38pt}
\bpi(63,0)
\put(13,3){\circle{16}}
\put(21,3){\circle*{4}}
\put(21,3){\line(1,0){13}}
\put(34,3){\circle*{4}}
\put(46,3){\circle{24}}
\put(46,-9){\circle*{4}}
\put(46,15){\circle*{4}}
\put(46,-9){\line(0,1){24}}
\epi
+\frac{3}{4}
\rule[-16pt]{0pt}{40pt}
\bpi(50,0)
\put(13,3){\circle{16}}
\put(37,3){\circle{16}}
\put(13,-5){\circle*{4}}
\put(13,11){\circle*{4}}
\put(37,-5){\circle*{4}}
\put(37,11){\circle*{4}}
\put(25,-3){\oval(24,16)[b]}
\put(25,11){\oval(24,16)[t]}
\epi
+\frac{1}{2}
\rule[-14pt]{0pt}{36pt}
\bpi(34.78,0)(-3.39,0)
\put(15,3){\circle{24}}
\put(15,3){\circle*{4}}
\put(15,15){\circle*{4}}
\put(4.61,-3){\circle*{4}}
\put(25.39,-3){\circle*{4}}
\put(15,3){\line(0,1){12}}
\put(15,3){\line(5,-3){10}}
\put(15,3){\line(-5,-3){10}}
\epi
\nn\\
&&
+\frac{5}{8}
\rule[-10pt]{0pt}{34pt}
\bpi(68,0)
\put(13,3){\circle{16}}
\put(21,3){\circle*{4}}
\put(21,3){\line(1,0){26}}
\put(34,3){\circle*{4}}
\put(34,11){\circle{16}}
\put(47,3){\circle*{4}}
\put(55,3){\circle{16}}
\epi
+\frac{5}{4}
\rule[-10pt]{0pt}{26pt}
\bpi(71,0)
\put(13,3){\circle{16}}
\put(21,3){\circle*{4}}
\put(21,3){\line(1,0){13}}
\put(34,3){\circle*{4}}
\put(42,3){\circle{16}}
\put(50,3){\circle*{4}}
\put(58,3){\circle{16}}
\epi
+\frac{5}{6}
\rule[-14pt]{0pt}{34pt}
\bpi(65,0)
\put(13,3){\circle{16}}
\put(21,3){\circle*{4}}
\put(21,3){\line(1,0){37}}
\put(34,3){\circle*{4}}
\put(46,3){\circle{24}}
\put(58,3){\circle*{4}}
\epi
+\frac{5}{4}
\rule[-16pt]{0pt}{38pt}
\bpi(50,0)
\put(13,3){\circle{16}}
\put(21,3){\circle*{4}}
\put(33,3){\circle{24}}
\put(33,-9){\circle*{4}}
\put(33,15){\circle*{4}}
\put(33,-9){\line(0,1){24}}
\epi
+\frac{5}{4}
\rule[-16pt]{0pt}{36pt}
\bpi(38,0)
\put(19,3){\circle{24}}
\put(7,-9){\oval(24,24)[rt]}
\put(31,-9){\oval(24,24)[lt]}
\put(19,-9){\circle*{4}}
\put(7,3){\circle*{4}}
\put(31,3){\circle*{4}}
\epi
\nn\\
&&
+\frac{1}{2}
\rule[-10pt]{0pt}{26pt}
\bpi(58,0)
\put(13,3){\circle{16}}
\put(21,3){\circle*{4}}
\put(29,3){\circle{16}}
\put(37,3){\circle*{4}}
\put(45,3){\circle{16}}
\epi
+\frac{1}{6}
\rule[-12pt]{0pt}{34pt}
\bpi(38,0)
\put(19,3){\circle{24}}
\put(19,3){\oval(24,8)}
\put(7,3){\circle*{4}}
\put(31,3){\circle*{4}}
\epi
\eea
and therefore, using (\ref{n1dngd}),
\bea
\label{w30}
\rule[-18pt]{0pt}{42pt}
\bpi(42,0)
\put(21,3){\circle{32}}
\put(21,3){\makebox(1,-2){$\ba{c}3\\0\ea$}}
\epi
&=&
\frac{1}{16}
\rule[-10pt]{0pt}{26pt}
\bpi(84,0)
\put(13,3){\circle{16}}
\put(21,3){\circle*{4}}
\put(21,3){\line(1,0){13}}
\put(34,3){\circle*{4}}
\put(42,3){\circle{16}}
\put(50,3){\circle*{4}}
\put(50,3){\line(1,0){13}}
\put(63,3){\circle*{4}}
\put(71,3){\circle{16}}
\epi
+\frac{1}{48}
\rule[-19pt]{0pt}{53pt}
\bpi(57.18,0)(-13.59,0)
\put(15,3){\circle*{4}}
\put(15,3){\line(0,1){10}}
\put(15,13){\circle*{4}}
\put(15,21){\circle{16}}
\put(15,3){\line(-5,-3){10}}
\put(6.34,-2){\circle*{4}}
\put(-0.59,-6){\circle{16}}
\put(15,3){\line(5,-3){10}}
\put(23.66,-2){\circle*{4}}
\put(30.59,-6){\circle{16}}
\epi
+\frac{1}{8}
\rule[-16pt]{0pt}{38pt}
\bpi(63,0)
\put(13,3){\circle{16}}
\put(21,3){\circle*{4}}
\put(21,3){\line(1,0){13}}
\put(34,3){\circle*{4}}
\put(46,3){\circle{24}}
\put(46,-9){\circle*{4}}
\put(46,15){\circle*{4}}
\put(46,-9){\line(0,1){24}}
\epi
+\frac{1}{16}
\rule[-16pt]{0pt}{40pt}
\bpi(50,0)
\put(13,3){\circle{16}}
\put(37,3){\circle{16}}
\put(13,-5){\circle*{4}}
\put(13,11){\circle*{4}}
\put(37,-5){\circle*{4}}
\put(37,11){\circle*{4}}
\put(25,-3){\oval(24,16)[b]}
\put(25,11){\oval(24,16)[t]}
\epi
+\frac{1}{24}
\rule[-14pt]{0pt}{36pt}
\bpi(34.78,0)(-3.39,0)
\put(15,3){\circle{24}}
\put(15,3){\circle*{4}}
\put(15,15){\circle*{4}}
\put(4.61,-3){\circle*{4}}
\put(25.39,-3){\circle*{4}}
\put(15,3){\line(0,1){12}}
\put(15,3){\line(5,-3){10}}
\put(15,3){\line(-5,-3){10}}
\epi
\nn\\
&&
+\frac{1}{16}
\rule[-10pt]{0pt}{34pt}
\bpi(68,0)
\put(13,3){\circle{16}}
\put(21,3){\circle*{4}}
\put(21,3){\line(1,0){26}}
\put(34,3){\circle*{4}}
\put(34,11){\circle{16}}
\put(47,3){\circle*{4}}
\put(55,3){\circle{16}}
\epi
+\frac{1}{8}
\rule[-10pt]{0pt}{26pt}
\bpi(71,0)
\put(13,3){\circle{16}}
\put(21,3){\circle*{4}}
\put(21,3){\line(1,0){13}}
\put(34,3){\circle*{4}}
\put(42,3){\circle{16}}
\put(50,3){\circle*{4}}
\put(58,3){\circle{16}}
\epi
+\frac{1}{12}
\rule[-14pt]{0pt}{34pt}
\bpi(65,0)
\put(13,3){\circle{16}}
\put(21,3){\circle*{4}}
\put(21,3){\line(1,0){37}}
\put(34,3){\circle*{4}}
\put(46,3){\circle{24}}
\put(58,3){\circle*{4}}
\epi
+\frac{1}{8}
\rule[-16pt]{0pt}{38pt}
\bpi(50,0)
\put(13,3){\circle{16}}
\put(21,3){\circle*{4}}
\put(33,3){\circle{24}}
\put(33,-9){\circle*{4}}
\put(33,15){\circle*{4}}
\put(33,-9){\line(0,1){24}}
\epi
+\frac{1}{8}
\rule[-16pt]{0pt}{36pt}
\bpi(38,0)
\put(19,3){\circle{24}}
\put(7,-9){\oval(24,24)[rt]}
\put(31,-9){\oval(24,24)[lt]}
\put(19,-9){\circle*{4}}
\put(7,3){\circle*{4}}
\put(31,3){\circle*{4}}
\epi
\nn\\
&&
+\frac{1}{16}
\rule[-10pt]{0pt}{26pt}
\bpi(58,0)
\put(13,3){\circle{16}}
\put(21,3){\circle*{4}}
\put(29,3){\circle{16}}
\put(37,3){\circle*{4}}
\put(45,3){\circle{16}}
\epi
+\frac{1}{48}
\rule[-12pt]{0pt}{34pt}
\bpi(38,0)
\put(19,3){\circle{24}}
\put(19,3){\oval(24,8)}
\put(7,3){\circle*{4}}
\put(31,3){\circle*{4}}
\epi.
\eea

\section{Graphs for $\Ga^{(3,0)}$}
\label{ga30example}
To demonstrate the use of the recursion relations for the Feynman
diagrams constituting $\Ga$, we compute here $\Ga^{(3,0)}$.
Eq.\ (\ref{gaissbid1e}) gives
\beq
\label{ga21}
\rule[-18pt]{0pt}{42pt}
\bpi(42,0)
\put(21,3){\circle{32}}
\put(21,3){\makebox(1,-2){$\ba{c}2\\1\ea$}}
\epi
=
\rule[-18pt]{0pt}{42pt}
\bpi(68.23,0)(-28.23,0)
\put(19,3){\circle{32}}
\put(19,3){\makebox(1,-2){$\ba{c}2\\0\ea$}}
\put(5.77,12){\circle*{4}}
\put(5.77,-6){\circle*{4}}
\put(5.77,3){\oval(28,18)[l]}
\put(-8.23,3){\circle*{4}}
\put(-19.23,2){\line(1,0){11}}
\put(-19.23,4){\line(1,0){11}}
\put(-21.23,3){\circle{4}}
\epi
+\frac{1}{3}
\rule[-18pt]{0pt}{42pt}
\bpi(68.23,0)(-28.23,0)
\put(19,3){\circle{32}}
\put(19,3){\makebox(1,-2){$\ba{c}1\\1\ea$}}
\put(5.77,12){\circle*{4}}
\put(5.77,-6){\circle{4}}
\put(3.77,3){\oval(24,18)[l]}
\put(-8.23,3){\circle*{4}}
\put(-19.23,2){\line(1,0){11}}
\put(-19.23,4){\line(1,0){11}}
\put(-8.23,3){\line(1,0){11.23}}
\put(3,3){\circle*{4}}
\put(-21.23,3){\circle{4}}
\epi
=
\frac{1}{4}
\rule[-10pt]{0pt}{26pt}
\bpi(57,0)
\put(7,3){\circle{4}}
\put(9,2){\line(1,0){11}}
\put(9,4){\line(1,0){11}}
\put(20,3){\circle*{4}}
\put(28,3){\circle{16}}
\put(36,3){\circle*{4}}
\put(44,3){\circle{16}}
\epi
+\frac{1}{6}
\rule[-14pt]{0pt}{34pt}
\bpi(51,0)
\put(7,3){\circle{4}}
\put(9,2){\line(1,0){11}}
\put(9,4){\line(1,0){11}}
\put(20,3){\line(1,0){24}}
\put(20,3){\circle*{4}}
\put(32,3){\circle{24}}
\put(44,3){\circle*{4}}
\epi
+\frac{1}{4}
\rule[-16pt]{0pt}{38pt}
\bpi(49,0)
\put(7,3){\circle{4}}
\put(9,2){\line(1,0){11}}
\put(9,4){\line(1,0){11}}
\put(20,3){\circle*{4}}
\put(32,3){\circle{24}}
\put(32,-9){\circle*{4}}
\put(32,15){\circle*{4}}
\put(32,-9){\line(0,1){24}}
\epi,
\eeq
while (\ref{gaissbid2vac}) gives
\bea
\rule[-18pt]{0pt}{42pt}
\bpi(53.23,0)(-13.23,0)
\put(19,3){\circle{32}}
\put(19,3){\makebox(1,-2){$\ba{c}3\\0\ea$}}
\put(5.77,12){\circle*{4}}
\put(5.77,-6){\circle*{4}}
\put(5.77,3){\oval(28,18)[l]}
\epi
&=&
\frac{1}{2}
\rule[-18pt]{0pt}{42pt}
\bpi(55.23,0)(-15.23,0)
\put(19,3){\circle{32}}
\put(19,3){\makebox(1,-2){$\ba{c}2\\1\ea$}}
\put(5.77,12){\circle*{4}}
\put(5.77,-6){\circle{4}}
\put(3.77,3){\oval(24,18)[l]}
\put(-8.23,3){\circle*{4}}
\put(-8.23,3){\line(1,0){11.23}}
\put(3,3){\circle*{4}}
\epi
+
\rule[-18pt]{0pt}{42pt}
\bpi(69.23,0)(-29.23,0)
\put(19,3){\circle{32}}
\put(19,3){\makebox(1,-2){$\ba{c}2\\0\ea$}}
\put(5.77,12){\circle*{4}}
\put(5.77,-6){\circle*{4}}
\put(5.77,3){\oval(28,18)[l]}
\put(-8.23,3){\circle*{4}}
\put(-16.23,3){\circle{16}}
\epi
+\frac{1}{3}
\rule[-18pt]{0pt}{42pt}
\bpi(54.78,0)(-14.78,0)
\put(19,3){\circle{32}}
\put(19,3){\makebox(1,-2){$\ba{c}2\\0\ea$}}
\put(5.22,11.4){\circle*{4}}
\put(10.87,17.1){\circle*{4}}
\put(5.22,-5.13){\circle*{4}}
\put(10.87,-10.78){\circle*{4}}
\put(10.87,3){\oval(41.3,27.88)[l]}
\put(5.22,3){\oval(30,16.53)[l]}
\put(-9.78,3){\circle*{4}}
\epi
\nn\\
&&
+
\rule[-18pt]{0pt}{42pt}
\bpi(96,0)(-2,0)
\put(19,3){\circle{32}}
\put(19,3){\makebox(1,-2){$\ba{c}2\\0\ea$}}
\put(32.23,12){\circle*{4}}
\put(32.23,-6){\circle*{4}}
\put(46,12){\circle*{4}}
\put(32.23,12){\line(1,0){13.77}}
\put(32.23,-6){\line(1,0){25.54}}
\put(57.8,12){\oval(23.6,8)[l]}
\put(57.8,16){\line(1,0){5.87}}
\put(63.67,16){\circle*{4}}
\put(57.8,8){\circle*{4}}
\put(73,3){\circle{32}}
\put(73,3){\makebox(1,-2){$\ba{c}1\\1\ea$}}
\put(59.77,-6){\circle{4}}
\epi
+\frac{1}{3}
\rule[-18pt]{0pt}{42pt}
\bpi(96,0)(-2,0)
\put(19,3){\circle{32}}
\put(19,3){\makebox(1,-2){$\ba{c}1\\1\ea$}}
\put(32.23,-6){\circle{4}}
\put(46,12){\circle*{4}}
\put(34.23,-6){\line(1,0){23.54}}
\put(34.2,12){\oval(23.6,8)[r]}
\put(34.2,16){\line(-1,0){5.87}}
\put(28.33,16){\circle*{4}}
\put(34.2,8){\circle*{4}}
\put(57.8,12){\oval(23.6,8)[l]}
\put(57.8,16){\line(1,0){5.87}}
\put(63.67,16){\circle*{4}}
\put(57.8,8){\circle*{4}}
\put(73,3){\circle{32}}
\put(73,3){\makebox(1,-2){$\ba{c}1\\1\ea$}}
\put(59.77,-6){\circle{4}}
\epi
\nn\\
&=&
\frac{1}{4}
\rule[-10pt]{0pt}{26pt}
\bpi(58,0)
\put(13,3){\circle{16}}
\put(21,3){\circle*{4}}
\put(29,3){\circle{16}}
\put(37,3){\circle*{4}}
\put(45,3){\circle{16}}
\epi
+\frac{1}{12}
\rule[-12pt]{0pt}{34pt}
\bpi(38,0)
\put(19,3){\circle{24}}
\put(19,3){\oval(24,8)}
\put(7,3){\circle*{4}}
\put(31,3){\circle*{4}}
\epi
+\frac{5}{8}
\rule[-16pt]{0pt}{36pt}
\bpi(38,0)
\put(19,3){\circle{24}}
\put(7,-9){\oval(24,24)[rt]}
\put(31,-9){\oval(24,24)[lt]}
\put(19,-9){\circle*{4}}
\put(7,3){\circle*{4}}
\put(31,3){\circle*{4}}
\epi
+\frac{5}{8}
\rule[-16pt]{0pt}{38pt}
\bpi(50,0)
\put(17,3){\circle{24}}
\put(17,-9){\circle*{4}}
\put(17,-9){\line(0,1){24}}
\put(17,15){\circle*{4}}
\put(29,3){\circle*{4}}
\put(37,3){\circle{16}}
\epi
+\frac{3}{8}
\rule[-16pt]{0pt}{40pt}
\bpi(50,0)
\put(13,3){\circle{16}}
\put(37,3){\circle{16}}
\put(13,-5){\circle*{4}}
\put(13,11){\circle*{4}}
\put(37,-5){\circle*{4}}
\put(37,11){\circle*{4}}
\put(25,-3){\oval(24,16)[b]}
\put(25,11){\oval(24,16)[t]}
\epi
+\frac{1}{4}
\rule[-14pt]{0pt}{36pt}
\bpi(34.78,0)(-3.39,0)
\put(15,3){\circle{24}}
\put(15,3){\circle*{4}}
\put(15,15){\circle*{4}}
\put(4.61,-3){\circle*{4}}
\put(25.39,-3){\circle*{4}}
\put(15,3){\line(0,1){12}}
\put(15,3){\line(5,-3){10}}
\put(15,3){\line(-5,-3){10}}
\epi
\nn\\
\eea
and therefore
\beq
\label{ga30}
\rule[-18pt]{0pt}{42pt}
\bpi(42,0)
\put(21,3){\circle{32}}
\put(21,3){\makebox(1,-2){$\ba{c}3\\0\ea$}}
\epi
=
\frac{1}{16}
\rule[-10pt]{0pt}{26pt}
\bpi(58,0)
\put(13,3){\circle{16}}
\put(21,3){\circle*{4}}
\put(29,3){\circle{16}}
\put(37,3){\circle*{4}}
\put(45,3){\circle{16}}
\epi
+\frac{1}{48}
\rule[-12pt]{0pt}{34pt}
\bpi(38,0)
\put(19,3){\circle{24}}
\put(19,3){\oval(24,8)}
\put(7,3){\circle*{4}}
\put(31,3){\circle*{4}}
\epi
+\frac{1}{8}
\rule[-16pt]{0pt}{36pt}
\bpi(38,0)
\put(19,3){\circle{24}}
\put(7,-9){\oval(24,24)[rt]}
\put(31,-9){\oval(24,24)[lt]}
\put(19,-9){\circle*{4}}
\put(7,3){\circle*{4}}
\put(31,3){\circle*{4}}
\epi
+\frac{1}{8}
\rule[-16pt]{0pt}{38pt}
\bpi(50,0)
\put(17,3){\circle{24}}
\put(17,-9){\circle*{4}}
\put(17,-9){\line(0,1){24}}
\put(17,15){\circle*{4}}
\put(29,3){\circle*{4}}
\put(37,3){\circle{16}}
\epi
+\frac{1}{16}
\rule[-16pt]{0pt}{40pt}
\bpi(50,0)
\put(13,3){\circle{16}}
\put(37,3){\circle{16}}
\put(13,-5){\circle*{4}}
\put(13,11){\circle*{4}}
\put(37,-5){\circle*{4}}
\put(37,11){\circle*{4}}
\put(25,-3){\oval(24,16)[b]}
\put(25,11){\oval(24,16)[t]}
\epi
+\frac{1}{24}
\rule[-14pt]{0pt}{36pt}
\bpi(34.78,0)(-3.39,0)
\put(15,3){\circle{24}}
\put(15,3){\circle*{4}}
\put(15,15){\circle*{4}}
\put(4.61,-3){\circle*{4}}
\put(25.39,-3){\circle*{4}}
\put(15,3){\line(0,1){12}}
\put(15,3){\line(5,-3){10}}
\put(15,3){\line(-5,-3){10}}
\epi.
\eeq

\end{document}